\documentclass[a4paper,10pt]{article}

\usepackage{theorem}
\usepackage{amsmath,amssymb}

\newtheorem{theorem}{Theorem}[section]
\newtheorem{proposition}[theorem]{Proposition}

\newtheorem{lemma}[theorem]{Lemma}
\numberwithin{equation}{section} \allowdisplaybreaks[1]

\allowdisplaybreaks

\newcount\driver
\newcount\bozza

scaled\magstep1 \font\msytw=msbm10 scaled\magstep1
\font\msytww=msbm8 scaled\magstep1 
scaled\magstep1

{\count255=\time\divide\count255 by 60
\xdef\hourmin{\number\count255}
        \multiply\count255 by-60\advance\count255 by\time
   \xdef\hourmin{\hourmin:\ifnum\count255<10 0\fi\the\count255}}

\let\a=\alpha \let\b=\beta    \let\g=\gamma     \let\d=\delta     \let\e=\varepsilon
\let\z=\zeta  \let\h=\eta     \let\th=\vartheta \let\k=\kappa     \let\l=\lambda
\let\m=\mu    \let\n=\nu      \let\x=\xi        \let\p=\pi        \let\r=\rho
\let\s=\sigma \let\t=\tau            \let\c=\chi
\let\ps=\psi   \let\o=\omega     
 \let\D=\Delta       \let\L=\Lambda    
             
\let\O=\Omega

\def\PP{{\cal P}}\def\EE{{\cal E}}\def\MM{{\cal M}}\def\VV{{\cal V}}
\def\CC{{\cal C}}\def\WW{{\cal W}}
\def\TT{{\cal T}}\def\NN{{\cal N}}\def\BB{{\cal B}}
\def\RR{{\cal R}}\def\LL{{\cal L}}
\def\DD{{\cal D}}\def\SS{{\cal S}}

\def\LLL{{\bf L}}

\def\RRR{\mathbb{R}} \def\ZZZ{\mathbb{Z}}  

       \def\CCC{\mathbb{C}} 

       \def\ZZZ{\hbox{\msytw Z}}
\def\zzzz{\hbox{\msytww Z}}       
\def\TTT{\mathbb{T}}  

\def\pp{{\bf p}}\def\xx{{\bf x}}
 
\def\yy{{\bf y}}\def\kk{{\bf k}}\def\nn{{\bf n}}
\def\zz{{\bf z}}\def\uu{{\bf u}}\def\vv{{\bf v}}

\def\bT{{\bf T}}\def\bP{{\bf P}}\def\bA{{\bf A}}

\def\oo{{\underline \omega}} \def\usp{{\underline s}} \def\ub{{\underline b}}
       \def\uo{{\underline \omega}}
\def\un{{\underline \nu}}          
          \def\ux{{\underline\xx}}
\def\uk{{\underline \kk}}          
\def\ual{{\underline \a}}

\def\uxi{{\underline \xi}}         
           \def\ue{{\underline \e}}
\def\uy{{\underline\yy}}           \def\uz{{\underline \zz}}
          \def\uo{{\underline \o}}

       \def\hv{{\widehat v}}

\def\hp{{\widehat \ps}}

       \def\hg{{\widehat g}}

\def\bk{{\bar \kk}}
\def\tv{{\tilde v}}

\def\be#1\ee{\begin{equation}#1\end{equation}}
\def\bsp#1\esp{\begin{split}#1\end{split}}
\def\bal#1\eal{\begin{align}#1\end{align}}
\def\bald#1\eald{\begin{aligned}#1\end{aligned}}
\def\ba#1#2\ea{\begin{array}{#1}#2\end{array}}

\def\bea{\begin{eqnarray}}   \def\eea{\end{eqnarray}}
\def\bean{\begin{eqnarray*}} \def\eean{\end{eqnarray*}}
\def\bfr{\begin{flushright}} \def\efr{\end{flushright}}
\def\bc{\begin{center}}      \def\ec{\end{center}}
\def\bd{\begin{description}} \def\ed{\end{description}}
\def\bv{\begin{verbatim}}

\def\lft{\left}                  \def\rgt{\right}
\def\la{{\langle}}               \def\ra{{\rangle}}

\def\erp{{\perp}}
\def\arp{{\parallel}}

\def\nn{\nonumber}
\def\Halmos{\hfill\vrule height10pt width4pt depth2pt \par\hbox to \hsize{}}
\def\pref#1{(\ref{#1})}
\def\virg{\;,\quad}
\def\qed{\raise1pt\hbox{\vrule height5pt width5pt depth0pt}}
\let\dpr=\partial

\let\bs=\backslash
\let\==\equiv

\let\io=\infty
\let\0=\noindent

\def\*{\vspace{.2cm}}

\def\tilde#1{{\widetilde #1}}

\def\Tr{\rm Tr}

\def\ins#1#2#3{\vbox to0pt{\kern-#2 \hbox{\kern#1 #3}\vss}\nointerlineskip}

\newdimen\xshift \newdimen\xwidth \newdimen\yshift
\newcount\griglia

\def\insertplot#1#2#3#4#5#6{%
\xwidth=#1pt \xshift=\hsize \advance\xshift by-\xwidth \divide\xshift by 2%
\begin{figure}[ht]
\vspace{#2pt} \hspace{\xshift}
\begin{minipage}{#1pt}
#3 \ifnum\driver=1 \griglia=#6
\ifnum\griglia=1 \openout13=griglia.ps \write13{gsave .2
setlinewidth} \write13{0 10 #1 {dup 0 moveto #2 lineto } for}
\write13{0 10 #2 {dup 0 exch moveto #1 exch lineto } for}
\write13{stroke} \write13{.5 setlinewidth} \write13{0 50 #1 {dup 0
moveto #2 lineto } for} \write13{0 50 #2 {dup 0 exch moveto #1
exch lineto } for} \write13{stroke grestore} \closeout13
\includegraphics{griglia.ps} \fi
\includegraphics{#4.ps}\fi%
\ifnum\driver=2 \fi
\end{minipage}
\caption{#5}
\end{figure}
}

\def\lb#1{\label{#1}}

\ifnum\bozza=1 
\usepackage[notref,notcite]{showkeys}\fi

\begin{document}

\title{Universality of one-dimensional Fermi systems, I.\\
Response functions and critical exponents.}

%

\author{G. Benfatto$^1$ \and P. Falco$^2$ \and V. Mastropietro$^3$}

\maketitle

\footnotetext[1]{\normalsize Dipartimento di Matematica, Universit\`{a} di Roma
``Tor Vergata", 00133 Roma, Italy.}
\footnotetext[2]{\normalsize Department
of Mathematics, California State university, Northridge, CA 91330.}
\footnotetext[3]{\normalsize Dipartimento di Matematica F.Enriquez,
Universit\`{a} di Milano, Via Saldini 50, Milano, Italy.}

\begin{abstract}  The critical behavior of one-dimensional interacting Fermi systems
is expected to display universality features, called Luttinger liquid
behavior. Critical exponents and certain thermodynamic quantities are
expected to be related among each others by model-independent formulas. We
establish such relations, the proof of which has represented a challenging
mathematical problem, for a general  model of spinning fermions on a one
dimensional lattice; interactions are short ranged and satisfy a positivity
condition which makes the model critical at zero temperature. Proofs are
reported in two papers: in the present one, we demonstrate that the zero
temperature response functions in the thermodynamic limit are Borel summable
and have anomalous power-law decay with multiplicative logarithmic
corrections. Critical exponents are expressed in terms of convergent
expansions and depend on all the model details. All results are valid for the
special case of the Hubbard model.
\end{abstract}


\section{Main Results}
\subsection{Introduction}\lb{sec1.1}

The charge carriers in metals are described by a gas of non relativistic
quantum particles (fermions). In the absence of interactions their
thermodynamic properties can be computed and provide a good understanding of
the physical properties of several systems. However, the free gas description
fails in many important cases and cannot explain phenomena, such as the
superconductivity, which are of the greatest importance both from the
applicative and theoretical point of view, providing a dramatic manifestation
of quantum physics at the macroscopic scales.

The analytic study of the properties of interacting fermions at zero
temperature and in the thermodynamic limit is an extremely difficult
task, and in several important cases even a convincing qualitative picture is
lacking. From the point of view of mathematical physics, to this date only in
two cases the ground state properties of a gas of weakly interacting fermions
in dimensions greater than one has been constructed with full mathematical
rigor by using Renormalization Group methods coming from Constructive Quantum
Field Theory: the case of non symmetric Fermi surface \cite{FKT04} and the
case of fermions on the honeycomb lattice at half-filling\cite{GM010_1}. Both
cases are rather special, as the interaction does not qualitatively modify
the physical properties even at zero temperature. The rigorous
study of an interacting model with a non trivial behavior in two or
three dimensions is still a challenging problem.

The situation is analytically more accessible for a one dimensional gas of
interacting fermions, where the interaction produces a number of remarkable
effects which are believed to have a counterpart even at higher dimensions,
in some special cases \cite{A097}. In addition to this role as a benchmark
for higher dimensions, the rapid progress of technology is producing
materials which are a physical realization of such systems. One dimensional
fermion gases have been extensively analyzed in the physical literature in
the last forty years by a variety of methods. Their behavior is radically
different with respect to the free gas, and the physical picture which is
commonly accepted is the so-called {\it Luttinger liquid conjecture} proposed
by Haldane \cite{Ha080} (extending previous ideas by Kadanoff \cite{KW071},
and Luther and Peschel \cite{LP075}): according to such conjecture, the low
energy properties at zero temperature of a wide class of interacting many
body fermion systems in one dimension are characterized by: a) {\it
anomalous dimensions}, that is the presence of critical model dependent
exponents in the correlations decay; b) {\it universality}, in the sense that
the exponents and other thermodynamic quantities verify a set of model
independent relations.

The universality property is particularly remarkable; in experiments we have
a poor knowledge of the microscopic parameters, but the validity of the
universal relations imply that one can predict exact and parameter-free
relations among exponents which could be experimentally measured. The
universal Luttinger liquid relations are verified in a special solvable spinless
models, the {\it Luttinger model}, which is the prototype of Luttinger liquid
behavior. Its exact solvability relies on the absence of the spin and on the
linear dispersion relation of the fermions, two features allowing for the mapping
in a model of free bosons by Mattis and Lieb \cite{ML065}. Non relativistic
fermions have a non linear dispersion relation, but Haldane \cite{Ha081}
provided arguments that, at least in some cases, the relations can be true
even if the mapping to free bosons is lost. The conjecture was partially
verified in a solvable model, the XYZ spin chain, which is equivalent to a
system of spinless fermions on the lattice with a nearest neighbor
interaction, whose ground state energy can be computed by the Bethe ansatz.
The solvability relies however on special and non generic peculiarity of
certain models, and traditional methods cannot say too much on the validity
of the Luttinger liquid relations in generic non solvable models. For
instance, Field theoretic Renormalization Group analysis \cite{So079}
confirms the existence of anomalous exponents and shows that the
contributions from the non-linear part of the dispersion relation is {\it
irrelevant} in the Renormalization Group sense; however such irrelevant
terms, which contribute to the exponents, are simply discarded in this
approach so that nothing can be concluded on the validity of the universal
relations. In recent times indeed a caveat for a not too extensive
application of the Luttinger liquid picture has been emerged; in particular,
it appeared that non linear bands surely affect the finite temperature and
the dynamical properties, see e.g. \cite{PWA09}. In addition, the possibility
of {\it different} physical properties, at least for the finite temperature
and the dynamical properties, between integrable and non integrable 1D models
has been extensively investigated, especially regarding the conduction
properties.

All the above considerations surely provide a strong motivation for a
mathematical proof of the Luttinger liquid relations, and we will provide
here such a proof for a standard (generically non solvable) model of a gas of
spinning fermions on a one dimensional lattice with a short range interaction
satisfying a {\em positivity condition}, to be defined later. We will call
such system {\it extended Hubbard model}, as it reduces to the (solvable)
Hubbard model in the special case of ultralocal interaction. As it will clear
by our analysis, the proof will be independent from the details of the model
considered, and it could be generalized to a wider class of systems. However,
for definiteness and sake of simplicity, we will not try to consider the most
general class of models. We use non-perturbative Renormalization Group
methods implemented with Ward Identities at each Renormalization Group step,
using a technique introduced in \cite{BM002,BM005}. Such methods have
provided for the first time the self-consistent construction (that is,
without resorting to properties found by exact solutions as was done in
previous works \cite{BGPS994}) of an interacting non solvable many body model
with a non trivial behavior (that is, where the interaction produces a
different behavior with respect to the free case); namely a system of weakly
interacting spinless fermions in one dimension. Subsequently, by such methods
the Luttinger liquid relations \cite{BM010,BFM010} for this {\it spinless}
case were proven.

The analysis of the {\it spinning} case, which is discussed in the
present work, present considerable new difficulties. Indeed the
fact that the inclusion of the spin in one dimensional physics
produces new phenomena, such as the  spin charge separation,
logarithmic corrections and the possibility of metal-insulator
transitions, is well known in the physical literature, see e.g.
\cite{A097,Gi004,Ts004}. The approximations leading to the
solvable Luttinger model in the spinless case, namely the
linearization of the dispersion relation, in the spinning case
lead to a {\it non solvable model}. Power law decay with  anomalous
exponents are found only for {\it repulsive} interactions and in
the non half filled band case; besides
the power law decay has multiplicative logarithmic corrections. Despite such
features, we can establish for the first time the validity of a
number of universal Luttinger liquid relations connecting the
exponents and other thermodynamical quantities in a generic non
solvable model of 1D spinning fermions on a lattice.

The proof is split in two papers. In the present one we present the
Renormalization Group construction of the model, which allows us to analyze
the asymptotic behavior of the correlations, to prove the existence of
critical exponents and logarithmic corrections and to establish their Borel
summability,
assuming the validity of a property called {\em asymptotic
vanishing of the Beta function}. The exponents and the other physical
quantities are expressed by sophisticated expansions, and while the validity of the
universal relations can be checked at lowest order, a direct verification at
all orders from the expansions look essentially impossible. Therefore, in the
subsequent paper \cite{BFM012_2}, we introduce an effective model verifying a
several extra symmetries (which are only asymptotic in the lattice model);
by fine tuning of its parameters one can show that its exponents are the same
as in the original model, and on the other hand such symmetries imply Ward
Identities, from which the asymptotic vanishing of the Beta function and the
universal relations can be derived. This method is a way to implement the
concept of {\it emerging symmetries} in a rigorous mathematical setting.

\subsection{Extended Hubbard Model and Physical Observables}

The Hamiltonian of a standard model of spinning fermions on a one
dimensional lattice (also called extended Hubbard model) is
\be\lb{1.1} H=-{1\over 2}\sum_{x\in \CC\atop s=\pm}
(a^+_{x,s}a^-_{x+1,s}+a^+_{x,s}a^-_{x-1,s}) + \bar\m \sum_{{x\in \CC}\atop{s=\pm }}
a^+_{x,s}a^-_{x,s} +\l \sum_{x,y\in \CC\atop s,s'=\pm }
v_L(x-y)a^+_{x,s}a^-_{x,s} a^+_{y,s'}a^-_{y,s'} \ee
where
\begin{enumerate}
\item $\CC=\{-[L/2] \le x \le [(L-1)/2]\}$ is a one dimensional lattice
    of step $1$ and $L$ sites;
\item $a^\pm_{x,s}$ are fermion creation and annihilation
operators at site $x$ with spin $s$, verifying
\be \{a^+_{x,s},a^-_{x',s'}\}=\d_{x,x'}\d_{x,x'}\quad\quad
\{a^+_{x,s},a^+_{x',s'}\}=\{a^+_{x,s},a^+_{x',s'}\}=0 \ee
and such that $a^\pm_{-[L/2],s}=a^\pm_{-[(L-1)/2]+1,s}$ (periodic
boundary conditions);
\item $v_L(x)$ is a function on $\ZZZ$, periodic of period $L$, such that
    $v_L(x)=v(x)$ for $x \in \CC$, $v(x)$ being an even function on
    $\ZZZ$ satisfying the short range condition $|v(x)|\le C e^{-\k
    |x|}$;
\item $-\bar\m \in (-1,+1)$ is the {\it chemical potential}.
\end{enumerate}

The results of this paper are only valid under the following condition on the
potential $v(x)$, that we call the {\em positivity condition}:
\be\lb{pos_cond} \l\,\hat v(2\,\arccos(\bar\m)) \ge 0\ee

The model is $SU(2)$ symmetric, as the Hamiltonian is invariant under the
transformation $a^\pm_{x,s}\to \sum_{s'}M_{s,s'}a^\pm_{x,s'}$, with $M\in
SU(2)$, and includes the standard (exactly solvable, \cite{LW068}) and the
U-V Hubbard models, corresponding to the interactions $\l v(x-y)=U \d_{x,y}$
and $\l v(x-y)=U \d_{x,y} + \frac12 V\d_{|x-y|,1}$, respectively: in the
former case the positive condition is $U\ge 0$.

We consider the operators $a_{\xx,s}^{\pm}=e^{x_0 H}
a_x^{\pm}e^{-H x_0}$, with
\be \xx=(x,x_0)\;,\quad 0\le x_0 < \b\ \ee
for some $\b>0$ ($\b^{-1}$ is the temperature); on $x_0$
antiperiodic boundary conditions are imposed, that is, if
$a_{\xx,s}^{\pm}=a_{x,x_0,s}^{\pm}$, then
$a_{x,\b,s}^{\pm}=-a_{x,0,s}^{\pm}$. Defining
\be\lb{trc} \la \cdot\ra_{L,\b} := {{\rm Tr}[e^{-\b H}\cdot]\over
{\rm Tr}[e^{-\b H}]} \ee
and $\la \cdot\ra_{L,\b}^T$ the corresponding truncated expectation,
the energy of the thermal ground state is
\be\lb{fe} E(\l):=-\lim_{\b\to \io}\lim_{L\to \io} (L\b)^{-1}\log
{\rm Tr}[e^{-\b H}]\;, \ee
The {\it Schwinger functions} are defined as
\be\lb{1.6} S_n^{\b,L}(\xx_1,\e_1,s_1;...;\xx_{n},\e_n,s_n)=\la {\bf
T}\{a^{\e_1}_{\xx_1,s_1}\cdots a_{\xx_n,s_n}^{\e_1} \}\ra_{\b,L}^T
\ee
where ${\bf T}$ is the operator of time ordering, acting on a
product of fermion fields as:
\be\lb{1.4} {\bf T}( a^{\e_1}_{\xx_1,s_1}... a^{\e_n}_{\xx_n,s_n})=
(-1)^\p a^{\e_{\p(1)}}_{\xx_{\p(1)},\s_{\p(1)}}...
a^{\e_{\p(n)}}_{\xx_{\p(n)},s_{\p(n)}}\ee
where $\p$ is a permutation of $\{1,\ldots,n\}$, chosen in such a way that
$x_{\p(1)0}\ge\cdots\ge x_{\p(n)0}$, and $(-1)^\p$ is its sign. [If some of
the time coordinates are equal each other, the arbitrariness of the
definition is solved by ordering each set of operators with the same time
coordinate so that creation operators precede the annihilation operators.]
Note that $S_n^{\b,L}$ is $L$-periodic in each $x_i$, $\b$-antiperiodic in
$x_{0,i}$ and is identically zero if $\sum_{i=1}^n\e_i\neq 0$.

We will introduce also the densities $\r^\a_{\xx}$:
\bal \r^C_{\xx}&= \sum_{s=\pm }a^+_{\xx,s}a^-_{\xx,s} &{\rm
(charge\
density)}\nn\\
\r^{S_i}_{\xx}&= \sum_{s,s'=\pm } a^+_{\xx,s}\s^{(i)}_{s,s'}
a^-_{\xx,s'}
&{\rm (spin\ densities)}\nn\\
\lb{rho}\nn\\[-30pt]
\\
\r^{SC}_{\xx}&= \frac12 \sum_{s=\pm \atop \e=\pm} s\, a^\e_{\xx,s}
a^\e_{\xx,-s} & {\rm (singlet\ Cooper\ density)}\nn\\
\r^{TC_i}_ \xx&= \frac12 \sum_{s,s'=\pm \atop \e=\pm} a^\e_{\xx,s}
\tilde \s^{(i)}_{s,s'} a^\e_{\xx+{\bf e},s'}\virg {\bf e}=(1,0)
&{\rm (triplet\ Cooper\ densities)}\nn \eal
where $i=1,2,3$ and
\bal \s^{(1)} &= \begin{pmatrix} 0 &1\\ 1& 0 \end{pmatrix}
&\s^{(2)} &= \begin{pmatrix} 0 &-i\\ i& 0 \end{pmatrix}
&\s^{(3)} &= \begin{pmatrix} 1 &0\\ 0& -1 \end{pmatrix}\nn\\
\tilde\s^{(1)} &= \begin{pmatrix} 1 &0\\ 0& 0 \end{pmatrix}
&\tilde\s^{(2)} &= \begin{pmatrix} 0 &1\\ 1& 0 \end{pmatrix}
&\tilde\s^{(3)} &= \begin{pmatrix} 0 &0\\ 0& 1 \end{pmatrix}\nn
\eal
%
%
%
The {\it response functions} are defined by the following truncated
correlations:
\be \O_{\a,\b,L}(\xx-\yy):= \la {\bf T}\r^\a_{\xx}
\r^\a_{\yy}\ra^T_{\b,L}:=\la {\bf T}\r^\a_{\xx} \r^\a_{\yy}\ra_{\b,L}- \la
\r^\a_{\xx}\ra_{\b,L} \la\r^\a_{\yy}\ra_{\b,L}\lb{ss111} \ee
%
%
where, if $O_\xx$ is quadratic in the fermion operators, ${\bf
T} O_{\xx}O_{\yy}=O_{\xx} O_{\yy}$ if $x_0\ge y_0$ and
$O_{\yy}O_{\xx}$ if $x_0\le y_0$.
If $\xx-\yy=(\x,\t)$, the response functions are defined in
$(-L,L)\times [-\b,\b]$ and are $\b$-periodic in $\t$ and
$L$-periodic in $\x$. If $F_{\b,L}$ is any function of this type,
we define its Fourier transform as
\be \hat F_{\b,L}(\pp)= \int_{-{\b\over 2}}^{\b\over 2}
dx_0\sum_{x\in\CC} e^{i\pp \xx}\;F_{\b,L}(\xx)\ee
where $\pp=(p, p_0)$, with  $p= {2\pi \over L}n$, $-[L/2]\le n\le [(L-1)/2]$
and $p_0\in {2 \pi\over \b}\ZZZ$.

%
%
%
%

In the following we will be interested in the zero temperature limit of the
Schwinger functions and response functions, calculated in the thermodynamic
limit. We shall denote these functions by the same symbols, without the $\b$
and $L$ labels; for example, we shall write: $\lim_{\b\to\io}\lim_{L\to\io}
\hat\O_{L,\b,\a}(\pp)\equiv \hat\O_{\a}(\pp)$. Note that the thermodynamic
limit $L\to\io$ is taken before the zero temperature limit $\b\to\io$; this
allows us to derive properties of the {\it thermal ground state}. To shorten
the notation, in the following we shall use the definition
\be \lim_{\b,L\to\io} \= \lim_{\b\to\io}
\lim_{L\to\io}\label{def}\ee

%
%
%
%

\subsection{The non interacting case}

In absence of interaction, the Hamiltonian is
\be\lb{1.1aa}  H_0=-{1\over 2}\sum_{x\in \CC\atop s=\pm}
(a^+_{x,s}a^-_{x+1,s}+a^+_{x,s}a^-_{x-1,s}) + \m \sum_{x\in
\CC\atop s=\pm 1} a^+_{x,s}a^-_{x,s}\ee
Being $H_0$ quadratic, the $2n$-point (not truncated) correlation functions
of the $a^{\pm}_{\xx,s}$ operators satisfy the Wick rule, i.e.
\bea&& \la {\bf T}\{a_{\xx_1,s_1}^-\cdots a_{\xx_n,s_n}^-
a_{\yy_1,s_1'}^+\cdots a_{\yy_n,s_n'}^+\}\ra_{\b,L}
=\det G\;,\nn\\
&& G_{ij}=\d_{s_i,s_j'}\la{\bf T}\{a^-_{\xx_i,s_i}
a^+_{\yy_j,s'_j}\}\ra_{\b,L} \;. \lb{A.2}\eea
Therefore, all the $n$--point Schwinger function
$S_n^{\b,L}(\xx_1,\e_1,s_1;\ldots;\xx_n,\e_n,\s_n)$ (truncated by definition)
are identically zero for any $n>2$,
and, in order to construct the whole set of response functions, it is
enough to compute the $2$--point function $g^{\b,L}(\xx-\yy)=\la{\bf
T}\{a^-_{\xx,s} a^+_{\yy,s}\}\ra_{\b,L}$, which is equal to
\be\lb{defo}
\bsp &g^{\b,L}(\xx-\yy)= {{\rm Tr} \left[e^{-\b H_0} {\bf T} (
a^-_\xx a^+_\yy)\right] \over {\rm Tr} [e^{-\b H_0}]} =
{1\over L} \sum_{k\in {\cal D}_L} e^{-ik(x-y)} \hat g^{\b,L}(k,x_0-y_0)=\\
&={1\over L} \sum_{k\in {\cal D}_L} e^{-ik(x-y)} \left\{ {e^{-(x_0-y_0) e(k)}
\over 1+e^{-\b e(k)}}{\bf I}(x_0-y_0>0) - {e^{-(\b+x_0-y_0) e(k)} \over
1+e^{-\b e(k)}}{\bf I}(x_0-y_0\le 0) \right\} \esp\ee
where ${\bf I}(t)$ is the indicator function, ${\cal D}_L = \{(2\p n)/L, n\in\ZZZ\}$ and
\be e(k)=\m-\cos k\ee
The function $\hat g^{\b,L}(k,\t)$ is defined only for $-\b<\t< \b$, but we can
extend it periodically over the whole real axis. This periodic
extension is smooth in $\t$ for
$\t\not= n\b, n\in \ZZZ$, but has a jump discontinuity at $\t=n\b$
equal to $(-1)^n$. It follows that $g^{\b,L}(x,x_0)$ is smooth in $x_0$ for
$x_0\not= n\b, n\in \ZZZ$, with a jump discontinuity at $x_0=n\b$
equal to $(-1)^n \d_{x,y}$; hence, it is discontinuous only at $\xx=(0,n\b)$

The function $\hat g^{\b,L}(k,\t)$ is antiperiodic in $\t$ of period $\b$; hence
its Fourier series is of the form
\be
\hat g^{\b,L}(k,\t)={1\over\b}\sum_{k_0={2\pi\over\b}(n_0+{1\over 2})}
\hat g^{\b,L}(k_0,k)e^{-i k_0 \t}\ee
with
\be \hat g^{\b,L}(\kk)=\int_0^\b d\t e^{i \t k_0}{e^{-\t e(k)} \over 1+e^{-\b
e(k)}}= {1\over -i k_0+e(k)} \ee
It is a classical result that, because of the jump discontinuities, this series is not absolutely
convergent; however, if we call $g_N^{\b,L}(k,\t)$ the sum over the terms
with $|k_0|\le N$, $g_N^{\b,L}(k,\t)$ is pointwise convergent
and the limit is given by $\hat g^{\b,L}(k,\t)$
at the continuity points, while at the discontinuities it is given by
the mean of the right and left limits. Hence, if $\xx-\yy\not= (0,n\b)$, we can write
\be\lb{limN} g^{\b,L}(\xx-\yy) = \lim_{N\to \io} \frac1{\b L} \sum_{k\in {\cal D}_{L,\b},|k_0|\le N}
\frac{e^{-i \kk(\xx-\yy)}}{-i k_0+e(k)}\ee
with ${\cal D}_{L,\b}:={\cal D}_L \times {\cal D}_\b$, ${\cal
D}_L:=\frac{2\p}{L}\CC$, ${\cal
D}_\b:=\frac{2\p}{\b}(\zzzz+\frac12)$.

It is convenient, for reasons that will appear clear below, to
slightly modify the representation \pref{limN} in the following
way. Let us take a smooth even compact support function $\chi_0(t)$,
equal to $1$ for $|t|<1$ and equal to $0$ if $|t|\ge \g$, for a
given {\sl scaling parameter} $\g>1$, fixed throughout the paper.
In App. \ref{app0} we prove that \pref{limN} is completely
equivalent to the representation
\be g^{\b,L}(\xx-\yy)=\lim_{M\to\io}{1\over \b L}\sum_{\kk\in
D_{\b,L}}\chi_0(\g^{-M}k_0){e^{-i\kk(\xx-\yy)}\over -i k_0+e(k)} \lb{ch} \ee
In particular, the above equality is not true for $\xx-\yy=(0,n\b)$, where
the propagator is equal, to $g^{L,\b}(0,0^-) \to_{\b,L\to\io} -p_F/\p$  while
the r.h.s. is equal to
\be\lb{1.26}{ g^{\b,L}(0,0^+)+ g^{\b,L}(0,0^-)\over 2}\to_{\b,L\to\io}
-\frac{p_F}{\p}+{1\over 2}\ee
where $p_F=\cos^{-1}\m$ is the {\it Fermi momentum}. The Fermi
momentum appears in the period of the oscillations of the large
distance behavior of the propagator; for $|\xx|$ large,
\be\lb{free1} \lim_{\b,L\to\io} g^{\b,L} (\xx)\equiv g(\xx)\sim
\sum_{\o=\pm } \frac{e^{-i\o p_F x}} {v_F x_0+i\o x}\virg
v_F\=\sin p_F \ee
where $\sim$ means up to faster decaying terms; $v_F$ is usually called the
{\em Fermi velocity}.

\subsection{The interacting case }
\lb{sec1.4} The first step of our construction consists in
computing the large distance behavior at zero temperature and in
the thermodynamic limit of the two-points Schwinger function and
of the response functions, proving the presence of anomalous
critical exponents and logarithmic corrections.

\begin{theorem}\lb{th1.1}
Let us consider the Schwinger and response functions, \pref{1.6} and
\pref{ss111}, with Hamiltonian \pref{1.1}. If $\bar\m \not=0$ and $\hat
v(2\arccos(\bar\m))>0$, there exists $\l_0>0$ such that, if $0\le \l \le
\l_0$, it is possible to find a continuous function $p_F\equiv
p_F(\bar\m,\l)=\arccos(\bar\m)+O(\l)$ verifying the conditions
\be\lb{ma} p_F \not=0,\p/2,\p \virg \hat v(2 p_F) > 0\ee
such that, setting $v_F=\sin p_F$ and defining
\be \bsp \tilde\xx:=(x,v_F x_0)&\virg L(\xx):=1+ b\l\hv(2\bar p_F)
\log |\xx| \virg
b=2(\pi \sin p_F)^{-1}\\
\bar\O_0(\xx) := \frac{x_0^2 - x^2}{x_0^2+ x^2}&\virg \bar
S_0(\xx):=\frac{v_F x_0\cos p_F -  x\sin p_F} {|\tilde\xx|} \esp
\ee
in the limit $\b,L\to\io$ \pref{def}, the large $|\xx|$ asymptotic
behavior of the two-points Schwinger function $S_2(\xx) \=
S_2(\xx,-,s; {\bf 0},+,s)$ is of the form
\be\lb{s2x} S_2(\xx) \sim \lft[\bar S_0(\xx) + R_2(\xx)\rgt]
{L(\xx)^{\z_z}\over |\tilde \xx|^{1+\h}} \ee
where $R_2(\xx)$ is a continuous function of $\l$ and $\xx$, such that, for
any $\th<1$ and a suitable positive constants $C_\th$, $|R_2(\xx)|\le C_\th
\l^{1-\th}$; the sign $\sim$ means up to terms bounded by $C|\xx|^{-1-\th}$.
Moreover, the large $|\xx|$ asymptotic behavior of the correlations is of the
form
\bal {\rm for}\quad \a=C,S_i\qquad &\O_{\a}(\xx)\sim
{\bar\O_0(\tilde\xx) + R_\a(\xx)\over \p^2|\tilde\xx|^2} + \cos[2
p_F x] {L(\xx)^{\z_\a}\over
\pi^2|\tilde\xx|^{2 X_\a}} \lft[1 + \tilde R_\a(\xx)\rgt]\nn\\
{\rm for}\quad \a=SC \qquad &\O_{\a}(\xx) \sim - \lft[
\bar\O_0(\tilde\xx) + \tilde R_\a(\xx) \rgt] \cos(2 p_F x)
{L(\xx)^{\tilde \z_\a} \over \pi^2|\tilde\xx|^{2 \tilde X_\a}}  -
{1\over \pi^2}{L(\xx)^{\z_\a} \over
|\tilde\xx|^{2 X_\a}} \lft[1 + R_\a(\xx)\rgt]\nn\\
\lb{asymp} {\rm for}\quad \a=TC_i \qquad &\O_{\a}(\xx) \sim -
{v_F^2\over \pi^2}{L(\xx)^{\z_\a}\over |\tilde\xx|^{2 X_\a}}
\lft[1 + R_\a(\xx)\rgt] \eal
with the functions $R_\a(\xx)$ and $\tilde R_\a(\xx)$ having the
same properties of $R_2(\xx)$; the sign $\sim$ means up to terms
bounded by $C|\xx|^{-2-\th}$.

The critical exponents $\h$ and $X_\a$ are continuous functions of $\l$, such
that $\h(0)=X_\a(0)-1=0$ and $\h/\l^2>0$, while the exponents $\tilde\z_{SC}$
and $\z_\a$ of the logarithmic corrections could also depend on $\xx$ (we can
not exclude it), but satisfy the bounds $|\tilde \z_{SC}|\le C\l$ and
$|\z_\a-\bar\z_\a|\le C\l$, for a suitable constant $C$, with
\be\lb{zalfa} \bar\z_z = 0 \virg \bar\z_C=-{3\over 2}\virg
\bar\z_{S_i}={1\over 2}\virg \bar\z_{SC}=-{3\over 2}\virg
\bar\z_{TC_i}= \frac12\ee

Finally, given $\d\in (0,\p/2)$, there exists $\e\=\e(\d)>0$, such
that the free energy, the two-points Schwinger functions and the
density correlations are analytic in the set
\be\lb{dom} D_{\e,\d}=\{\l\in\CCC: 0<|\l|< \e, |\text{Arg } \l|< \p-\d \} \ee
continuous in the closure $\bar D_{\e,\d}$ and Borel summable in $\l=0$.
\end{theorem}
This Theorem will be proved in \S\ref{sec2.5}. It is completely based on
the multi-scale analysis of the Grassmannian functional representation of the
model, which is discussed in \S\ref{sec2}. In this analysis
we choose, for technical reasons, to fix the Fermi momentum $p_F$ of the
interacting model by adding to the chemical potential a finite counterterm
$\n(\l,p_F)$, which is uniquely determined by the condition that the
multi-scale expansion is well defined; in \S\ref{sec2.5aa} we prove that the
relation between $p_F$ and $\n$ can be inverted, so determining the function
$p_F(\bar\m,\l)$.

\medskip
\0{\bf Remarks.}
\begin{enumerate}
\item If $\bar\m=0$, a different behavior is expected, as proved in
    \cite{LW068} for the (exactly solvable) Hubbard model.

\item In the free $\l=0$ case the response functions decay for large
    distance with power law of exponent equal to $2$. The interaction
    partially removes such degeneracy by producing {\it anomalous
    exponents} which are (in general) non trivial functions of the
    coupling.

\item While the presence of non universal exponents in the model
    \pref{1.1} is a common feature with the Luttinger model, both in the
    spinless \cite{ML065} and spinning case \cite{Ma064}, the presence of
    {\it logarithmic corrections} is a striking difference. Such
    corrections remove the degeneracy in the response of charge and spin
    densities, present in the spinning Luttinger model.

\item The exponents of the non oscillating part of charge or spin density
    correlations are the same as in the free case; also logarithmic
    corrections are excluded.

\item In the Luttinger model the exponents, as function of the coupling,
    are analytic in a complex disk around $\l=0$, both in the spinless
    and spinning case. This property is valid also for the a general
    spinless model with short range interaction \cite{BGPS994,BM001}, but
    in the present spinning case the perturbative expansion in $\l=0$ is
    only Borel summable.

\item Our analysis could be extended to the generic $2l$-point Schwinger
    function, by using the same strategy used in \S2.3 of \cite{BFM007}
    to analyze the corresponding tree expansion in the case of the
    Thirring model.
\end{enumerate}

\section{RG  Analysis for the extended-Hubbard Model}\lb{sec2}

\subsection{Functional integral representation}\lb{ss2.5a}

The analysis of the Hubbard model correlations is done by a rigorous
implementation of the RG techniques. To begin with, we need a {\it functional
integral representation} of the model, because the RG techniques are
optimized for that.

We find convenient (even if not necessary) to fix the value of the
singularities of two-point function Fourier transform $\hat S^{\b,L}_2(\kk)$
(that is, of the Fermi momentum $p_F$) by writing the chemical potential
$\bar\m$ in \pref{1.1} in the form
\be\label{ss} \bar\m=\m+\n^{\b,L}(\l) \virg \m=\cos p_{F,L}\ee
where $p_{F,L} ={2\pi\over L}(n_F+{1\over 2})$, with $n_F=[(p_F L)/(2\p)]$;
then we show that it is possible to choose $\n^{\b,L}(\l)$, uniquely up to
corrections of order $\min \{L^{-1},\b^{-1}\}$, so that the interacting Fermi
momentum is indeed $p_F$, in the limit $\b,L\to\io$ \pref{def}. Our results
can be translated in the form of Theorem \ref{th1.1}, because we can show
that the equation \pref{ss} has a unique solution $p_F=p_F(\bar\m,\l)$ in a
right interval of $\l=0$, small enough (how small depending on $p_F$).

The choice, at finite $L$, of $p_{F,L}$ in place of $p_F$ is motivated by
technical reasons, see \S\ref{sec2.4a} below; this choice does not affect the
infinite volume limit, since it changes $\m$ for terms of order $1/L$ and
$\n^{\b,L}(\l)$ is defined up to terms of the same order.

The main object we will study is the functional $\WW(J,\h)$ (depending on
$M$, $L$ and $\b$), defined by
\bal\lb{1z} \WW(J,\h) = -L\b e_C + &r_C\int d\xx J^{C}_\xx + \log \int
P(d\psi) \exp \Big\{ -\VV^{(M)}(\psi) + \BB^{(M)}(\psi,J,\h) \Big\}\\
\lb{1zv} -\VV^{(M)}(\psi) &= -\VV(\psi)- \n\NN(\psi) - \n_C\NN(\psi)\\
\lb{1zb} \BB^{(M)}(\psi,J,\h) &=  \sum_\a \int d\xx J^{\a}_\xx \r^{\a}_\xx  +
\sum_s\int d\xx [\h^+_{\xx,s} \psi^-_{\xx,s} + \ps^+_{\xx,s}
\h^-_{\xx,s}]\eal
where $\psi^\pm_{\xx,s}$ and $\h^\pm_{\xx,s}$ are Grassmann variables and the
fermion density operators $\r^{\a}_\xx$ are defined as in \pref{rho}, with
$\psi^\pm_{\xx,s}$ in place of $a^\pm_{\xx,s}$, $J^{\a}_\xx$ are commuting
variables, $\int d\xx$ is a short form for $\sum_{x\in\CC}\int_{-\b/2}^{\b/2}
dx_0$, $P(d\psi)$ is a Grassmann-valued  Gaussian measure in the field variables
$\psi^\pm_{\xx,s}$ with covariance (the free propagator) given by
\bal
&\int P(d\ps)\; \ps^\e_{\xx,s}\ps^\e_{\yy,s'}=0\;,
\qquad \int P(d\ps)\; \ps^-_{\xx,s}\ps^+_{\yy,-s}=0\;, \cr
&\bar g_{\b,L,M}(\xx-\yy) := \int P(d\ps)\; \ps^-_{\xx,s}\ps^+_{\yy,s}=
{1\over\b L}\sum_{\kk\in\DD_{L,\b}} {\chi_0(\g^{-M} k_0)
e^{-i\kk(\xx-\yy)} \over -i k_0+ (\cos p_{F,L} -\cos k)}\;.
\eal
In the above formulae, $\chi_0(t)$ and $\DD_{L,\b}$ are defined as in \pref{ch},
\be \VV(\psi)=\l \sum_{s,s'=\pm}\int d\xx d\yy\;
\psi_{\xx,s}^+\psi_{\xx,s}^-
 v(\xx-\yy) \psi_{\yy,s'}^+
\psi_{\yy,s'}^- \ee
with $v(\xx-\yy)=\d(x_0-y_0)v_L(x-y)$,
\be \NN(\ps)=
\sum_{s=\pm}\int d\xx\; \psi_{\xx,s}^+\psi_{\xx,s}^- \ee
and
\be\lb{lll} \bsp \n_C &= 2\l \hat v_L(0) r_C := \l \hat v_L(0)
[g^{\b,L}(0,0^+)-g^{\b,L}(0,0^-)] \\ -e_C &:= -\l \hat v_L(0) r_C^2 +\n
r_C\esp \ee
Note that, while the presence in the interaction of the term $\n\NN(\ps)$ is
needed, as explained above, to fix the Fermi momentum of the measure, the
terms $\n_C\NN(\ps)$ and $r_C \int d\xx J^C_\xx$ and the constant $e_C$ have
the role to correct the value of the free propagator at the discontinuity
points, in the limit $M=\io$, where this correction is important. To better
explain this point, let us define the {\em free energy at finite $M$} as
$E^{M,\b,L}= \log W(0,0)$, the {\em Schwinger functions at finite $M$}
as
\be\lb{f2}
S_n^{M,\b,L}(\xx_1,s_1,\e_1;....;\xx_n,s_n,\e_n)
={\partial^n\over\partial\h^{-\e_1}_{\xx_1,s_1}...
\partial\h^{-\e_n}_{\xx_n,s_n}} \WW(J,\h)\Big|_{0,0}\\
\ee
and the {\em response functions at finite $M$} as
\be\lb{f2r} \O^M_{\a,\b,L}(\xx-\yy) ={\partial^2\over\partial
J^\a_\xx \; \dpr J^\a_\yy} \WW(J,\h)\Big|_{0,0}\\
\ee
and recall that one can express their perturbative expansion in terms of
connected Feynmann graphs. Each Feynmann graph $G$ is defined by a set of
{\em internal points} $\uy=(\yy_1, \ldots, \yy_n)$, associated with one
of the three terms in \pref{1zv}, a set of {\em external points} $\ux
=(\xx_1, \ldots, \xx_m)$, associated with one of the three terms in
\pref{1zb}, and a set of lines $l=(\uu_l,\zz_l)$, with $\uu_l,\zz_l\in
\ux\cup\uy$, and has a value proportional to an integral of the form
$\int d\uy \prod_{l\in G} \bar g^{\b,L,M}(\uu_l-\zz_l)$. The same claim is
true for the perturbative expansion of the Schwinger functions $S_n^{\b,L}$,
defined in \pref{1.6}, with the only difference that one has to substitute
everywhere $\bar g^{\b,L,M}$ with $g^{\b,L}$, defined as in \pref{defo}. Now,
the possibility to study our model in terms of the functional $W(J,\h)$ is of
course related with the fact that the perturbative expansions coincide for
$M\to\io$. This would be trivial if no Feynmann graph had a {\em tadpole},
that is a line with $\uu_l=\zz_l$, or a line connecting two coinciding
external points. In fact, one can see easily that, for any graph $G$,
$$\lim_{N\to\io} \int d\uy \prod_{l\in G} \bar g^{\b,L,M}(\uu_l-\zz_l)
= \int d\uy \prod_{l\in g} \bar g^{\b,L}(\uu_l-\zz_l)$$
with $\bar g^{\b,L}(\xx)$ defined as in \pref{ch}, hence equal to
$g^{\b,L}(\xx)$ for $\xx\not=(0,n\b)$; it follows that, if the graph $G$ has
no tadpole and there are no coinciding external points, we can substitute
everywhere $\bar g^{\b,L}$ with $g^{\b,L}$, by changing the integrand in a
set of zero measure. Note that the lines connecting two external points can
be present only in the graphs of the response functions and in the trivial
graph connecting two $\h$ fields with a free propagator; in any case, let us
suppose, from now on, that there are no coinciding points. Hence,
there is a problem only if there are tadpoles and, in such case, their
contribution is a constant $\prod_{l\; \text{tadpole}} \bar g^{\b,L}(0,0^-)$,
which is different from $\prod_{l\; \text{tadpole}} g^{\b,L}(0,0^-)$.

Note now that, if we consider
the graphs contributing to the Schwinger functions (those with at least two external
lines), any tadpole can only be obtained by contracting the two fields based on one of
the two vertices of a $\l$ term, while the other two fields are contracted
with two other fields based on two other (possibly coinciding) vertices;
hence, the presence of a tadpole implies that in the value of $G$ there is
a factor of the form
$$\bar g^{\b,L}(\xx_1-\xx)\, (2 \n_T)\, \bar g^{\b,L}(\xx-\xx_2) \virg
2\n_T := -\l\hat v_L(0) [g^{L,\b}(0,0^+)+g^{L,\b}(0,0^-)]$$
where we used \pref{1.26} and the fact that there are two ways to choose the
couple of fields contracted in the tadpole. On the other hand, given a graph
$G$ of this type, there is another graph $\tilde G$, which differs from it
only because, in place of the term $\VV(\psi)$ which produced the tadpole,
there is a vertex $\n_C\NN(\psi)$. If we sum the values of $G$ and $\tilde
G$, we get a number which is equal to the value of $G$, with $2\n_T+\n_C=-2
\l\hat v_L(0) g^{L,\b}(0,0^-)$ in place of $2\n_T$. By iterating this
argument, we see that the sum over all the graph can be rewritten as the sum
over the graph obtained by putting $\n_C=0$ and $\bar g^{\b,L}(\xx_l-\yy_l)
=g^{\b,L}(\xx_l-\yy_l)$ everywhere.

The previous procedure is not sufficient to ``correct'' completely the
perturbative expansion of the free energy. In fact, in this case there is a
graph of first order in $\n$, whose value is $\n
[g^{L,\b}(0,0^+)+g^{L,\b}(0,0^-)]/2$, and two graphs of first order in $\l$,
one with a $\l$-vertex and two tadpoles, whose value is $-(\l\hat
v_L(0)/4)[g^{L,\b}(0,0^+)+g^{L,\b}(0,0^-)]^2$, the other with a $\n_C$ vertex
and one tadpole, whose value is $(\n_C/2)[g^{L,\b}(0,0^+)+g^{L,\b}(0,0^-)]$.
Their sum is different from the correct value $\n g^{L,\b}(0,0^-)-\l\hat
v_L(0)[g^{L,\b}(0,0^-)]^2$, but the difference is compensated by the constant
$e_C$.

As concerns the functional derivatives containing at least one derivative
with respect to the external fields $J^\a_\xx$, the only graph which is not
``corrected '' by the counterterm $\n_C \NN(\psi)$, is the graph with one
vertex $J$ and no $\l$ or $\n$ vertex. This graph has a value different from
$0$ only if $\a=C$ and, in that case, is corrected by the term $r_C \int d\xx
J^C_\xx$.

Another important remark is that, for $M$ finite, the integrand in the r.h.s.
of \pref{1z} can be seen as a polynomial in the Grassmann variables $\hat\ps^+_{\kk,s}$,
defined as the Fourier transform of the field $\ps^+_{\xx,s}$:
\be\lb{2.7}
\ps^+_{\xx,s} = \frac{1}{L\b} \sum_{\kk\in\DD_{L,\b}} e^{-i\kk\xx}\, \hat\ps^+_{\kk,s} \ee
In fact, thanks to the {\em ultraviolet cutoff} (UV cutoff) on $k_0$, only a finite set of the variables
$\hat\ps^+_{\kk,s}$, those such that $\chi_0(\g^{-M}k_0)\not=0$, may give a contribution to the Grassmann
integral, and these variables are anticommuting. Hence, the structure of the interaction implies
that the integral is a polynomial in $\l$ and $\n$ and that the functions
$S_n^{M,\b,L}(\xx_1,s_1,\e_1;....;\xx_n,s_n,\e_n)$ are analytic in $\l$ and $\n$
at least in a small set around $\l=\n=0$.

We can now prove that the Grassmann integral \pref{1z} can be used to
compute the thermodynamical properties of the model with
Hamiltonian \pref{1.1}. This follows from the following
proposition.

\begin{proposition}\lb{p2.1}
Assume that, for any finite $\b$ and $L$, there is a function $\n^{\b,L}(\l)$
such that $\n^{\b,L}(0)=0$ and both $\n^{\b,L}(\l)$ and the Schwinger functions
at finite $M$ $S_n^{M,\b,L}$, see \pref{f2}, with $\n=\n^{\b,L}(\l)$,
see \pref{lll}, are analytic and bounded in
\be\lb{2.12a}
D=\{\l, |\l|< c\e_0 \max \{(\log\b)^{-1}, (\log
L)^{-1}\}\} \bigcup \{|\l| < \e_0, |\arg \l|<{\pi\over 2}+\d\}\ee
with $c, \e_0>0$, $0<\d<\pi/2$ independent of $\b$ and $L$, and that
they are uniformly convergent as $M\to\io$. Then, if $\l\in D$,
\be\lb{trace} S_n^{\b,L}(\xx_1,s_1,\e_1;....;\xx_n,s_n,\e_n) =\lim_{M\to\io}
S_n^{M,\b,L}(\xx_1,s_1,\e_1;....;\xx_n,s_n,\e_n)\;; \ee
where $S_n^{\b,L}$ is defined as in \pref{1.6}, with $H$ given by \pref{1.1}
with $\bar\m=\m+\n^{\b,L}(\l)$.

A similar statement is true for the thermal ground state energy
and the response functions.
\end{proposition}
\0 {\bf Proof} - The main point, strictly related with the fact that we are
treating a fermion problem, is that, for $L$ and $\b$ finite,
$S_n^{\b,L}$ is the ratio of the traces of two
matrices whose coefficients are entire functions of $\l$ and $\n$, hence it
is the ratio of two entire functions of $\l$ and $\n$.
On the other hand, the hypotheses on $\n^{\b,L}(\l)$ and $S_n^{M,\b,L}$ and the
Weierstrass theorem imply that $\n^{\b,L}(\l)$ and $\lim_{M\to\io} S_n^{M,\b,L}$
are analytic in $D$. It follows, in particular, that
$S_n^{\b,L}$, calculated with $\n=\n^{\b,L}(\l)$,
is the ratio of two functions analytic in $D$; hence, it may have a
singularity in a point $\l_0\in D$ only if $\Tr[e^{-\b H}]$ vanishes there, which certainly
does not happen in a neighborhood of $\l=0$ small enough (how small
possibly depending on $L,\b$), since $\n(\l)$ is of order $\l$.
Moreover, also the r.h.s. of \pref{trace} is
analytic in a small neighborhood of $\l=0$ and, as we have explained
above, its power expansion in $\l$ and $\n$, hence also its power
expansion in $\l$ for $\n=\n^{\b,L}(\l)$ coincide with that of $S_n^{\b,L}$;
hence, the two functions coincide in
a disk $\tilde D_{L,\b}$ with center in $\l=0$ and radius $\e_{\b,L}$
possibly vanishing as $\b,L\to\io$.
However, $S_n^{\b,L}$, being the ratio of two functions analytic in $D$,
may have only isolated poles in $D\backslash \tilde D_{L,\b}$; hence, if
$E$ is the set of poles, $S_n^{\b,L}$ is analytic in
$D\backslash E$ and necessarily coincide with the r.h.s. of \pref{trace}
in this set, since the two functions coincide in $\tilde D_{L,\b}\subset
D\bs E$. It follows that, if $E$ were not empty, $S_n^{\b,L}$ would be
unbounded in $D\bs E$, while this is not of course true for the other function.

A similar argument can be used for the response
functions and the thermal ground state energy.\Halmos

\vspace{.3cm}

The RG analysis will allow us to prove that the analyticity domain of
the r.h.s. of \pref{trace} is indeed of the form $D$ and this
allows us to extend this result to all the physical quantities
studied in this paper and to prove all results described before.

The proof of Theorem 1.1 is done in two steps; first we write $\bar\m=\m+\n$,
$\m=\cos p_{F,L}$ and we show that it is possible to choose $\n(\m,\l)$ so
that the expansions are convergent in the zero temperature and infinite
volume limit, if $\l\in D$; the second step is to prove that $|\dpr
\n(\m,\l)/\dpr\m| \le C\e_0$ in $D$, so that, if $C\e_0\le 1/2$, the equation
\pref{ss} can be uniquely solved with respect to $p_F$ and the solution is of
the form $p_F=\arccos(\bar\m)+O(\l)$ (with $p_F$ real for $\l$ real
positive), that is the interacting Fermi momentum is a well defined function
of the parameters in the Hamiltonian, as expected.

\subsection{The ultraviolet integration}\lb{sec2.2b}


In the following, to simplify the notation, we shall in general drop the superscripts
$M$, $\b$ and $L$. Moreover, we shall denote
$\TTT$ the one dimensional torus $[0,2\p]$, $\|k-k'\|_\TTT$
the usual distance between $k$ and $k'$ in $\TTT $ and
$\|k\|_\TTT=\|k-0\|_\TTT$. Analogously $\|\xx-\yy\|$ will denote
the distance on the the space-time $\CC \times [-\b,\b]$, with periodic
boundary conditions.

We introduce a positive function
$\c(\kk') \in C^{\io}(\TTT \times \RRR)$, $\kk'=(k',k_0)$,
such that $\c(\kk') = \c(-\kk')$ and $\c(\kk') = 1$, if $|\kk'| <t_0 = a_0
v_F/\g$, and $\c(\kk')=0$ if $|\kk'|>a_0 v_F$, where $v_F=\sin p_F$, $a_0=
\min \{{p_F\over 2}, {\p- p_F\over 2}\}$ and
$|\kk'|=\sqrt{k_0^2+v^2_F \|k'\|_{\TTT }^2}$. The above definition
is such that the supports of $\c(k-p_F,k_0)$ and $\c(k+p_F,k_0)$
are disjoint and the $C^\io$ function on $\TTT \times R$
\be \lb{f1} \hat f_{u.v.} (\kk) := 1- \c(k-p_F,k_0) -
\c(k+p_F,k_0) \ee
is equal  to $0$, if $v_F^2\|\big[|k|-p_F\big]\|_{\TTT }^2
+k_0^2<t_0^2$. We want to apply this identity with $k\in\DD_L$; hence
$k'=k\pm p_F \in \DD'_L =\frac{2\p}{L}(\CC+\frac12)$, since $p_F=\frac{2\p}{L}(n_F+\frac12)$.
It follows that, if $\DD'_{L,\b}=\DD'_L \times \DD_\b$, we can write the fermion
propagator in the following way:
\be\lb{gdefaa0} g(\xx-\yy) = g^{(u.v)}(\xx-\yy) + \sum_{\o=\pm}
e^{-i\o p_F(x-y)} g^{(i.r.)}_\o(\xx-\yy) \ee
\bal g^{(u.v.)}(\xx-\yy) &= {1\over\b L}
\sum_{\kk\in\DD_{L,\b}} e^{-i\kk(\xx-\yy)}
{\hat f_{u.v.}(\kk)\chi_0(\g^{-M} k_0)\over -i k_0 + (\cos p_F -\cos k)}\\
g^{(i.r.)}_\o(\xx-\yy) &= {1\over\b L} \sum_{\kk'\in\DD'_{L,\b}}
e^{-i\kk'(\xx-\yy)} {\chi(k',k_0)\over -i k_0+ E_\o(k')} \eal
\be \lb{dE} E_\o(k') = \o v_F \sin k' + \cos p_F (1-\cos k') \ee
The properties of Grassmann integration imply the following identity for the
functional integral in the r.h.s. of \pref{1z}:
\be\lb{2z} \bsp e^{\WW(J,\h)} &= e^{-L\b e_C + r_C\int d\xx J^{C}_\xx} \int
P(d\psi^{(i.r.)}) \int P(d\psi^{(u.v.)})\, e^{ -\VV^{(M)}(\psi) +
\BB^{(M)}(\psi,J,\h)}=\\
&=e^{-L\b E_0 + \SS_0(J,\h)}\; \int P(d\psi^{(i.r.)}) \,
e^{-\VV^{(0)}(\psi^{(i.r.)})+\BB^{(0)}(\psi^{(i.r.)},J,\h)} \esp \ee
where $\psi^\pm_{\xx,s}=\sum_\o e^{\pm i\o p_F x}
\psi_{\xx,\o,s}^{(i.r.)}+\psi^{(u.v.)}_{\xx,s}$ and $P(d\psi^{(u.v.)})$ and
$P(d\psi^{(i.r.)})$ are the Grassmann gaussian integrations with propagator
$g^{(u.v.)}(\xx)$ and $g^{(i.r.)}(\xx)$ respectively; moreover
\be\lb{2.14} \bsp &-L\b E_0 + \SS_0(J,\h) - \VV^{(0)}(\psi^{(i.r.)}) +
\BB^{(0)}(\psi^{(i.r.)},J,\h)=\\
&=\sum_{n\ge 1}{1 \over n!}\EE^T_{u.v.} \Bigg(
-\VV^{(M)}(\psi^{(i.r.)}+\psi^{(u.v.)}) +
\BB^{(M)}(\psi^{(i.r.)}+\psi^{(u.v.)},J,\h);n \Bigg) \esp\ee
where $\VV^{(0)}(\psi)$ is fixed by the condition $\VV^{(0)}(0)=0$ and, given
a function $F(\psi)$ on the Grassmann algebra, which is a polynomial in the
variables $\hat\psi_{\kk,s}$ (see remark around \pref{2.7}, the {\it
truncated expectation} $\EE^T_{u.v.} [F(\psi^{(i.r.)}+\psi^{(u.v.)})]$ is a
polynomial in the variables $\psi^{(i.r.)}_{\kk,s}$, defined as
\be \EE^T_{u.v.} [F(\psi^{(i.r.)}+\psi^{(u.v.)})]=\frac{\dpr^n}{\dpr\l^n} \log\int
P(d\psi^{(u.v.)})e^{\l F(\psi^{(i.r.)}+\psi^{(u.v.)})]}
\Big|_{\l=0}\lb{2.15}\ee
We will see that, if we put $\ux=(\xx_1,\ldots,\xx_{2l})$,
$\oo=(\o_1,\ldots,\o_{2l})$, $\usp=(s_1,\ldots,s_{2l})$ and
$\psi_{\ux,\oo,\usp} = \prod_{i=1}^l$ $\psi^+_{\xx_i,\o_i,s_i}
\prod_{i=l+1}^{2l} \psi^-_{\xx_i,\o_i,s_i}$, the {\it effective potential}
$\VV^{(0)}(\ps)$ can be represented as
\be\lb{2.17b} \VV^{(0)}(\psi^{(i.r.)})= \sum_{l\ge 1} \sum_{\oo,\usp} \int
d\ux\; W^{(0)}_{\usp, 2l}(\ux) e^{i p_F \sum_{i=1}^{2l} \e_i\o_i x_i}
\psi^{(i.r.)}_{\ux,\oo,\usp} \ee
where $W^{(0)}_{\usp, 2l}(\ux)=\sum_{\z\in A} W_\z(\ux)$, with $A$ a finite
set, and $W_\z(\ux) = F_\z(\ux) \prod_{(i,j)\in L_\z} \d(\xx_i-\xx_j)$, where
$F_\z(\ux)$ is a smooth function and $L_\z$ is a subset of the couples
$(i,j)$. In the following we shall use the notation
\be\lb{L1_norm} \int d\ux |W^{(0)}_{\usp, 2l}(\ux)| := \sum_{\z\in A} \int
d\ux |F_\z(\ux)| \prod_{(i,j)\in L_\z} \d(\xx_i-\xx_j)\ee

A similar representation can be written for the functional
$\BB^{(0)}(\psi^{(i.r.)},J,\h)$ (containing all terms which are of order
greater than $0$ both in the external fields $J,\h$ and in $\psi$), for
$\SS^{(0)}(J,\h)$ (containing only the terms without $\psi$ external fields),
and $L\b E_0$ (containing the terms without external fields). In all cases,
the corresponding kernels are called $W^{(0)}_{\ual,\ue,\usp,
m_\psi,m_J,m_\h}(\ux)$, with $m_\psi$, $m_J$, $m_\h$ the number of $\psi$
fields, $J$ fields, $\h$ fields, respectively,
$\ux=(\xx_1,\ldots,\xx_{m_\psi+m_J+m_\h})$, $\ual=(\a_1, \ldots, \a_{m_J})$
(the $\a$ indices of the $J^\a$ fields), $\ue=(\e_1,\ldots,\e_{m_\psi+m_\h})$
(the $\e$ indices of the $\psi^\e$ and $\h^\e$ fields in a fixed arbitrary
order); note that $m_\psi + m_\h$ has to be even, hence we shall also define
$m_\psi + m_\h=2l$. We shall also use the notation $W^{(0)}_{\ual,\ue,\usp,
2l,0,0}(\ux) = W^{(0)}_{\usp, 2l}(\ux)$. Moreover we shall define the Fourier
transform $\hat W^{(0)}_{\ual,\ue,\usp, m_\psi,m_J,m_\h}(\uk)$, $\uk=(\kk_1,
\ldots,\kk_{m_\psi+m_J+m_\h-1}$), so that, if $m^*=m_\psi+m_J+m_\h$
\be W^{(0)}_{\ual,\ue,\usp, m_\psi,m_J,m_\h} = \frac1{(L\b)^{\bar
m-1}}\sum_{\uk} e^{i\sum_{j=1}^{m^*-1} \e_j\kk_j(\xx_j-\xx_{m^*})} \hat
W^{(0)}_{\ual,\ue,\usp, m_\psi,m_J,m_\h}(\uk) \ee
where, if $m_J>0$, $\e_j=+1$ for the indices corresponding to the $J$ fields.

\begin{lemma}\lb{p2.2} The constant $E_{0}$ and the kernels
$W^{(0)}_{\ual,\ue,\usp, m_\psi,m_J,m_\h}$ are given by power series in $\l$
and $\n$, convergent for $|\l|,|\n|\le \e_0$, for $\e_0$ small enough and
independent of $\b,L,M$. They satisfy the following bounds:
\be |E_0|\le C\e_0\virg \int d\ux \big|W^{(0)}_{\ual,\ue,\usp,
m_\psi,m_J,m_\h}(\ux)\big| \le \b L C^{l+m} \e_0^{k_{l,m}}\;,\lb{2.17}\ee
for some constant $C>0$ and $k_{l,m}=\max\{1,l-1\}$, if $m_J+m_\h=0$,
otherwise $k_{l,m}=\max\{0,l-1\}$.

Moreover, $\lim_{M\to\io}E_{0}$ and $\lim_{M\to\io} \hat
W^{(0)}_{\ual,\ue,\usp, m_\psi,m_J,m_\h}(\uk)$ do exist and are reached
uniformly in $M$, so that, in particular, the limiting functions are analytic
in the same domain.
\end{lemma}

The proof of Lemma \ref{p2.2} is quite standard, but we present it here with
some details, as this will allow us to introduce in a simple case a number of
techniques and concepts we will use throughout the paper. Note that the proof
could be generalized without any problem to the multi-dimensional Hubbard
model.

\subsection{Proof of Lemma \ref{p2.2}}\lb{sec2.2c}

We start writing
\be g^{(u.v.)}(\xx)=\sum_{h=1}^{M} g^{(h)}(\xx)\;,\lb{B.1}\ee
where
\be g^{(h)}(\xx)={1\over \b L}\sum_{\kk\in\DD_{\b,L}}
e^{-i\kk\xx} {\hat f_{u.v.}(\kk)\, H_h(k_0)\over -i k_0 + (\cos p_F -\cos k)}
\lb{B.2}\ee
with $H_1(k_0)=\c_0(\g^{-1}|k_0|)$ and, if $h\ge 2$, $H_h(k_0)=
\c_0(\g^{-h}|k_0|)- \c_0(\g^{-h+1}|k_0|)$. We shall use also the notation
$g^{[h_1,h_2]} := \sum_{h=h_1}^{h_2} g^{(h)}$.

Note that, for any integer $K\ge 0$, $g^{(h)}(\xx)$ satisfies the bound
\be |g^{(h)}(\xx)|\le {C_K\over 1+ (\g^h |x_0|_\b + |x|_L)^K}
\;,\lb{B.3}\ee
where $|\cdot|_\b$ is the distance on the one dimensional torus of
size $\b$ and $|\cdot|_L$ is the distance on the periodic lattice of
size $L$. Moreover, $g^{(h)}(\xx)$ admits a Gram representation:
$g^{(h)}(\xx-\yy)=\int d\zz\, A_h^*(\xx-\zz)\cdot B_h(\yy-\zz)$,
with
\bea A_h(\xx)&=&{1\over \b L}\sum_{\kk\in\DD_{\b,L}} \sqrt{f_{u.v.}(\kk)
H_h(k_0)}\frac{e^{i\kk\xx}}{ k_0^2+(\cos p_F-\cos k)^2}\;,
\nn\\
B_h(\xx)&=&{1\over \b L}\sum_{\kk\in\DD_{\b,L}} \sqrt{f_{u.v.}(\kk)
H_h(k_0)}\,e^{i\kk\xx}(ik_0+\cos p_F-\cos k)\lb{B.4a}\eea
and
\be ||A_h||^2=\int d\zz |A_h(\zz)|^2\le C\g^{-3h}\;,\quad\quad
||B_h||^2\le C \g^{3h}\;,\lb{B.5}\ee
for a suitable constant $C$. Moreover
\be \int d\xx |g^{(h)}(\xx)|\le C\g^{-h} \lb{cc1}\ee

The decomposition of the UV propagator \pref{B.1} allows us to make the
decomposition of the measure $P(d\psi^{(u.v.)})=\prod_{h=1}^M P(d\psi^{(h)})$
and the corresponding decomposition of the field $\psi^{(u.v.)}_{\xx,s}
=\sum_{h=1}^M \psi^{(h)}_{\xx,s}$. Hence, we can integrate iteratively the
fields $\psi^{(M)},\psi^{(M-1)},...,\psi^{(h)}$ with $h\ge 1$ and, if we
define $\psi^{(\le 0)} = \psi^{i.r.}$ and $\psi^{(\le h)} = \psi^{i.r.} +
\sum_{j=1}^{h} \psi^{(j)}$, if $h\ge 0$, we get:
\be\lb{2za} e^{\WW(J,\h)} =e^{-L\b E_h + \SS_h(J,\h)}\; \int P(d\psi^{\le h})
\, e^{-\VV^{(h)}(\psi^{(\le h)})+\BB^{(h)}(\psi^{(\le h)},J,\h)} \ee
This definition agrees with \pref{1z}, if we put
\be\lb{1zM} E_M=e_C \virg \SS_M(J,\h)=r_C\int d\xx J^C_\xx\ee

Let us consider first the {\em effective potentials on scale $h$},
$\VV^{(h)}(\psi^{(\le h)})$ and $\BB^{(h)}(\psi^{(\le h)},J,\h)$. We want to
show that they can be expressed as sums of terms, each one associated to an
element of a family of labeled trees; we shall call this expansion {\em the
tree expansion}. The tree definition can be followed looking at Fig.
\ref{h2}.

\insertplot{300}{150}{ \ins{30pt}{85pt}{$r$}
\ins{50pt}{85pt}{$v_0$} \ins{130pt}{100pt}{$v$}
\ins{35pt}{-5pt}{$h$} \ins{52pt}{-5pt}{$h+1$}
\ins{135pt}{-5pt}{$h_{v}$} \ins{235pt}{-5pt}{$M$}
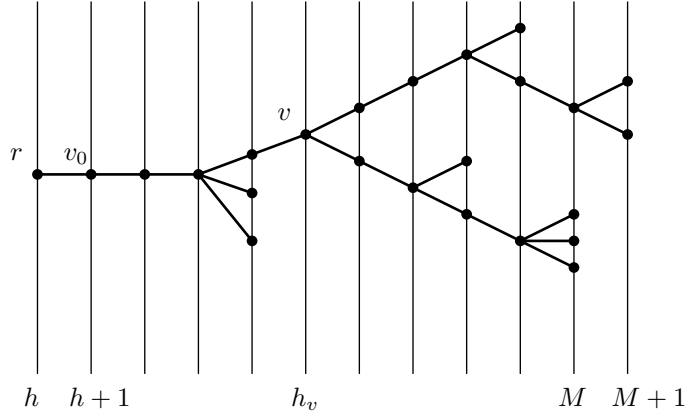
\ins{255pt}{-5pt}{$M+1$}}{fig50}{A tree appearing in
the tree expansion of $\scriptstyle \VV^{(h)\lb{h2}}$}{0}

Let us consider the family of all trees which can be constructed by joining a
point $r$, the {\it root}, with an ordered set of $\bar n\ge 1$ points, the
{\it endpoints} of the {\it unlabeled tree}, so that $r$ is not a branching
point. $\bar n$ will be called the {\it order} of the unlabeled tree and the
branching points will be called the {\it non trivial vertices}. The unlabeled
trees are partially ordered from the root to the endpoints in the natural
way; we shall use the symbol $<$ to denote the partial order. Two unlabeled
trees are identified if they can be superposed by a suitable continuous
deformation, so that the endpoints with the same index coincide. It is then
easy to see that the number of unlabeled trees with $\bar n$ end-points is
bounded by $4^{\bar n}$. We shall also consider the set $\TT_{M,h,n,m}$ of
the {\it labeled trees} with $n+m$ endpoints (to be called simply trees in
the following); they are defined by associating some labels with the
unlabeled trees, as explained in the following items.

\0 1) We associate a label $V$, $J$ or $\h$ to each endpoint, so that there
are $n$ endpoints with label $V$, to be called {\em normal endpoints}, and
$m=m_J+m_\h$ endpoints, $m_J$ with label $J$ and $m_\h$ with label $\h$, to
be called {\em special endpoints}. We shall also call $\TT_{M,h,n,m_J,m_\h}$
the family of trees with fixed values od $m_J$ and $m_\h$.

\0 2) We associate a label $h\le M$ with the root. Moreover, we introduce a
family of vertical lines, labeled by an integer taking values in $[h,M+1]$,
and we represent any tree $\t\in\TT_{M,h,n,m}$ so that, if $v$ is an endpoint
or a non trivial vertex, it is contained in a vertical line with index
$h_v>h$, to be called the {\it scale} of $v$, while the root $r$ is on the
line with index $h$. In general, the tree will intersect the vertical lines
in set of points different from the root, the endpoints and the branching
points; these points will be called {\it trivial vertices}. The set of the
{\it vertices} will be the union of the endpoints, of the trivial vertices
and of the non trivial vertices; note that the root is not a vertex. Every
vertex $v$ of a tree will be associated to its scale label $h_v$, defined, as
above, as the label of the vertical line whom $v$ belongs to. Note that, if
$v_1$ and $v_2$ are two vertices and $v_1<v_2$, then $h_{v_1}<h_{v_2}$.

\0 3) There is only one vertex immediately following the root, which will be
denoted $v_0$; its scale is $h+1$. If $v_0$ is an endpoint, the tree is
called the {\em trivial tree}; this can happen only if $n+m=1$.

\0 4) Given a vertex $v$ of $\t\in\TT_{M,h,n,m}$ that is not an endpoint, we
can consider the subtrees of $\t$ with root $v$, which correspond to the
connected components of the restriction of $\t$ to the vertices $w\ge v$; the
number of endpoint of these subtrees will be called $n_v$. If a subtree with
root $v$ contains only $v$ and one endpoint on scale $h_v+1$, it will be
called a {\it trivial subtree}.

\0 5) Given an end-point, the vertex $v$ preceding it is surely a non trivial
vertex, if $n+m>1$.

\vspace{.3cm}

Our expansion is build by associating a value to any tree $\t\in
\TT_{M,h,n,m}$ in the following way.

First of all, given a normal endpoint $v\in\t$ with $h_v=M+1$, we associate
to it one of the three terms contributing to the potential $\VV^{(M)}(\psi)$
in \pref{1zv}, that is $-\VV(\psi^{(\le M)})$, $-\n \NN(\psi^{(\le M)})$ or
$-\n_C \NN(\psi^{(\le M)})$, while, if $h_v\le M$, we associate to it one of
the four terms appearing in the following expression:
\be\lb{sssa}
\bsp -\VV(\psi^{(< h_v)}) &-\n \NN(\psi^{(< h_v)})
-2\l \sum_{s}\int d\xx d\yy v(\xx-\yy)g^{[h_v,M]}(\xx-\yy) \psi^{+(<
h_v)}_{\xx,s}\psi^{-(< h_v)}_{\yy,s}+\\
&+ \left(-\n_C +2\l
\hat v(0) g^{[h_v,M]}({\bf 0})\right) \sum_{s} \int d\xx \psi^{+(<
h_v)}_{\xx,s}\psi^{-(< h_v)}_{\xx,s}
\esp\ee
If $v$ is a special endpoint, we associate to it one of the terms
contributing to the potential $\BB^{(M)}(\psi,J,\h)$ in \pref{1zb}, with
$\psi^{(< h_v)}$ in place of $\psi$.

All these possible choices will be distinguished by a label $a$ in a set
$A_\t$, depending on $\t$. Moreover, for any $a\in A_\t$, we introduce a {\it
field label} $f$ to distinguish the field variables appearing in the
different terms associated to the endpoints and a {\em source label} $\a_v$ for each
special endpoint; the set of field labels
associated with the endpoint $v$ will be called $I_v$.
Analogously, if $v$
is not an endpoint, we shall call $I_v$ the set of field labels associated
with the endpoints following the vertex $v$ and $S_v$ the set of special endpoints
following $v$; $\xx(f)$, $\e(f)$, $s(f)$ and
$\o(f)$ will denote the space-time point, the $\e$ index, the $s$ index and
the $\o$ index, respectively, of the Grassmann field variable with label $f$.

The previous definitions imply that, if $0\le h< M$, the following iterative
equations are satisfied:
\be -\VV^{(h)}(\psi^{(\le h)}) + \BB^{(h)}(\psi^{(\le h)},J,\h)  -\b L e_h +
s_h(J,\h)= \sum_{n=1}^\io \sum_{\t\in\TT_{M,h,n,m}\atop a\in A_\t}
\bar\VV_J^{(h)}(\t,a,\psi^{(\le h)})\;,\lb{B.12}\ee
where, if $v_0$ is the first vertex of $\t$ and $\t_1,\ldots,\t_s$, $s\ge 1$,
are the subtrees with root in $v_0$,
\be \bar\VV_J^{(h)}(\t,a,\psi^{(\le h)})={(-1)^{s+1}\over s!} \EE^T_{h+1}
\big[\bar\VV_J^{(h+1)}(\t_1,a_1,\psi^{(\le h+1)}); \ldots;\bar\VV_J^{(h+1)}
(\t_{s},a_s,\psi^{(\le h+1)})\big]\;,\lb{B.13}\ee
where $\bar\VV_J^{(h+1)}(\t_i,\psi^{(\le h+1)})$ is equal to
$\bar\VV_J^{(h+1)}(\t_i,\psi^{(\le h+1)})$ if the subtree $\t_i$ contains
more than one end-point, otherwise it is given by one of the terms
contributing to the potentials in \pref{1zv} or \pref{1zb}, if $h_v=N+1$, or
one of the addends in \pref{sssa}, if $h_v\le M$, the choice depending on the
label $a$.

The identity \pref{B.12} implies, in particular, that the constant $E_{h}$
and the functional $\SS_h(J)$ defined in \pref{2za} are given by
\be\lb{Eh} E_h=\sum_{j=h}^M e_j \virg \SS_h(J,\h)= \sum_{j=h}^M s_h(J,\h)\ee
with $E_M=e_M$ and $\SS_M(J,\h)=s_M(J,\h)$ given by \pref{1zM}.

Note that
\be {1\over L\b} \int d\xx d\yy
\big| v(\xx-\yy)g^{[h,M]}(\xx-\yy) \big|\le C\g^{-h}\ee
and
\be \left|-\n_C +2\l \hat v(0) g^{[h,M]}({\bf 0})\right| \le C|\l|\ee
with a constant $C$ independent of $M$ and $h$.

The above definitions imply, in particular, that, if $n+m>1$ and $v$ is not
an endpoint, then $n(v)>1$, with $n(v)$ denoting the number of endpoints
following $v$ on $\t$; in fact the vertex preceding an end-point is
necessarily non trivial, if $n+m>1$.

Using its inductive definition, the right hand side of \pref{B.12}
can be further expanded, and in order to describe the resulting
expansion we need some more definitions.

We associate with any vertex $v$ of the tree a subset $P_v$ of $I_v$, the
{\it external fields} of $v$, and the set $\xx_v$ of all space-time points
associated with one of the end-points following $v$; moreover, we shall
denote $\xx^J_v\subset \xx_v$ and $\xx^\h_v\subset \xx_v$ the set of all
space time points associated with the special endpoints following $v$ of type
$J$ and $\h$, respectively. The subsets $P_v$ must satisfy various
constraints. First of all, $|P_v|\ge 2$, if $v>v_0$; moreover, if $v$ is not
an endpoint and $v_1,\ldots,v_{s_v}$ are the $s_v\ge 1$ vertices immediately
following it, then $P_v \subseteq \cup_i P_{v_i}$; if $v$ is an endpoint,
$P_v=I_v$. If $v$ is not an endpoint, we shall denote by $Q_{v_i}$ the
intersection of $P_v$ and $P_{v_i}$; this definition implies that $P_v=\cup_i
Q_{v_i}$. The union ${\cal I}_v$ of the subsets $P_{v_i}\setminus Q_{v_i}$
is, by definition, the set of the {\it internal fields} of $v$, and is non
empty if $s_v>1$. Given $\t\in\TT_{M,h,n,m}$, there are many possible choices
of the subsets $P_v$, $v\in\t$, compatible with all the constraints. We shall
denote ${\cal P}_\t$ the family of all these choices and ${\bf P}$ the
elements of ${\cal P}_\t$.

With these definitions, we can rewrite $\bar\VV_J^{(h)}(\t,a,\Psi^{(\le
h)})$ in the r.h.s. of \pref{B.12} as:
\bea &&\bar\VV_J^{(h)}(\t,a,\Psi^{(\le h)})=\sum_{{\bf P}\in{\cal
P}_\t} \bar\VV_J^{(h)}(\t,a,{\bf P})\;,\nn\\
&&\bar\VV_J^{(h)}(\t,a,{\bf P})=\int d\xx_{v_0} \widetilde\Psi^{(\le
h)}(P_{v_0}) \tilde J(S_{v_0}) \tilde \h(H_{v_0}) K_{\t,{\bf
P}}^{(h+1)}(\xx_{v_0})\;,\lb{2.43a}\eea
where $S_v$ and $H_v$ denote the set of endpoints of type $J$ and $\h$,
respectively, following $v$ and
\be \widetilde\psi^{(\le h)} (P_{v})=\prod_{f\in P_v}\psi^{ (\le
h)\e(f)}_{\xx(f),s(f)} \virg \tilde J(S_v) = \prod_{v\in S_v}
J^{\a_v}_{\xx_v}\virg \tilde \h(H_v) = \prod_{v\in H_v} \h^{\e_v}_{\xx_v}
\lb{2.44}\ee
and $K_{\t,{\bf P}}^{(h+1)}(\xx_{v_0})$ is defined inductively by the
equation, valid for any $v\in\t$ which is not an endpoint,
\be K_{\t,{\bf P}}^{(h_v)}(\xx_v)={1\over s_v !} \prod_{i=1}^{s_v}
[K^{(h_v+1)}_{v_i}(\xx_{v_i})]\; \;\EE^T_{h_v}[
\widetilde\psi^{(h_v)}(P_{v_1}\setminus Q_{v_1}),\ldots,
\widetilde\psi^{(h_v)}(P_{v_{s_v}}\setminus
Q_{v_{s_v}})]\;,\lb{2.45}\ee
where $\widetilde\Psi^{(h_v)}(P_{v_i}\setminus Q_{v_i})$ has a definition
similar to (\ref{2.44}). Moreover, if $v_i$ is an endpoint,
$K^{(h_v+1)}_{v_i}(\xx_{v_i})$ is equal to the kernel of one of the terms
contributing to the potential in \pref{1zv} or \pref{1zb}, if $h_{v_i}=N+1$,
or one of the four terms in \pref{sssa}, if $h_{v_i}\le N$; if $v_i$
is not an endpoint, $K_{v_i}^{(h_v+1)}=K_{\t_i,{\bf P}_i}^{(h_v+1)}$, where
${\bf P}_i=\{P_w, w\in\t_i\}$.

In order to get the final form of our expansion, we need a convenient
representation for the truncated expectation in the r.h.s. of (\ref{2.45}).
Let us put $s=s_v$, $P_i\=P_{v_i}\setminus Q_{v_i}$; moreover we order in an
arbitrary way the sets $P_i^\pm\=\{f\in P_i,\e(f)=\pm\}$, we call
$f_{ij}^\pm$ their elements and we define $\xx^{(i)}=\cup_{f\in
P_i^-}\xx(f)$, $\yy^{(i)}=\cup_{f\in P_i^+}\yy(f)$, $\xx_{ij}=\xx(f^-_{ij})$,
$\yy_{ij}=\xx(f^+_{ij})$. Note that $\sum_{i=1}^s |P_i^-|=\sum_{i=1}^s
|P_i^+|\=k$, otherwise the truncated expectation vanishes. A couple
$l\=(f^-_{ij},f^+_{i'j'})\=(f^-_l,f^+_l)$ will be called a line joining the
fields with labels $f^-_{ij},f^+_{i'j'}$. Then, by using the {\it
Brydges-Battle-Federbush} formula (see \cite{Bry84,Le087}), we get, if $s>1$,
\be \EE^T_{h}(\widetilde\psi^{(h)}(P_1),\ldots,
\widetilde\psi^{(h)}(P_s))=\sum_{T}\prod_{l\in T}
\big[g^{(h)}(\xx_l-\yy_l)\big]\, \int dP_{T}({\bf t})\; {\rm
det}\, G^{h,T}({\bf t})\;,\lb{2.46a}\ee
where $T$ is a set of lines forming an {\it anchored tree graph} among the
clusters of points $\xx^{(i)}\cup\yy^{(i)}$, that is $T$ is a set of lines,
which becomes a tree graph if one identifies all the points in the same
cluster. Moreover ${\bf t}=\{t_{ii'}\in [0,1], 1\le i,i' \le s\}$,
$dP_{T}({\bf t})$ is a probability measure with support on a set of ${\bf t}$
such that $t_{ii'}={\bf u}_i\cdot{\bf u}_{i'}$ for some family of vectors
${\bf u}_i\in \RRR^s$ of unit norm. Finally $G^{h,T}({\bf t})$ is a
$(k-s+1)\times (k-s+1)$ matrix, whose elements are given by
\be G^{h,T}_{ij,i'j'}=t_{ii'} \d_{s_{ij},s_{i'j'}}
\,g^{(h)}(\xx_{ij}-\yy_{i'j'}) \lb{2.48a}\ee
with $(f^-_{ij}, f^+_{i'j'})$ not belonging to $T$ and $s_{ij},s_{i'j'}$
the corresponding spin variables. In the
following we shall use (\ref{2.46a}) even for $s=1$, when $T$ is
empty, by interpreting the r.h.s. as equal to $1$, if $|P_1|=0$,
otherwise as equal to ${\rm det}\,G^{h}=
\EE^T_{h}(\widetilde\psi^{(h)}(P_1))$.

The l.h.s. of \pref{B.12} can also be written in the form, analogous to
\pref{2.17b},
\be\lb{2.17c}\bsp
& \VV^{(h)}(\psi^{(\le h)}) + \BB^{(h)}(\psi^{(\le h)},J,\h) +\b L e_h +
s_h(J,\h)=\\
& \sum_{m_\psi,m_J,m_\h \ge 0} \sum_{\ual,\ue,\usp} \int d\ux d\uy d\uz
W^{(h,M)}_{\ual,\ue,\usp, m_\psi,m_J,m_\h}(\ux,\uy,\uz) \psi^{(\le
h)}_{\ux,\usp} J_{\ual,\uy} \h_{\ue,\uz} \esp\ee
where $\ux=(\xx_1,\ldots,\xx_{m_\psi})$, $\uy=(\yy_1,\ldots,\yy_{m_J})$,
$\uz=(\zz_1,\ldots,\zz_{m_\h})$. The kernels $W^{(h,M)}$ admit a tree
expansion that can be  easily obtained from the previous discussion. Note that these
kernels coincide, for $h=0$, with those of Lemma \ref{p2.2}, only if
$m_\psi>0$, otherwise they are the kernels of the terms $\b L e_h$ and
$s_h(J)$, which have to be summed up over $h$ to get the corresponding
kernels, see \pref{Eh}.

If $\e_0=\max \{|\l|,|\n|\}$, by using \pref{2.45} and \pref{2.46a}, we get the bound
\be\lb{B.17}\bsp &{1\over \b L}\int d\ux d\uy d\uz |W^{(h,M)}_{\ual,\ue,\usp,
m_\psi,m_J,m_\h}(\ux,\uy,\uz)| \le \sum_{n\ge k_{l,m}} (C \e_0)^n\sum_{\t\in
{\cal T}_{M,h,n,m_J,m_\h}\atop a\in A_\t} \sum_{\bP\in
\PP_\t\atop |P_{v_0}|=m_\psi}\cdot\\
&\cdot\sum_{T\in{\bf T}} \int\prod_{l\in T} d(\xx_l-\yy_l)
\cdot\Bigg[\prod_{v\ {\rm not}\ {\rm e.p.}}{1\over s_v!} \max_{{\bf
t}_v}\big|{\rm det}\, G^{h_v,T_v}({\bf t}_v)\big| \prod_{l\in T_v}
\big|g^{(h_v)}(\xx_l-\yy_l)\big|\Bigg]\esp\ee
where, given the tree $\t$, $\bf T$ is the family of all tree graphs joining the
space-time points associated to the endpoints, which are obtained by taking,
for each non trivial vertex $v$, one of the anchored tree graph $T_v$ appearing
in \pref{2.46a}, and by adding the lines connecting the two vertices
associated to non local endpoints.

A standard application of Gram--Hadamard inequality, combined with
\pref{B.5}, see \cite{Le087,BGPS994,BM001}, implies the dimensional
bound (without factorials):
\be |{\rm det} G^{h_v,T_v}({\bf t}_v)| \le
C^{\sum_{i=1}^{s_v}|P_{v_i}|-|P_v|-2(s_v-1)}\;.\lb{2.54a}\ee
By the decay properties of $g^{(h)}(\xx)$ given by (\ref{cc1}), it
also follows that
\be\lb{2.55a} \prod_{v\ {\rm not}\ {\rm e.p.}} {1\over s_v!}\int \prod_{l\in
T_v} d(\xx_l-\yy_l)\, ||g^{(h_v)}_{\o_l}(\xx_l-\yy_l)||\le C^{n+m} \prod_{v\
{\rm not}\ {\rm e.p.}} {1\over s_v!} \g^{-h_v(s_v-1)}\ee
We can now perform the sum $\sum_{T\in{\bf T}}$, which erases the $1/s_v!$ up
to a $C^n$ factor. Then, by using the identity $\sum_{v'\ge v} (s_{v'}-1) =
n_v-1$ and the bound $\sum_{v\ge v_0}
[\sum_{i=1}^{s_v}|P_{v_i}|-|P_v|-2(s_v-1)] \le 4n+2m-2(n+m-1)$, we easily get
the final bound
\be\lb{B.18a} \bsp &{1\over \b L}\int d\ux d\uy d\uz
|W^{(h,M)}_{\ual,\ue,\usp, m_\psi,m_J,m_\h}(\ux,\uy,\uz)|\le \sum_{n\ge
k_{l,m}} C^{n+m} \e_0^n \sum_{\t\in {\cal T}_{M,h,n,m_J,m_\h}
\atop a\in A_\t} \cdot\\
&\cdot \sum_{\bP\in{\cal P}_\t\atop |P_{v_0}|=m_\psi}
\g^{-h(n-1)}\Big[\prod_{v\ \text{not trivial}}
\g^{-(h_v-h_{v'})(n_v-1)}\Big]\esp\ee
where $v'$ is the non trivial vertex immediately preceding $v$ or $v_0$. This
bound is suitable to control the expansion, if $n+m>1$, since $n_v>1$ for any
non trivial vertex, see above, and there is in such case at least one non
trivial vertex. If $n+m=1$, the resummation implicit in the definition
\pref{sssa} of the terms associated to the endpoints implies that the allowed
trees have only one endpoint of scale $h+1$, hence there is no problem.

Note that $\sum_{T\in{\bf T}}$ can be bounded by $\prod_v s_v!
C^{\sum_{i=1}^{s_v}|P_{v_i}|-|P_v|-2(s_v-1)} \le c^{n+m}\prod_v s_v!$, see
again \cite{Le087,BGPS994,BM001}. In order to bound the sum over $\t$
and $a\in A_\t$, note that the number of unlabeled trees is $\le 4^n$ and
that, given $\t$, $|A_\t| \le C^M$; moreover, as $n(v)\ge 2$ and, if $v>v_0$,
$2\le |P_v|\le 4n_v -2(n_v-1)$, so that $n_v-1 \ge |P_v|/6$,
\be \Big[\prod_{v\ \text{not trivial}} \g^{-(h_v-h_{v'})(n(v)-1)}\Big]\le
\Big[\prod_{v\ \text{not trivial}} \g^{- {2\over 5}(h_v-h_{v'})}\Big]
\Big[\prod_{v\ {\rm not}\ {\rm e.p.}}\g^{- {|P_v|\over 10}}\Big]\lb{B.18b}
\ee
The factor $\g^{- {2\over 5}(h_v-h_{v'})}$ can be used to bound the sum over
the scale labels of the tree; moreover, see \cite{BGPS994},
\be \sum_{\bP\in{\cal P}_\t} \g^{- {|P_v|\over 10}}\le C^{n+m}\lb{B.18c}\ee
Since the constant $C$ is independent of $M,\b,L$, the bounds above imply
analyticity of the kernels in $\l$ and $\n$, if $\e_0$ is small enough.

Finally in order to prove the uniform convergence as $M\to\io$, we shall
first consider the case $l\ge 1, m=0$ and we prove that, if $M'>M$
and $0<\th<1$, there is a constant $C_\th$ such that
\be\int d\ux |W^{(0,M')}_{\usp,2l}(\ux) - W^{(0,M)}_{\usp,2l}(\ux)|\le C_\th
\e_0^{\max\{1,l-1\}}\g^{-\th M}\;,\lb{B.30}\ee
In order to prove this bound, we note that the tree expansion of
$W^{(0,M')}_{\usp,2l}(\ux)$ differs from that of $W^{(0,M)}_{\usp,2l}(\ux)$
only for two reasons:

\0 1) The trees contributing to $W^{(0,M')}_{\usp,2l}(\ux)$ are the same
contributing to $W^{(0,M)}_{\usp,2l}(\ux)$ plus a set of trees with at least
one endpoint of scale $h_0>M+1$. It is easy to see that the sum over the
values associated to these trees satisfies a bound like \pref{B.30}, which
differs from the overall bound \pref{2.17} only for a factor $\g^{-\th M}$.
This factor is obtained by taking, for each tree $\t$ of this type, an
arbitrary endpoint $v_0$ of scale $h_0\ge M+1$ and by extracting from the
bound \pref{B.18a} a factor $\g^{-\th (h_v-h_{v'})}$ for each line connecting
two non trivial vertices on the path which connects $v_0$ with the root on
$\t$. This operation changes the factors $\g^{-(h_v-h_{v'})(n(v)-1)}$
associated to these lines in $\g^{-(h_v-h_{v'})(n(v)-1-\th)}$, which is still
good enough for the bounds following \pref{B.18a}, since $n(v)\ge 2$ and
$\th<1$.

\0 2) Note that the single scale propagator \pref{B.2} is independent of $M$,
for any $h\le M$. Hence, the other trees contributing to
$W^{(0,M')}_{\usp,2l}(\ux)$ differ from the corresponding trees contributing
to $W^{(0,M')}_{\usp,2l}(\ux)$ only because, for some choices of the label
$\a\in A_\t$, the potentials associated to some endpoints, those depending on
$g^{[h_v,M']}$, are substituted with $g^{[h_v,M]}$. For all these labels, the
difference between the corresponding tree values can be written, if $n$ is
the order of the tree $\t$, as the some over at most $n$ terms, such that
there is at least one endpoint whose associated potential contains
$g^{[h_v,M']}-g^{[h_v,M]}$. On the other hand, by \pref{B.2},
$$g^{[h_v,M']}-g^{[h_v,M]}(\xx) = \frac1{L}\sum_{k\in\DD_L}
e^{-ikx} \D_{M,M'}(k,x_0)$$
$$\D_{M,M'}(k,x_0) = \frac1{\b} \sum_{k_0\in\DD_\b}
e^{-i k_0 x_0} {\hat f_{u.v.}(\kk)\, [H_{M'}(k_0)-H_{M+1}(k_0)]
\over -i k_0 + (\cos p_F -\cos k)}$$
so that, by proceeding as in the proof of \pref{ch} in App. \ref{app0}, we
can easily prove that $|g^{[h_v,M']}(\xx)-g^{[h_v,M]}(\xx)| \le C \g^{-M}$.
It follows that the bound \pref{B.30} is verified also by the sum over the
values associated to these trees.

The bound \pref{B.30} implies that, for any $\uk=(\kk_1,\ldots,\kk_{2l})$,
the sequence of functions $F_M(\l,\n):= \hat W^{(0,M)}_{\usp,2l}(\uk)$, $M\ge
0$, is a Cauchy sequence, uniformly in $\uk$ and in the domain $|\l|,|\n|\le
\e_0$, where the $F_M(\l,\n)$ are analytic. Hence, by Weierstrass theorem,
the kernels $\hat W^{(0,M)}_{\usp,2l}(\uk)$ admit a limit $\hat
W^{(0)}_{\usp,2l}(\uk)$ as $M\to\io$; the limit is analytic in $|\l|,|\n|\le
\e_0$ and its Taylor coefficients are the limits of the coefficients of $\hat
W^{(0,M)}_{\usp,2l}(\uk)$.

\*

Let us now consider the constant $E_0$, which can be written as in \pref{Eh}.
We can write $e_j$ in terms of a tree expansion, which can be described
exactly as before, the only difference being that the root has scale $j$ and
$|P_{v_0}|=0$. The bound \pref{B.18a} implies that $|e_j|\le C\e_0 \g^{-j}$,
hence $|E_0| \le C\e_0$. The claim about $\lim_{M\to \io} E_0$ is proved
exactly as before.

A similar argument applies to the kernels $W^{(0,M)}_{\ual,\ue,\usp,
m_\psi,m_J,m_\h}(\ux,\uy,\uz)$, with $m_\psi=0$ and $m=m_J+m_\h>0$, if we
write them as in \pref{Eh} and use the bound \pref{B.18a}. \Halmos

\subsection{Infrared integration}\lb{sec2.4a}

If $\c(\kk')$ is the function defined in \S\ref{sec2.2b}, we put, for any
integer $h\le 0$,
\be f_h(\kk')= \c(\g^{-h}\kk')-\c(\g^{-h+1}\kk') \ee
which has support $t_0 \g^{h-1}\le |\kk'|\le t_0 \g^{h+1}$ and
equals 1 at $|\kk'| =t_0\g^h$; then
\be \c(\kk') = \sum_{h=h_{L,\b}}^0 f_h(\kk') \ee
where
\be h_{L,\b} :=\min \lft\{h:t_0\g^{h+1} > |\kk_{\rm
m}|\rgt\}\qquad {\rm for}\quad\kk_{\rm m}=(\p/\b,\p/L)\;.
\ee
For $h\le 0$ we also define
\be \hat f_h(\kk) = f_h(k-p_F,k_0) +f_h(k+p_F,k_0) \ee
(for $h=1$ the definition is \pref{f1}). This definition implies that, if
$h\le 0$, the support of $\hat f_h(\kk)$ is the union of two disjoint sets,
$A_h^+$ and $A_h^-$. In $A_h^+$, $k$ is strictly positive and
$\|k-p_F\|_{\TTT }\le t_0\g^h \le t_0$, while, in $A_h^-$, $k$ is strictly
negative and $\|k+p_F\|_{\TTT }\le t_0\g^h$. The label $h$ is called the {\sl
scale} or {\sl frequency} label. Note that, if we redefine $\hat f_1(\kk)$
the function $\hat f_{u.v.}(\kk)$ of \pref{f1}, we have
\be 1=\sum_{h=h_{L,\b}}^1 \hat f_h(\kk) \ee
We can write the infrared propagator introduced in \pref{gdefaa0} in the
following way
\be\lb{gdefaa} g^{(i.r.)}_\o(\xx-\yy) = \sum_{h=h_{L,\b}}^0
g^{(h)}_\o(\xx-\yy) \ee
where
\be\lb{2.58} g^{(h)}_\o(\xx-\yy) = {1\over\b L} \sum_{\kk'\in\DD'_{L,\b}}
e^{-i\kk'(\xx-\yy)} {f_h(\kk')\over -i k_0+ E_\o(k')} \ee

The integration of the infrared scales $h\le 0$ is done iteratively in the
following way. Suppose that we have integrated the scales $0,-1,-2,..,j+1$,
obtaining
\be\lb{61} e^{\WW(J,\h)}=e^{-L\b E_j + \SS_j(J,\h)} \int
P_{Z_j,C_j}(d\psi^{\le j})e^{-\VV^{(j)}(\sqrt{Z_j}\psi^{\le
j})+\BB^{(j)}(\sqrt{Z_j}\psi^{\le j},J,\h)} \ee
where, if we put $C_j(\kk')^{-1}=\sum_{h=h_{L,\b}}^j f_h(\kk')$,
$P_{Z_j,C_j}$ is the Grassmann integration with propagator
\be\lb{62}
{1\over Z_j}\, g^{(\le j)}_\o(\xx-\yy)= {1\over
Z_j}{1\over\b L} \sum_{\kk\in\DD'_{L,\b}} e^{-i\kk(\xx-\yy)}{C_j^{-1}(\kk)
\over -i k_0+E_\o(k')}
\ee
$\VV^{(j)}(\psi)$ is of the form
\be\lb{3.2aaax} \VV^{(j)}(\psi)= \sum_{l\ge 1} \sum_{\oo,\usp} \int d\ux
W^{(j)}_{\oo,\usp,2l}(\ux) \psi_{\ux,\oo,\underline s} \ee
while $\SS_j(J,\h)$ and $\BB^{(j)}(\psi^{\le j},J,\h)$ contain all terms
which are of order at least one in the source field $J$ and of order $0$ or
at least $2$, respectively, in the field $\psi^{\le j}$. For $j=0$, $Z_0=1$
and the functional $\VV^{(0)}$, $\SS_0$ and $\BB^{(0)}$ are those appearing
in \pref{2z}, with $W^{(0)}_{\uo,\ual,\ue,\usp, m_\psi,m_J,m_\h}(\ux)=
W^{(0)}_{\ual,\ue,\usp, m_\psi,m_J,m_\h}(\ux) e^{i p_F
\sum_{i=1}^{m_\psi+m_\h}\e_i\o_i x_i}$ (in other words, we have included in
the definition of the kernel the $e^{ip_F\o_i x_i}$ factors appearing in the
decomposition of the infrared field associated to \pref{gdefaa0}). We find
also convenient to write $\VV^{(j)}(\psi)$ as
\be\lb{2.62}\VV^{(j)}(\psi^{(\le h)}) = \sum_{l=1}^\io {1\over (L\b)^{2l}}
\sum_{\kk'_1,...,\kk'_{2l},\atop {\underline \e}, {\underline s},\oo}
\prod_{i=1}^{2l} \hat\psi^{(\le j)\e_i}_{\kk'_i,\o_i,s_i}\hat W_{2l, \ue,
\usp, \oo}^{(j)}(\kk'_1,...,\kk'_{2l-1})
\d(\sum_{i=1}^{2l}\e_i(\kk'_i+\pp_F))\ee
where
\be \d(\kk)=\d(k)\d(k_0)\;,\quad\d(k)=L \sum_{n\in\zzzz} \d_{k,2\p
n}\;,\quad \d(k_0)=\b\d_{k_0,0}\ee
%
In order to integrate $\psi^{(j)}$ we split $\VV^{(j)}$ as
$\LL \VV^{(j)}+\RR \VV^{(j)}$, where $\RR=1-\LL$ and $\LL$, the {\it
localization operator}, is a linear operator on functions of the field
defined in the following way by its action on the kernels $\hat
W_{2l,\ue,\usp,\oo}^{(j)}$.

\* 1) If $2l=4$, then
\be\lb{2.75a} \LL \hat W_{4,\ue,\usp,\oo}^{(j)}(\kk'_1,\kk'_2,\kk'_3)= \d_{\sum_{i=1}^4
\e_i\o_i p_F,0} \hat W_{4,\ue,\usp,\oo}^{(j)}(\bk_{++},\bk_{++},\bk_{++})\ee
where \be \bk_{\h{\h'}} = \left(\h{\p\over
L},\h'{\p\over\b}\right)\ee

\* 2) If $2l=2$ and $s_1=s_2$, $\o_1=\o_2$, $\e_1+\e_2=0$ (otherwise
$W_{2,\ue,\usp,\oo}^{(j)}=0$, by spin symmetry and the compact support
properties of the propagators $g^{(\le j)}$),
\be \LL \hat W_{2,\ue,\usp,\oo}^{(j)}(\kk')= {1\over 4} \sum_{\h,\h'=\pm 1}
\hat W_{2,\ue,\usp,\oo}^{(h)}(\bk_{\h{\h'}})\left\{ 1+ \h {L\over \p}
\left(b_L+a_L{E(k')\over v_F}\right) + \h'{\b\over \p} k_0
\right\}\label{2.75}\ee
where
\be a_L {L\over \p} \sin {\p \over L}=1\;,\qquad {\cos p_F\over
v_0}(1-\cos {\p \over L}) +b_L  {L\over \p} \sin {\p \over L}=0
\ee
In order to better understand this definition, note that, if
$L=\b=\io$,
\be \LL \hat W_{2,\ue,\usp,\oo}^{(j)}(\kk')= \hat
W_{2,\ue,\usp,\oo}^{(j)}(0)+ {E(k')\over v_F} {\dpr \hat
W_{2,\ue,\usp,\oo}^{(j)}\over \dpr k'}(0) + k_0 {\dpr \hat
W_{2,\ue,\usp,\oo}^{(j)}\over \dpr k_0}(0)\lb{2.76} \ee
Hence, $\LL \hat W_{2,\ue,\usp,\oo}^{(h)}(\kk')$ has to be understood as a
discrete version of the Taylor expansion up to order $1$. Since
$a_L=1+O(L^{-2})$ and $b_L=O(L^{-2})$, this property would be true also if
$a_L=1$ and $b_L=0$; however the choice \pref{2.75} has the advantage to
share with \pref{2.76} another important property, that is $\LL^2 \hat
W_{2,\ue,\usp,\oo}^{(h)}(\kk')=\LL \hat W_{2,\ue,\usp,\oo}^{(h)}(\kk')$.

\* 3) In all the other cases
\be \LL \hat W_{2l,\ue,\usp,\oo}^{j}(\kk'_1,\ldots,\kk'_{2l-1})=0\ee
\vspace{.2cm} \*

Note that the operator $\LL$ satisfies the relation $\RR \LL =0$. By the
above definition we get
\be\lb{rel}
\bsp
\LL \VV^{(j)}(\sqrt{Z_j}\psi) &= \g^j n_j F_\n(\sqrt{Z_j}\psi)+a_j
F_\a (\sqrt{Z_j}\psi)+z_j F_z(\sqrt{Z_j}\psi)\\
&+ l_{1,j} F_1(\psi) + l_{2,j} F_2(\sqrt{Z_j}\psi) + l_{4,j}
F_4(\sqrt{Z_j}\psi)
\esp
\ee
where
\bal
F_\n &= \sum_{\o,s} \int d\xx\, \psi^+_{\xx,\o,s}\psi^-_{\xx,\o,s}\virg &F_1 &=
\frac12 \sum_{\o,s,s'} \int d\xx\, \psi^+_{\xx,\o,s}\psi^-_{\xx,-\o,s}
\psi^+_{\xx,-\o,s'}\psi^-_{\xx,\o,s'}\nn\\
F_\a &= \sum_{\o,s} \int d\xx\, \psi^+_{\xx,\o,s}\DD\psi^-_{\xx,\o,s}\virg &F_2
&= \frac12 \sum_{\o,s,s'} \int d\xx\, \psi^+_{\xx,\o,s} \psi^-_{\xx,\o,s}
\psi^+_{\xx,-\o,s'} \psi^-_{\xx,-\o,s'}\lb{ff11}\\
F_z &= \sum_{\o,s}\int d\xx\, \psi^+_{\xx,\o,s}\partial_0
\psi^-_{\xx,\o,s}\virg &F_4 &= \frac12 \sum_{\o,s} \int d\xx\,
\psi^+_{\xx,\o,s} \psi^-_{\xx,\o,s} \psi^+_{\xx,\o,-s} \psi^-_{\xx,\o,-s}\nn
\eal
and $\DD\psi_{\xx,\o,s}={1\over L\b}\sum_{\kk'}e^{i\kk'\xx}\; E_\o(k')
\psi^+_{\kk',\o,s}$ (see definition \pref{dE}). Note that
\be\lb{init}
l_{4,0}=2\l \hat v(0)+O(\l^2)\quad l_{2,0}=2\l \hat v(0)+O(\l^2)\quad
l_{1,0}=2\l\hat v(2p_F)+O(\l^2)
\ee
and in writing \pref{rel}  the $SU(2)$ spin symmetry has been used. In the
case of local interactions, $\hat v(p)=1$. We will call $F_1$ in \pref{ff11}
{\it backward interaction} and $F_2,F_4$ are the {\it forward interactions};
the {\it umklapp interaction} is not present in $\LL \VV^{(j)}$, as well as
other terms quadratic in the fields. The reason is that the condition
$p_F\neq 0, \frac\p2, \p$ says that such terms are vanishing for $j$ smaller
than a suitable constant (depending on $|p_F-\p/2|$), because they cannot
satisfy the conservation of the momentum, so there is no need to localize
them (more details are in \cite{M005}).

Let us now consider $\BB^{(j)}(\sqrt{Z_j}\psi,J,\h)$. The following analysis
shows that it is necessary to localize only the terms with $m_J=1$ and
$m_\psi=2$. Hence we define
\be\lb{2.20} \LL \BB^{(j)}(\sqrt{Z_j}\psi,J,\h)= \int d\xx \; J^{(\a)}_\xx
\Bigg[\sum_{\a\not=TC_i} \frac{Z^{(1,\a)}_j}{Z_j}
O^{(1,\a)}_\xx(\sqrt{Z_j}\psi) + \sum_\a \frac{Z^{(2,\a)}_j}{Z_j}
O^{(2,\a)}_\xx(\sqrt{Z_j}\psi)\Bigg] \ee
where $ O^{(1,\a)}$ are
%
\bal O^{(1,C)}_\xx &= \sum_{\o,s} \psi^+_{\xx,\o,s}
\psi^-_{\xx,\o,s}\nn\\
\lb{op1} O^{(1,S_i)}_\xx &= \sum_{\o,s,s'}
\psi^+_{\xx,\o,s}\s^{(i)}_{s,s'}\psi^-_{\xx,\o,s'}\\
O^{(1,SC)}_\xx &= \sum_{\e,\o,s} s\, e^{2i \e\o p_F x} \psi^\e_{\xx,\o,s}
\psi^\e_{\xx,\o,-s}\nn
\eal
while $O^{(2,\a)}$ are
\bal O^{(2,C)}_\xx &= \sum_{\o,s} e^{2i \o p_F x}
\psi^+_{\xx,\o,s} \psi^-_{\xx,-\o,s}\nn\\
O^{(2,S_i)}_\xx &= \sum_{\o,s,s'}
e^{2i \o p_F x} \psi^+_{\xx,\o,s}\s^{(i)}_{s,s'}\psi^-_{\xx,-\o,s'}\nn\\
\lb{op2} \nn\\[-30pt]
\\
O^{(2,SC)}_\xx &= \sum_{\e,\o,s} s\,
\psi^\e_{\xx,\o,s} \psi^\e_{\xx,-\o,-s}\nn\\
O^{(2,TC_i)}_\xx &= \sum_{\e,\o,s,s'} e^{-i\e \o p_F} \psi^\e_{\xx,\o,s}
\tilde\s^{(i)}_{s,s'} \psi^\e_{\xx,-\o,s'}\nn
\eal
These definitions are such that the difference between $-\VV^{(j)} + \BB^{(j)}$
and $-\LL \VV^{(j)} + \LL \BB^{(j)}$ is made of irrelevant terms.

Note that the factor $e^{-i\e \o p_F}$ in the definition of $O^{(2,TC_i)}_\xx$
comes from the fact that the two $a^\e$ operators in the definition \pref{rho}
of the triplet Cooper density are located in two different lattice
sites
(otherwise the density would vanish). Moreover,
there is no local operator $O^{(1,TC_i)}_\xx$ because
$\sum_{s,s'}\psi^\e_{\xx,\o,s} \tilde\s^{(i)}_{s,s'} \psi^\e_{\xx,\o,s'}\=0$ by
anticommutation of the fermion fields.

We then renormalize the integration measure, by moving to it part
of the quadratic terms of the effective potential, that is $-z_j
(\b L)^{-1} \sum_{\o,s} \sum_{\kk} [-i k_0 + E_\o(k)]
\psi^+_{\kk,\o,s} \psi^-_{\kk,\o,s}$; equation \pref{61} takes the
form:
\be\lb{61a} e^{\WW(J,\h)}=e^{-L\b (E_j+t_j)+\SS_j(J,\h)} \int P_{\tilde
Z_{j-1},C_j} (d\psi^{(\le j)}) e^{-\tilde\VV^{(j)}(\sqrt{Z_j} \psi^{\le j}) +
\BB^{(j)}(\sqrt{Z_j} \psi^{\le j},J,\h)} \ee
where $\tilde\VV^{(j)}$ is the remaining part of the effective interaction,
$P_{\tilde Z_{j-1},C_j}(d\psi^{\le j})$ is the measure whose propagator is
obtained by substituting in \pref{62} $Z_j$ with
\be\lb{2.77}
\tilde Z_{j-1}(\kk) =Z_j [1+z_j C_j(\kk)^{-1}]
\ee
and $t_j$ is a constant coming from the normalization of the measure. It is
easy to see that we can decompose the fermion field as $\psi^{(\le j)} =
\psi^{(\le j-1)} + \psi^{(j)}$, so that
\be
P_{\tilde Z_{j-1},C_j}(d\psi^{\le j}) = P_{ Z_{j-1}, C_{j-1}}(d\psi^{(\le
j-1)}) P_{ Z_{j-1}, \tilde f_j^{-1}}(d\psi^{(j)})
\ee
where $\tilde f_j(\kk)$ (see eq. (2.90) of \cite{BM001}) has the same support
and scaling properties as $f_j(\kk)$. Hence, if we make the field rescaling
$\psi\to [\sqrt{Z_{j-1}}/ \sqrt{Z_j}]\psi$ and call
$\hat\VV^{(j)}(\sqrt{Z_{j-1}} \psi^{\le j})$ the new effective potential, we
can write the integral in the r.h.s. of \pref{61a} in the form
\be \int P_{Z_{j-1},C_{j-1}} (d\psi^{(\le j-1)}) \int P_{ Z_{j-1}, \tilde
f_j^{-1}}(d\psi^{(j)}) e^{-\hat\VV^{(j)}(\sqrt{Z_{j-1}} \psi^{(\le j)}) +
\hat\BB^{(j)}(\sqrt{Z_{j-1}} \psi^{(\le j)},J,\h)}\nn \ee
If we perform the integration over $\psi^{(j)}$ and we call
\be\lb{blue1} E_{j-1}=E_j+ t_j +\tilde E_j \virg \SS_{j-1}(J,\h) =
\SS_j(J,\h) + \tilde\SS_j(J,\h) \ee
the result, we finally get \pref{61}, with $j-1$ in place of $j$ and
In order to analyze the result of this iterative procedure, we note that $\LL
\hat\VV^{(j)}(\psi)$ can be written as
\be \LL \hat\VV^{(j)}(\psi) = \g^j\n_j F_\n(\psi) + \d_j
F_\a(\psi) + g_{1,j} F_1(\psi)+ g_{2,j} F_2(\psi)+ g_{4,j}
F_4(\psi)\lb{blue}\ee
where $\n_j=(\sqrt{Z_j}/ \sqrt{Z_{j-1}}) n_j$, $\d_j=(\sqrt{Z_j}/
\sqrt{Z_{j-1}}) (a_j-z_j)$ and $g_{i,j}=(\sqrt{Z_j}/ \sqrt{Z_{j-1}})^2
l_{i,j}$, $i=1,2,4$, are called the {\it running coupling constants} (r.c.c.)
on scale $j$ and denoted all together by $v_j$. Analogously, $\LL
\hat\BB^{(j)}(\psi,J)$ can be written as in \pref{2.20}, with
$Z^{(i,\a)}_j/Z_{j-1}$ in place of $Z^{(i,\a)}_j/Z_{j}$.

Let us now call $W^{(h)}_{\uo,\ual,\ue,\usp, m_\psi,m_J,m_\h}(\ux)$ the
kernels of the various terms contributing to\\
$\BB^{(h)}(\sqrt{Z_{h-1}} \psi^{(\le h-1)},J,\h)$ (in this case
$m_\psi\not=0$) or to $\tilde \SS_h(J,\h)$ (in this case $m_\psi=0$). We
shall prove the following Lemma, which follows from a careful dimensional
analysis of the tree expansion, similar to that used in many previous papers,
see for example \cite{BM001}.

\begin{lemma}\lb{p2.3} Assume that
\be\lb{ass} \max\{|\l|,\sup_{j> h} |v_j|\}\le \e_0 \virg \sup_{j> h} Z_j/Z_{j-1} \le
e^{c_1 \e_0^2} \virg  \sup_{j> h\atop i,\a} Z^{(i,\a)}_j/Z_{j-1}\le e^{c_1
\e_0}\ee
for some $c_1>0$. The constant $E_{h}$ and the $\LLL^1$ norm of the kernels
$W^{(h)}_{\uo,\ual,\ue,\usp, m_\psi,m_J,m_\h}$ (defined as in \pref{L1_norm})
are given by power series in $\{v_j\}_{j> h}$, convergent in the complex disc
$\sup_{j> h} |v_j|\le \e_0$, for $\e_0$ small enough and independent of $\b$
and $L$; moreover, if $2l=m_\psi+m_\h$, $m=m_J+m_\h$ and
$D_{m_\psi,m_J,m_\h}=-2+l+m_J (1+ c_1\e_0) + m_\h (1+ \frac12 c_1\e_0^2)$,
they satisfy the following bounds:
\be\lb{2.17aa} |E_h-E_{h+1}|\le C\e_0\g^{2h} \virg \int d\ux \big|
W^{(h)}_{\uo,\ual,\ue,\usp, m_\psi,m_J,m_\h}(\ux)\big| \le \b L C^{l+m}
\e_0^{k_{l,m}} \g^{-h D_{m_\psi,m_J,m_\h}} \ee
for some constant $C>0$ and $k_{l,m}=\max\{1,l-1\}$, if $m=0$, otherwise
$k_{l,m}=\max\{0,l-1\}$.
\end{lemma}

\subsection{Proof of Lemma 2.3}\lb{ss2.5}
The constants $\tilde E_h$ and the kernels $W^{(h)}_{\uo,\ual,\ue,\usp,
m_\psi,m_J,m_\h}$ can be written in terms of a tree expansion similar to that
used in \S\ref{sec2.2c}, but with some important differences, which we shall
describe with the help of Fig. \ref{h2a}.

\insertplot{300}{150}{ \ins{30pt}{85pt}{$r$} \ins{50pt}{85pt}{$v_0$}
\ins{130pt}{100pt}{$v$} \ins{35pt}{-5pt}{$h$} \ins{52pt}{-5pt}{$h+1$}
\ins{135pt}{-5pt}{$h_{v}$} \ins{215pt}{-5pt}{$0$} \ins{235pt}{-5pt}{$+1$}
\ins{255pt}{-5pt}{$+2$}} {fig51} {A renormalized tree for
$\VV^{(h)}$\lb{h2a}}{0}

\0 1) The scale index now is an integer taking values in $[h,2]$, $h$ being
the scale of the root. Moreover, there is only one vertex $v_0$ immediately
following the root, as before, but now it can not be an endpoint. The number
of endpoints is still $n+m$, but now $n_v$ will denote the number of normal
endpoints following $v$ and we introduce three new symbols $m_{J,v}$,
$m_{\h,v}$ and $m_v=m_{J,v} +m_{\h,v}$ to denote the number of special
endpoints following $v$ of type $J$, type $\h$ and both type, respectively.

\0 2) With each vertex $v$ of scale $h_v=+1$, which is not an endpoint, we
associate one of the terms contributing to $-\VV^{(0)}(\psi^{(\le
0)})+\BB^{(0)}(\psi^{(\le 0)},J,\h)$, in the limit $M=\io$, see \pref{2.14}.
The endpoints of scale $h=+2$ are associated with one of the terms
contributing to the potentials in \pref{1zv} and \pref{1zb}.

\0 3) With each endpoint $v$ of scale $h_v\le 1$ we associate one of local
terms that contribute to $\LL V^{(h_v-1)}$, see \pref{blue}, or $\LL
\BB^{(h_v-1)}$, see \pref{2.20}, or one of the two terms linear in $\psi$ and
$\h$ appearing in the \pref{1zb} (recall that they are not renormalized).
With each trivial or non trivial vertex $v>v_0$, which is not an endpoint, we
associate the $\RR=1-\LL$ operator, acting on the corresponding kernel.

\0 4) If $v$ is one endpoint of scale $h_v\le 1$, it is still true that its
scale is $h_{v'}+1$, if $v'$ is the non trivial vertex immediately preceding
it or $v_0$, but this property does not apply to the endpoints of scale
$h=+2$ involved in the localization procedure, that is those associated with
the non local potential $\VV(\psi)$ of \pref{1zb}; note that, in this
case,the trivial vertex preceding them carry an $\RR$ operator.

\0 5) If there is only one endpoint, the previous conditions imply that its
scale must be equal to $+2$ or $h+2$, if $h\le 0$. However, we need also to
define the {\em trivial tree}, which is the tree with one endpoint of scale
$h+1$; these trees do not belong to $\TT_{h,n,m}$ with $n+m=1$, if $h\le 0$,
and are associated with one of the terms in the local part of $\hat\VV$, see
\pref{blue}, or one of the terms in the r.h.s. of \pref{2.20}.

\*

The previous definitions imply that the following iterative equations,
similar to \pref{B.12}, are satisfied:
\be\bsp -\VV^{(h)}(\sqrt{Z_h}\psi^{(\le h)}) &+
\BB^{(h)}(\sqrt{Z_h}\psi^{(\le h)},J,\h) - L\b \tilde E_{h+1} +\tilde
\SS_h(J,\h)
=\\
& =\sum_{n=1}^\io\sum_{\t\in\TT_{h,n,m}\atop a\in A_\t} \bar
V_J^{(h)}(\t,a,\sqrt{Z_h}\psi^{(\le h)})\label{tr}\esp\ee
where, if $v_0$ is the first vertex of $\t$ and $\t_1,..,\t_s$
($s=s_{v_0}$)
are the subtrees of $\t$ with root $v_0$,\\
$\bar V_J^{(h)}(\t,a,\sqrt{Z_h}\psi^{(\le h)})$ is defined inductively by the
relation
\bea && \bar V_J^{(h)}(\t,a,\sqrt{Z_h}\psi^{(\le h)})=\\
&&{(-1)^{s+1}\over s!} \EE^T_{h+1}[\bar
V_J^{(h+1)}(\t_1,a_1,\sqrt{Z_{h}}\psi^{(\le h+1)});..; \bar
V_J^{(h+1)}(\t_{s},a_s,\sqrt{Z_{h}}\psi^{(\le h+1)})]\nn\eea
and $\bar V^{(h+1)}(\t_i,a_i,\sqrt{Z_{h}}\psi^{(\le h+1)})$

\0 a) is equal to $\RR\hat \VV^{(h+1)}(\t_i,a_i,\sqrt{Z_{h}}\psi^{(\le
h+1)})$ if the subtree $\t_i$ is not trivial;

\0 b) if $\t_i$ is trivial and $h\le -1$, it is equal to one of the terms
associated with the corresponding endpoint (of  scale $h+1$), as described in
item 3) above, or, if $h=0$, to one of the terms in the r.h.s. of \pref{1zv}
or \pref{1zb}.

The main difference with respect to the proof of Lemma \ref{p2.2} is in the
presence of the $\RR=1-\LL$ operators. Let us assume first that $\RR=1$, so
that, in particular, we do not perform the free measure renormalization; in
this case we can repeat exactly the analysis leading from \pref{2.43a} to
\pref{B.18a}, with $\e_0$ having the same meaning. The only difference is
just that $g_\o^{(h)}(\xx)$ admits a Gram representation:
$g_\o^{(h)}(\xx-\yy)=\int d\zz\, A_h^*(\xx-\zz)\cdot B_h(\yy-\zz)$, with
\be\lb{B.4} \bsp A_h(\xx)&={1\over \b L}\sum_{\kk'\in\DD'_{\b,L}}
\sqrt{\tilde f_h(\kk')}\frac{e^{i\kk'\xx}}{ k_0^2+E_\o(k')^2}\\
B_h(\xx)&={1\over \b L}\sum_{\kk'\in\DD'_{\b,L}} \sqrt{\tilde
f_h(\kk')}\,e^{i\kk'\xx}(i k_0+E_\o(k'))\esp\ee
and
\be ||A_h||^2=\int d\zz |A_h(\zz)|^2\le C\g^{-2h}\;,\quad\quad
||B_h||^2\le C \g^{4h}\;,\lb{B.5xx}\ee
for a suitable constant $C$. Therefore the Gram--Hadamard inequality implies
that
\be |{\rm det} G^{h_v,T_v}({\bf t}_v)| \le
c^{\sum_{i=1}^{s_v}|P_{v_i}|-|P_v|-2(s_v-1)}\cdot\;\g^{\frac{h_v}{2}
\left[\sum_{i=1}^{s_v}|P_{v_i}|-|P_v|-2(s_v-1)\right]}\;.\lb{2.54}\ee
By the decay properties of $g^{(h)}_\o(\xx)$, it also follows that
\be \prod_{v\ {\rm not}\ {\rm e.p.}} {1\over s_v!}\int \prod_{l\in T_v}
d(\xx_l-\yy_l)\, ||g^{(h_v)}_{\o_l}(\xx_l-\yy_l)||\le C^{n+m} \prod_{v\ {\rm
not}\ {\rm e.p.}} {1\over s_v!} \g^{-h_v(s_v-1)}\;.\lb{2.55b}\ee
Note now that $|I_v|=4 n_{4,v} + 2 n_{2,v} + m_{\h,v} + 2 m_{J,v}$, where
$n_{4,v}$ and $n_{2,v}$ are the number of normal endpoints with four and two
$\psi$ fields, respectively. Hence,
\be\bsp \sum_{\tv \ge v} \Bigg\{ \frac12 &\Big( \sum_{i=1}^{s_{\tv}}
|P_{v_i}|-|P_v| \Big) -2(s_{\tv}-1)\Bigg\} = \frac12 (|I_v|-|P_v|)
-2(n_v+m_v-1)=\\
&=2 -\frac12 |P_v| - n_{2,v} - m_{J,v} - \frac32 m_{\h,v}\esp\ee

Therefore, repeating the same steps leading from \pref{2.44} to
\pref{2.55a} we get, instead of \pref{B.18a}
\be\lb{B.18aa} \sum_{n\ge k_{l,m}} C^{n+m} \e_0^n \sum_{\t\in {\cal
T}_{h,n,m_J,m_\h}\atop a\in A_\t} \sum_{\bP\in{\cal P}_\t\atop
|P_{v_0}|=m_\psi}\sum_{T\in{\bf T}} \g^{- (D_{v_0}+n_{2,v_0})
h}\Big[\prod_{v\ {\rm not}\ {\rm e.p.}\atop v>v_0} \frac{1}{s_v!}
\g^{-(D_v+n_{2,v})}\Big]\ee
where
\be D_v=-2+ \frac12 |P_v| + m_{J,v} + \frac32 m_{\h,v} \ee
$D_v$ is called {\em scaling dimension}.

The fact that the scaling dimension $D_v$ can be negative or vanishing
prevents the possibility of performing the sum over the scales, as we did in
the equations leading from \pref{B.18a} to \pref{B.18c}. The action of the
$\RR$ operator \pref{rel} has the effect that instead of \pref{B.18aa} the
following bound is found
\be\lb{B.18caa} \bsp \sum_{n\ge k_{l,m}} &\sum_{\t\in {\cal T}_{h,n,m}\atop
a\in A_\t} \sum_{\bP\in{\cal P}_\t\atop |P_{v_0}|=2l} \sum_{T\in{\bf T}}
C^{n+m} \e_0^n
\Bigg[ \prod_{t=1}^{m_J} \frac{Z^{i_t,\a_t}_{h_{v_t}}}{Z_{h_{v_t}-1}} \Bigg]
\Bigg[ \prod_{s=1}^{m_\h} \frac1{\sqrt{Z_{h_{\bar v_s}-1}}} \Bigg]\cdot\\
&\cdot \g^{- D_{v_0} h}\Big[\prod_{v\ {\rm not}\ {\rm e.p.}\atop v>v_0}
\Big(\frac{Z_{h_v}}{Z_{h_v-1}}\Big)^{|P_v|/2} \frac{1}{s_v!}
\g^{-[D_v+z(|P_v|,m_v)]}\Big]\esp\ee
where $\e_0$ is defined as in \pref{ass}, $v_t$, $t=1,\ldots,m_J$, and $\bar
v_s$, $s=1,\ldots,m_\h$, are the special endpoints of type $J$ and $\h$,
respectively; moreover, $z(2,0)=2$, $z(4,2)=1$, $z(2,1)=1$ and $z(p,m)=0$
otherwise.

The proof of this bound is by now rather standard, but does not depend at all
on the details of the model, hence we address the reader to \S3 of \cite{BM001},
where a similar bound is obtained. In any case, the change of
the dimensional factors is easy to understand. First of all, $n_{2,v}$
disappears, because it is erased by the effect of the dimensional factor
$\g^j$ which multiplies the r.c.c. $\n_j$ in \pref{blue}. Moreover, the
presence of $z(|P_v|,m_v)$ is explained by the simple remark that the $\RR$
operation on the kernel $W^{(h_v)}(\kk'_1,\ldots,\kk'_{2l-1})$, associated to
the vertex $v$ of scale $h_v$, see \pref{2.62}, has roughly the effect of
substituting it with the rest of the Taylor expansion of order $z-1$ in at
least one of its variables, let us say $\kk'_1$, see \pref{2.76}. The
derivative of order $z$ acting on $\kk'_1$ will produce a "bad factor" at
most equal to $\g^{-z h_v}$, while the size of $|\kk'_1|$ gives a "good
factor" at least equal to $\g^{-z h_{\tilde v}}$, where $\tilde v< v$ is the
vertex where the external field of momentum $\kk'_1$ is contracted or $v_0$,
if it belongs to $P_{v_0}$.

Note that
\be\lb{2.92} D_v+z(|P_v|,m_v) >0 \virg \forall v>v_0\ee
except in the case $|P_v|=m_{\h,v}=1$ and $m_{J,v}=0$. However, thanks to
support properties of the single scale covariance in the $\kk$ variables,
this can happen only in the non trivial vertex where an endpoint of type $\h$
is connected to the tree, otherwise the tree value vanishes. It follows
immediately that this exception does not give any problem in the evaluation
of the sum over the scale indices, which is out of control only if one can
find an arbitrary long chain of tree vertices with non positive scale
dimension.

In order to bound in \pref{B.18caa} the sums over the scale labels and the
set $\PP_\t$, we first use \pref{ass}, by adding the hypothesis that
$c_1\e_0,c_1\e_0^2\le 1/16$; we get

\be\lb{3.111}\bsp &\Bigg[ \prod_{t=1}^{m_J}
\frac{Z^{i_t,\a_t}_{h_{v_t}}}{Z_{h_{v_t}-1}} \Bigg] \Bigg[ \prod_{s=1}^{m_\h}
\frac1{\sqrt{Z_{h_{\bar v_s}-1}}}\Bigg] \Big[\prod_{v\ {\rm not}\ {\rm
e.p.}\atop v>v_0} \Big(\frac{Z_{h_v}}{Z_{h_v-1}}\Big)^{|P_v|/2}
\g^{-[D_v+z(|P_v|,m_v)]}\Big]\le\\
&\le e^{m_J c_1 \e_0 h + \frac12 m_\h c_1\e_0^2} \Big[\prod_{v\ {\rm non}\
{\rm trivial}} \g^{-{1\over 40}(h_{v}-h_{v'})}\Big] \Big[\prod_{v\ {\rm not}\
{\rm e.p.}}\g^{-{|P_v|\over 40}}\Big]\esp\ee

Then we can continue as in the proof of Lemma \ref{p2.2}.\Halmos

\*

\0{\bf Remark 1} - An easy corollary of the above proof is that the bound for
the value associated to trees with root $h$ and at least one non trivial
vertex of scale $j$ can be improved by a factor $\g^{\th(j-h)}$ with
$0<\th<1$. It is sufficient to notice that, thanks to \pref{2.92}, one can
extract from the bound in the first line of \pref{3.111} one factor
$\g^{\th(h_v-h_{v'})}$ for each non trivial vertex on the path $\CC$
connecting the vertex $v^*$ of scale $j$ with $v_0$. Hence, there is
$\a_\th>0$ such that the bound in the second line of \pref{3.111} can be
substituted with the product of $e^{m_J c_1 \e_0 h + \frac12 m_\h c_1\e_0^2}$
times
\be\lb{3.111a} \g^{\th(j-h)} \prod_{\tilde v\in\CC\atop {\rm non trivial}}
\g^{-\a_\th(h_{\tilde v}-h_{\tilde v'})} \prod_{v\in\CC\atop {\rm not
e.p.}}\g^{-\a_\th|P_v|} \prod_{v\notin\CC\atop {\rm non trivial}}
\g^{-{1\over 40}(h_{v}-h_{v'})} \prod_{v\notin\CC\atop {\rm not
e.p.}}\g^{-{|P_v|\over 40}}\ee
This important property will be called in the following the {\em short memory
property}.

\*

\0{\bf Remark 2} - The tree expansion has another important property, that
will be used many times in the following to translate ``rough'' dimensional
arguments into rigorous dimensional bounds. Suppose that we make a small
change of one of the parameters of the model, so that the main objects
involved in the tree expansion, such as the r.c.c., the ren.c.'s or the
single scale propagators, are subject to a small perturbation. Then, by using
the ``stability'' of the Gram-Hadamard inequality \pref{2.54a} under a small
perturbation of the propagator and the short
memory property, one can see, by an iterative argument, that the sum over the
trees with $n$ endpoints is subject to a small variation, up to a $C^n$
factor in the bounds. This property, which is model independent, is explained
with enough details in \S4.6 of \cite{BM001} in a particular case. We shall
call it the {\em stability property of the tree expansion}.

\subsection{The flow of the running coupling constants}\lb{sec2.2}

In order to use Lemma \ref{p2.3}, we must show that the assumptions
\pref{ass} are verified for any $h>h_{L,\b}$. Let us consider first the
r.c.c. and define for them the following vector notations:
\be\lb{vn} \vec v_h\= (v_{1,h}, v_{2,h}, v_{4,h}, v_{\d,h}, v_{\n,h})
=(g_{1,h}, g_{2,h}, g_{4,h},\d_h, \n_h) \=(\vec g_{h}, \d_h, \n_h)\equiv
({\bf v}_h,\n_h)\;. \ee
The r.c.c.  satisfy a set of recursive equations, which can be obtained by
applying the localization operator to the r.h.s. of \pref{tr}; the result can
written in the form:
\be\lb{floweq0} v_{\a,j-1}=A_\a v_{\a,j} + \hat
\b_\a^{(j)}(v_j;...,v_0;\l,\n) \ee
with $A_\n=\g$, $A_\a=1$ for $\a\not=\n$. These equations have been already
analyzed in \cite{M005}, where it has been proved that, if $\l$ is real
positive and small enough, then it is possible to choose $\n$ so that, fixed
$\th<1$, $|\n_h|\le C\l\g^{\th h}$, $\forall h\le 0$, and $0< g_{1,h} < \l(1+
\bar a \l |h|)^{-1}$, for some $\bar a>0$, while the other r.c.c. stay bounded
by $C\l$ and converge for $h\to -\io$. In this paper, in order to proof Borel
summability of perturbation theory, we extend the proof to complex values of
$\l$, restricted to the set $D_{\e,\d}$ defined in \pref{dom}; this implies
that we need an analysis a bit more precise of the flow equations
\pref{floweq0}.

To begin with, we put $\n_1\=\n$ and we suppose that the sequence
$\{\n_h\}_{h \le 1}$ is made of known functions of $\l$, analytic in
$D_{\e,\d}$, such that
\be\lb{ne2} \sup_{j\le 1} \g^{-\th j}|\n_j|\le \x|\l|\ee
and study the flow equations of the other variables. The idea is that this
restricted flow has properties such that, by a fixed point argument, the
sequence $\{\n_h\}_{h\le 1}$, satisfying the last equation of \pref{floweq0},
can be uniquely determined, for $\x$ large enough. This point can be treated
in a way similar to that used in the spinless case (see App. 5 of
\cite{GM005_1} or \S4.3 of \cite{BM001}, where a different method is used);
we shall give the main details below, see \S\ref{sec2.nu}. Hence, from now
on, we shall consider the restriction of \pref{floweq0} to ${\bf v}_j$, see
\pref{vn}.

The next step is to extract from the functions $\hat \b_\a^{(j)}$ the leading
terms for $j\to-\io$. Observe that the propagator $\tilde g^{(j)}_\o$ of the
single scale measure $P_{ Z_{j-1}, \tilde f_j^{-1}}$, can be decomposed as
\be\lb{ne1}
\tilde g^{(j)}_\o(\xx)= {1\over Z_j}\,
g^{(j)}_{{\rm D},\o}(\xx)+r^{(j)}_\o(\xx)
\ee
where $g^{(j)}_{{\rm D},\o}$ is the  {\it Dirac propagator}
(with cutoff) and  describes the leading asymptotic behavior
\be\lb{gjth}
g^{(j)}_{{\rm D},\o}(\xx):= {1\over\b
L}\sum_{\kk\in\DD_{L,\b}}e^{-i\kk\xx} {\tilde f_j(\kk) \over -i k_0+\o
v_F k}\;,
\ee
while the {\em remainder}
$r^{(j)}_\o$ satisfies, for any $q>0$ and $0<\th<1$, the bound
\be\lb{2.30}
|r^{(j)}_\o(\xx)|\le {\g^{(1+\th)j}\over Z_j} {C_{q,\th}\over 1+(\g^j
|\xx|)^q}\;.
\ee
Let us now call $Z_{{\rm D},j}$ the values of $Z_j$ one would
obtain by substituting $\VV^{(0)}$ with $\LL \VV^{(0)}$ in
\pref{2z} and by using for the single scale integrations the
propagator \pref{ne1} with $r^{(i)}_\o(\xx)\=0$ for any $i\ge j$.
It can be proved by an inductive argument that, if all the r.c.c.
stay of order $\l$,
\be\lb{ne} \left| {Z_{j}\over Z_{j-1}}- {Z_{{\rm D}, j}\over Z_{{\rm D},
j-1}}\right|\le C \e_j^2\g^{\th j} \ee
where
$$\e_j := \max\{ |\l|,
\max_{0\ge h\ge j}|\vec g_h|, \max_{0\ge h\ge j}|\d_h| \}$$
It is then convenient to decompose the functions $\hat \b_\a^{(j)}$ as
\be\lb{bal} \hat \b_\a^{(j)}(\vec v_j;...,\vec v_0;\l,\n) =
\b_\a^{(j)}(\vv_j,...,\vv_0) +\bar \b_\a^{(j)}(\vec v_j;...,\vec v_0;\l,\n)
\ee
where $\b_\a^{(j)}(\vv_j,...,\vv_0)$ is given by the sum of all trees
containing only endpoints with r.c.c. $\d_h, \vec g_h$, $0\ge h\ge j$,
modified so that the propagators $g^{(h)}_\o$ and the wave function
renormalizations $Z_h$, $0\ge h\ge j$, are replaced by $g^{(h)}_{{\rm D},\o}$
and $Z_{{\rm D}, h}$; $\bar\b_\a^{(j)}$ contains the correction terms
together with the remainder of the expansion. \pref{ne} and \pref{ne2} imply
that there two constants $\bar c$ and $C$, such that
%
\be\lb{2.34}
|\bar \b_\a^{(j)}(\vec v_j;...,\vec v_0;\l)|\le
\begin{cases}
C \e_j^2 \g^{\th j} & \text{if $\a\not=\d$}\\
(\bar c\,\e_0 + C\e_j^2) \g^{\th j} & \text{if $\a=\d$}
\end{cases}
\ee

\*

\0{\bf Remark} Note that the constant $C$ in \pref{ne} and \pref{2.34}
depends on the constant $\x$ of \pref{ne2}. It is easy to see that, if we
call $C_1$ the constant appearing in \pref{B.18caa}, then $C=C_1
\max\{C_1,\x\}$ and all bounds of this section are verified only if, say,
$\max\{C_1,\x\}\e_0 \le 1/2$. In the following discussion, the only constant
which depends on $C$ under this smallness hypothesis, is the constant $b_2$
of \pref{vdiff0} below. Hence, all the other constants will be independent of
$\x$, if the first condition in \pref{cond1} is also verified.

\*

The leading term in \pref{bal}, that is $\b_\a^{(j)}$, can be further
decomposed as
\be\lb{2.43b}
\b_\a^{(j)}(\vv_j,...,\vv_0) = \tilde\b_\a^{(j)}(\vv_j) +
r_{\a,j}(\vv_j,...,\vv_0)
\ee
where $\tilde \b_\a^{(j)}(\vv) = \b_\a^{(j)}(\vv,...,\vv)$. We can write:
\be\lb{2.43c}
\tilde\b_\a^{(j)}(\vv_j) = \sum_{i=0,1} b_{\a,i}^{(j)}(\vv_j) + b_{\a,\ge
2}^{(j)}(\vv_j)
\ee
where $b_{\a,i}^{(j)}(\vv_j)$ is the contribution of order $i$ in $g_{1,j}$,
wile $b_{\a,\ge 2}^{(j)}(\vv_j)$ is the contributions of all trees with at
least two endpoints of type $g_1$. The crucial property is the following one,
called {\it partial vanishing of  the beta function}, whose proof is in
Appendix \ref{appB}
%
\be\lb{beta23} |b_{\a,i}^{(j)}(\vv_j)| \le C \e_j^2 [ \g^{\th j} +
\g^{-(j-h_{L,\b})} ] \virg i=0,1 \ee
%
Now, let us extract from $\tilde \b_\a^{(j)}(\vv_j)$ the second
order contributions, which all belong to $b_{\a,\ge 2}^{(j)}(\vv_j)$; we get:
\be\lb{2.46} \tilde \b_\a^{(j)}(\vv_j) = -a_\a g_{1,j}^2 + \sum_{i=0,1}
b_{\a,i}^{(j)}(\vv_j) +\tilde r_{\a,j}(\vv_j)\ee
with $a_1=a^{(j)}_{L,\b}>0$, $a_2=a^{(j)}_{L,\b}/2$, $a_4=a_\d=0$, and, for
some $b_1>0$,
\be\lb{2.47a} |\tilde r_{\a,j}(\vv_j)| \le b_1 \e_j |g_{1,j}|^2\;.
\ee
Note that, if we put
\be\lb{defa} a = 2 \lim_{h\to-\io} \frac{1}{|h|} \int \frac{dk}{(2\p)^2}
\hg_{{\rm D},+}^{(\ge h)}(\kk) \hg_{{\rm D},-}^{(\ge h)}(\kk) =
\frac{\log\g}{\p v_F} \ee
where $g_{{\rm D},\o}^{(\ge h)} \= \sum_{j=h}^0 g_{{\rm D},\o}^{(j)}$, then
\be\lb{a1L} |a^{(j)}_{L,\b}-a| \le C \g^{-(j-h_{L,\b})}\ee

Let us now analyze in more detail the functions
$r_\a^{(j)}(\vv_j,...,\vv_0)$, which appear in \pref{2.43b}. If we define,
for $j'\ge j+1$,
\be
D_\a^{(j,j')}(\vv_j,...,\vv_0) =
\b_\a^{(j)}(\vv_j,...,\vv_j,\vv_{j'},...,\vv_0)-
\b_\a^{(j)}(\vv_j,...,\vv_j,\vv_{j},...,\vv_0)
\ee
we can decompose $r_\a^{(j)}(\vv_j,...,\vv_0)$ in the following way:
\be\lb{2.50}
r_{\a,j}(\vv_j,...,\vv_0) = \sum_{j'=j+1}^0 D_\a^{(j,j')}(\vv_j,...,\vv_0)
\ee
Note that $D_\a^{(j,j')}(\vv_j,...,\vv_0)$ is obtained from
$\b_\a^{(j)}(\vv_j,...,\vv_0)$, by changing the values of the r.c.c. in the
following way: the r.c.c. of scale lower than $j'$ are put equal to the
corresponding r.c.c. of scale $j$; those of scale greater than $j'$ are left
unchanged; at least one of the r.c.c. $v_{r,j'}$ is substituted with
$v_{r,j'} - v_{r,j}$. By using the stability property (remark 2 after
\pref{3.111a}, we can show that , if $\e_j$ is small enough,
\be\lb{3.22} |D_\a^{(j,j')}(\vv_j,...,\vv_0)| \le b_3 \e_j \g^{-(j'-j)\th}
|\vv_{j'} -\vv_j| \ee
for some $b_3>0$. If we insert in the flow equation \pref{floweq0} the
equations \pref{bal}, \pref{2.43b}, \pref{2.46}, \pref{2.50} and use the
bounds \pref{2.34}, \pref{beta23}, \pref{2.47a}, \pref{a1L} and \pref{3.22},
we get, if $\e_j$ is small enough,
\be\lb{vdiff0} \bsp |\vv_{j-1} -\vv_j| &\le (a+ b_1\e_j) |g_{1,j}|^2 + (\bar
c\,\e_0 + b_2 \e_j^2) \g^{\th j}+ b_2 \e_j^2 \g^{-(j-h_{L,\b})} +\\
&+ b_3\e_j \sum_{j'=j+1}^0 \g^{-\th(j'-j)} |\vv_{j'} -\vv_j|\esp \ee
for some $b_2>0$. The form of this bound implies that, in order to control
the flow, it is sufficient to prove that $g_{1,j}$ goes to $0$ as $j\to -\io$
so fast that $|g_{1,j}|^2$ is summable on $j$. Hence, we have to look more
carefully to the flow equation of $g_{1,j}$. By proceeding as before, we can
write
\be\lb{2.53a}
g_{1,j-1} = g_{1,j} - a g_{1,j}^2 + \tilde r_{1,j} + r_{1,j} + \bar r_{1,j}
\ee
\be\lb{2.53b}\bsp |\tilde r_{1,j}| &\le b_1 \e_j |g_{1,j}|^2 \virg |r_{1,j}|
\le
b_3\e_j \sum_{j'=j+1}^0 \g^{-\th(j'-j)} |\vv_{j'} -\vv_j|\\
&|\bar r_{1,j}| \le b_2\e_j |g_{1,j}| [\g^{\th j} + \g^{-(j-h_{L,\b})}]\esp
\ee
where, in the bound of $\bar r_{1,j}$, we used the fact that, for symmetry
reasons, $b_{1,0}^{(j)}(\vv_j)=0$.

The proof that, if $g_{1,0}\in D_{\e_0,\d}$, $g_{1,j}$ goes to $0$ as $j\to
-\io$ so fast that $|g_{1,j}|^2$ is summable on $j$, uniformly in $L$ and
$\b$, would be rather simple if $\bar r_{1,j}=0$. This is not true, hence we
have to show that its contribution is in any case negligible; however, this
claim looks reasonable only if both $|j|$ and $j-h_{L,\b}$ are large enough.
To control the ``small" values of $j$, we use the remark that, as it is easy
to show, if $\e_0$ is small enough, there is a constant $c_4$, such that, if
$g_{1,0}\in D_{\e_0,\d}$ and $c_4 |j_0| |g_{1,0}|^2\le |g_{1,0}|^{2-\h}$,
$\h<1$, then, for $j\ge j_0$,
\be\lb{2.55} g_{1,j}\in D_{2\e_0,\d/2} \virg |g_{1,0}|/2 \le |g_{1,j}|\le
2|g_{1,0}| \virg \e_j\le 2\e_0 \ee
Hence we put $j_0 = -(c_4|g_{1,0}|^{1/2})^{-1}$ and suppose $\e_0$ so small
that
\be\lb{2.57c} \e_{j_0} \g^{\frac{\th}2 j_0} \le 2 c_5 |g_{1,j_0}|
\g^{\frac{\th}2 j_0} \le |g_{1,j_0}|^3\ee
where we also used the fact that, since $\hat v(2p_F)>0$, $\e_0\le c_5
|g_{1,0}|$, for some constant $c_5$.

\begin{lemma}\lb{lm2.4}
If $g_{1,0}\in D_{\e_0,\d}$  and $j\ge j_0$, then, if $\e_0$ is small enough,
\be\lb{vdiff1} |\vv_{j-1} -\vv_j| \le 2a |g_{1,j}|^2 + 2\bar c \e_0
\g^{\frac{\th}{2} j} + 2 b_2 \e_j^2 \g^{-(j-h_{L,\b})}\ee
\end{lemma}

\0{\bf Proof} - We shall proceed by induction. By \pref{2.55}, if $\e_0$ is
small enough, $\bar c\,\e_0 + b_2 \e_j^2 \le (3/2)\bar c\,\e_0$ and
$a+b_1\e_j \le 3a/2$; hence, \pref{vdiff1} is true for $j=0$. Let us suppose
that \pref{vdiff1} is verified for $j>h> 0$. By \pref{2.55}, if $j\ge h\ge
j_0$, $|g_{1,j}|/|g_{1,h}|\le 4$; hence, by using \pref{vdiff0} and
\pref{vdiff1}, we get:
\be\nn
\bsp
&|\vv_{h-1} -\vv_h| \le (3/2) a |g_{1,h}|^2 + (3/2) \bar c\,\e_0 \g^{\th h}  +
b_2 \e_h^2 \g^{-(h-h_{L,\b})}+\\
&b_3\e_h \sum_{j=h+1}^0 \g^{-\th(j-h)} (j-h) \lft\{ \max_{h< j' \le j}
\left[2a |g_{1,j'}|^2 + 2\bar c\e_0 \g^{\frac{\th}{2} j'} \right] +
2 b_2 \e_{j'}^2 \sum_{j'=h+1}^j \g^{-(j'-h_{L,\b})} \rgt\}\\
&\le |g_{1,h}|^2 \left[(3/2) a + 64 a b_3\e_0 \sum_{n=0}^\io n\, \g^{-\th n}
\right] + \g^{\frac{\th}{2} h} \e_0 \left[(3/2) \bar c + 4\bar c b_3\e_0
\sum_{n=0}^\io n \g^{-\frac{\th}{2} n}\right]+\\
& b_2 \e_h^2 \g^{-(h-h_{L,\b})} \lft[1 + 4b_3 \e_0 \sum_{n=1}^\io \g^{-n}
\rgt] \esp \ee
Hence, \pref{vdiff1} is verified also for $j=h$, if $\e_0$ is small
enough.\Halmos

The previous analysis implies that the flow is essentially trivial up to
values of $j$ of order $|g_{1,0}|^{-1/2}$ (or even $|g_{1,0}|^{-\h}$,
$0<\h<1$). Let us now consider the region $j\le j_0$, where the term
proportional to $\g^{\th j}$ in the bound of $\bar r_{1,j}$ is expected to be
negligible, thanks to the condition \pref{2.57c}, so that we can hope to
prove that $|g_{1,j}|$ is decreasing. However, since the the term
proportional to $\g^{-(j-h_{L,\b})}$ is not negligible for $j-h_{L,\b}$ ``too
small", we have to put some restriction on the values of $j$. We choose to
restrict the detailed analysis of the flow to the region
\be\lb{hstar} j\in [j_0,h^*_{L,\b}] \virg h^*_{L,\b} := \min\{ j< j_0:
\g^{-(j-h_{L,\b})} \le |g_{1,j}|^2 \}\ee
In this region we write \pref{2.53a} in the form
\be\lb{2.57b}
g_{1,j-1} = g_{1,j} - a_j\, g_{1,j}^2 \virg a_j\= a - \frac{\tilde r_{1,j} +
r_{1,j}+ \bar r_{1,j}}{g_{1,j}^2}
\ee
and we define $A_{j_0}=0$ and, for $j< j_0$,
\be A_j= \frac1{j_0-j} \sum_{j'=j+1}^{j_0} a_{j'}
\qquad
\lb{gtilde}
\tilde g_{1,j} = \frac{g_{1,j_0}}{1+A_j g_{1,j_0}(j_0-j)}
\ee

\begin{lemma}\lb{lm2.5}
There are constants $c_1, c_2, c_3$ such that, if $g_{1,0}\in D_{\e_0,\d}$
and it $\e_0$ is small enough, then the following bounds are satisfied, for
all $j\in[j_0,h^*_{L,\b})$.
\be\lb{bej} \e_j\le c_3\e_0\ee
\be\lb{vdiff}
|\vv_j -\vv_{j+1}| \le c_1 |g_{1,j+1}|^2
\ee
\be\lb{gerr}
|g_{1,j} - \tilde g_{1,j}| \le |\tilde g_{1,j}|^{3/2}
\ee
\be\lb{bAj} |a_j - a| \le c_2 |g_{1,j_0}|\ee

If $j\in[h^*_{L,\b}+1,h_{L,\b})$, we can only say that
\be\lb{overstar} |g_{1,j}| \le 2|g_{1,h^*_{L,\b}}| \virg \e_j\le 2 c_3\ee
\end{lemma}

\0{\bf Proof} - We shall proceed by induction. By using \pref{2.57c},
\pref{vdiff1}, \pref{hstar} and \pref{2.55}, we see that the bounds
\pref{bej} and \pref{vdiff} are satisfied for $j=j_0$, if $c_3\ge 2$, $c_1\ge
3a$ and $(\bar c/c_5) \e_0 + 8b_2\e_0^2\le a$. Moreover, $g_{1,j_0} = \tilde
g_{1,j_0}$ and, by proceeding as in the proof of Lemma \ref{lm2.4} and using
\pref{2.57c}, it is easy to prove that there is a constant $\bar c_2$, such
that
$$|a_{j_0}-a| \le \bar c_2 |g_{1,j_0}|$$
Hence, all the bounds are verified (for $\e_0$ small enough) for $j=j_0$, if
$c_1\ge 3a$, $c_2\ge \bar c_2$ and $c_3\ge 2$. Suppose that they are verified
for $j_0 \ge j\ge h$.

The validity of \pref{gerr} for $j=h-1$ follows from Prop. \ref{propA2},
which only rests on the bound \pref{bAj} for $j\ge h$. On the other hand,
\pref{gerr} implies that, if $\e_0$ is small enough, $2^{-1} |\tilde g_{1,j}|
\le |g_{1,j}| \le 2 |\tilde g_{1,j}|$; hence, using \pref{gtilde}, we get,
for $j>h$
\be\lb{Rjh} \left| \frac{g_{1,j}}{g_{1,h}} \right| \le 4 \frac{|1+A_h
g_{1,j_0}(j_0-h)|}{|1+A_j g_{1,j_0}(j_0-j)|} \ee
Let us now define, as in App. \ref{appA}, $A_j=\a_j+i\b_j$, $\a_j = \Re A_j$,
and suppose that
\be\lb{cond0} 2 c_2 \e_0 \le a/2 \ee
so that, by \pref{2.55}, $\a_j\ge a/2$, $|\b_j|\le 2 c_2 \e_0$, $|A_j|\le 3a/2$,
for $j>h$. By proceeding as in the proof of the bound \pref{AA4} in App.
\ref{appA}, we get, if $j>h$ and $|\text{Arg}\; g_{1,0}|\le \p-\d$, $\d>0$ (so
that $|\text{Arg}\; g_{1,j_0}|\le \p-\d/2$, see \pref{2.55}),
$$|1+g_{1,j_0} \a_j (j_0-j)| \ge \frac{1}{3} \sin (\d/2) [1+|g_{1,j_0}| \a_j (j_0-j)]$$
and, if we put $1+A_j g_{1,j_0}(j_0-j) = 1+\a_j g_{1,j_0}(j_0-j) + w_j$, we choose $\e_0$ so that
\be\lb{cond01}\frac{|w_j|}{|1+g_{1,j_0} \a_j (j_0-j)|} \le \frac{6 c_2\e_0
|g_{1,j_0}|(j_0-j)}{\sin (\d/2) |g_{1,j_0}| (a/2) (j_0-j)} = \frac{12
c_2\e_0}{a \sin (\d/2)} \le \frac12\ee
Then, by using \pref{Rjh}, we get
\be\lb{2.56a} \left| \frac{g_{1,j}}{g_{1,h}} \right| \le
\frac{24}{\sin(\d/2)} \frac{1+ (3a/2) |g_{1,j_0}|(j_0-h)|}{1+(a/2)
|g_{1,j_0}|(j_0-j)|} \le C_\d (j-h) \ee
for some constant $C_\d$, only depending on $\d$ and $a$.
Moreover, since $\e_h \le c_3\e_0$, then $\bar c\,\e_0 + b_2 \e_h^2 \le 2\bar
c\,\e_0$ and $a+b_1\e_j + b_2\e_j^2\le 2a$, if
\be\lb{cond1} b_2 c_3^2 \e_0 \le \bar c\virg \hbox{and\ } b_1 c_3 \e_0 + b_2
c_3^2 \e_0^2\le a\ee
Hence, by using the bounds \pref{vdiff0}, \pref{vdiff}, \pref{2.57c},
\pref{hstar}, \pref{cond1} and \pref{2.56a}, we get
\be\nn \bsp |\vv_{h-1} -\vv_h| &\le 2a |g_{1,h}|^2 + \frac{\bar c\,\e_0}{c_5}
\g^{-\th(j_0- h)} |g_{1,j_0}|^2 + c_1 b_3\e_h \sum_{j=h+1}^0 \g^{-\th(j-h)}
(j-h)
\max_{h< j' \le j} |g_{1,j'}|^2\\
&\le |g_{1,h}|^2 \left[ 2a + \frac{\bar c\,\e_0}{c_5} C_\d^2 \max_{n\ge 0}
\g^{-n\th} n^2 + c_1 \e_h b_3 C_\d^2 \sum_{n=0}^\io \g^{-\th n} n^3
\right]\esp \ee
It follows that \pref{vdiff} is satisfied also for $j=h$, if
\be\lb{cond2} 2a + \frac{\bar c\,\e_0}{c_5} C_\d^2 \max_{n\ge 0} \g^{-n\th}
n^2 + 2 c_1 c_3 \e_0 b_3 C_\d^2 \sum_{n=0}^\io \g^{-\th n} n^3 \le c_1 \ee
Moreover, by using \pref{vdiff} and $|g_{1,j}| \le 2 |\tilde g_{1,j}|$, we get,
for some $b_4>0$, only depending on $a$, under the condition \pref{cond0}:
$$\e_{h-1} \le \e_0 + \sum_{j=h}^0 |\vv_{j-1} -\vv_j| \le \e_0 + b_4 c_1 \e_0$$
so that $\e_{h-1} \le c_3\e_0$, if
\be\lb{cond3} 1+b_4 c_1 \le c_3\ee
The bound for $a_{h-1}-a$ can be done in the same way; it is easy to see that
\be |a_{h-1}-a| \le \left[ b_1 c_3 + b_2 c_3^2 \e_0 C_\d^2 \max_{n\ge 0}
\g^{-n\th} n^2 + 2 c_1 c_3 b_3
C_\d^2 \sum_{n=0}^\io \g^{-\th n} n^3\rgt] \e_0
\ee
Hence, \pref{bAj} is verified for $j=h-1$, if
\be\lb{cond4}
\tilde c_2 \= 2\bar c\, C_\d^2 \max_{n\ge 0} \g^{-n\th} n^2 + 2 c_1 c_3 b_3
C_\d^2 \sum_{n=0}^\io \g^{-\th n} n^3 \le c_2
\ee
The conditions \pref{cond0}, \pref{cond01}, \pref{cond1}, \pref{cond2},
\pref{cond3} and \pref{cond4} can be all satisfied, by taking, for example,
$c_1=4a$, $c_3=1+4a b_4$ and $c_2= \max\{\bar c_2, \tilde c_2\}$, if $\e_0$
is small enough.

We still have to analyze the flow in the region $j\in [h^*_{L,\b},
h_{L,\b}+1]$, in order to prove the bounds \pref{overstar}. We should again
proceed by iteration, but we prefer to explain the idea of the proof, which
can be by now easily translated in the longer formal proof.

Let us consider first the flow equation for $g_{1,j}$. In this region it is
not convenient to include the term bounded by $b_2 \e_j |g_{1,j}|
\g^{-(j-h_{L,\b})}$ (see \pref{2.53b}) in the definition of $a_j$; hence, we
decompose $\bar r_{1,j}$ as $\bar r_{1,j} = \bar r^{(\th)}_{1,j} + \bar
r^{(L,\b)}_{1,j}$ and we write \pref{2.53a} in form
\be\bsp g_{1,j-1} &= g_{1,j} - a'_j\, g_{1,j}^2 + \bar r^{(L,\b)}_{1,j} \virg
a'_j\= a - \frac{\tilde r_{1,j} + r_{1,j}+ \bar r^{(\th)}_{1,j}}{g_{1,j}^2}\\
&|\bar r^{(L,\b)}_{1,j}| \le b_2 \e_j |g_{1,j}| \g^{-(j-h_{L,\b})}\esp\ee
Note that, if $\e_j$ satisfies the second condition in \pref{overstar},
$a'_j$ satisfies a bound like \pref{bAj}, so that the term $- a'_j\,
g_{1,j}^2$ has still the effect to lower the value of $|g_{1,j}|$ as $j$
decreases. This remark can be translated easily in the claim that $|g_{1,j}|$
can be bounded by the solution of the flow equation
$$\bar g_{1,j-1} = \bar g_{1,j} [1+ 2 c_3 b_2\e_0 \g^{-(j-h_{L,\b})}] \virg
\bar g_{1,h^*_{L,\b}} = |g_{1,h^*_{L,\b}}|$$
whose solution satisfies, for $\e_0$ small enough, the bound
\be\lb{g1b} \bar g_{1,h} \le |g_{1,h^*_{L,\b}}| \exp\lft\{ 2 c_3 b_2\e_0
\sum_{j=h+1}^{h^*_{L,\b}} \g^{-(j-h_{L,\b})}\rgt\} \le 2|g_{1,h^*_{L,\b}}|\ee
Let us now consider the other couplings; even in this case we have to
separate the term proportional to $b_2 \e_j^2 \g^{-(j-h_{L,\b})}$ from the
others; however it is easy to see, by proceeding as before, that the only
consequence is that the bound \pref{vdiff} has to be modified as
\be\lb{vdiffL} |\vv_j -\vv_{j+1}| \le c'_1 |g_{1,j+1}|^2 + c''_1 \e_j^2
\g^{-(j-h_{L,\b})}\ee
so that, if $h\in [h^*_{L,\b}, h_{L,\b}+1]$
$$ |\vv_h| \le |\vv_{h^*_{L,\b}}| + c'_1\sum_{j=h+1}^{{h^*_{L,\b}-1}}
|g_{1,j+1}|^2 + c''_1 \sum_{j=h+1}^{{h^*_{L,\b}-1}}\e_j^2 \g^{-(j-h_{L,\b})}$$
By using \pref{g1b}, \pref{bej}, the inductive hypothesis that $\e_j\le 2
c_3\e_0$ and the fact that, by \pref{hstar}, $h^*_{L,\b}- h_{L,\b}-1\le \log
|g_{1,h^*_{L,\b}}|^{-2}$, we get
$$|\vv_h| \le c_3 \e_0 + 2 c'_1 |g_{1,h^*_{L,\b}}|^2 \log
|g_{1,h^*_{L,\b}}|^{-2} + 4 c_3^2 c''_1 \e_0^2 \sum_{n=1}^\io \g^{-1} \le
2 c_3 \e_0$$
if $\e_0$ is small enough.\Halmos

We finally show that the running coupling constants are well defined in the
zero temperature and thermodynamic limit.

\begin{lemma}\lb{lm2.5a}
For any fixed sequence $\n_h$, $h\in (h_{L,\b},1]$, satisfying \pref{ne2} and
any fixed $j\le 0$, $\lim_{\min\{\b,L\}\to\io} \vv_j=\bar \vv_j$ does exist;
moreover, $\lim_{j\to-\io}\bar \vv_j={\vv}_{-\io}$ with
\bal \lb{2.42} g_{2,-\io} &= g_{2,0}-{1\over 2}g_{1,0}+O(|\l|^{3/2})=
\left[2\hat v(0)-\hat v(2p_F) \right]\l+ O(|\l|^{3/2})\\
\lb{2.43} g_{4,-\io} &= g_{4,0} +O(\l^2) = 2\l\hat v(0)+O(\l^2) \cr \d_{-\io}
&= O(\l) \eal
\end{lemma}

\0{\bf Proof} - By applying the localization procedure (see \pref{2.75a},
\pref{2.75}) to the effective potential $\VV^{(0)}(\psi)$, we see that
\be\lb{hbb1} {\bf v}_{0}={\bf v}_{\io}
+\sum_{j=0}^\io\sum_{n=1}^\io\sum_{\t\in \TT_{j,n}} \sum_{{\bf P}:
|P_{v_0}|=m_\a}\int_{\L^M} d(\xx_{v_0}/\xx_0) K^{(j+1)}_{\LL,\t,L,\b,{\bf
P}}(\xx_{v_0})\ee
where $m_\a=4$, if $\a=1,2,4$, $m_\a= 2$, if $\a=\d$, $\L=\CC
\times(-\b/2,\b/2)$, $\xx_0$ is an arbitrary fixed point in the set
$\xx_{v_0}$, $M$ is the number of points in $\xx_{v_0}/\xx_0$, $\TT_{j-1,n}$
is the family of trees with scale root $j-1$, $n$ normal endpoints and no
special endpoint. Moreover, ${\bf v}_{\io}$ is the term of order $1$ in $\l$
and the kernels $K^{(j+1)}_{\LL,\t,{\bf P},L,\b}(\xx_{v_0})$ are obtained
from the kernels \pref{2.45} (where the dependence on $L$ and $\b$ was
hidden) by the procedure described after \pref{3.2aaax}. Let us now define
\be \bar{\bf v}_{0}=\bar{\bf v}_{\io}  +\sum_{j=1}^\io
\sum_{n=1}^\io\sum_{\t\in \TT_{j-1,n}} \sum_{{\bf P}: |P_{v_0}|=m_\a}
\int_{\L_\io^M} d(\xx_{v_0}/\xx_0) K^{(j+1)}_{\LL,\t,{\bf
P}}(\xx_{v_0})\lb{hbb2}\ee
where $\L_\io=\ZZZ\times\RRR$, $K^{(j+1)}_{\LL,\t,{\bf P}}(\xx_{v_0}):=
\lim_{\min\{\b,L\}\to\io} K^{(j+1)}_{\LL,\t,L,\b,{\bf P}}(\xx_{v_0})$,
$\bar{\bf v}_{\io} :=\lim_{\min\{\b,L\}\to\io} {\bf v}_{\io} =(2\l\hat v(0),
2\l \hat v(0), 2\l \hat v(2 p_F),\d_0)$. We want to prove that
\be\lb{d_v0} |{\bf v}_{0}- \bar {\bf v}_{0}|\le C_0|\l| \g^{h_{L,\b}}\ee
It is easy to see that a bound of this type is valid for $|{\bf v}_{\io} -
\bar{\bf v}_{\io}|$. Hence, if we call $T_{j,L,\b}$ and $T_{j,\io}$ the
contribution of the trees with scale root $j$ to the sum in \pref{hbb1} and
\pref{hbb2}, respectively, the bound \pref{d_v0} will be proved, if we prove
that
\be\lb{Tj} |T_{j,L,\b} - T_{j,\io}| \le C |\l|\g^{-j} {\g^{-(j-h_{L,\b})}}\ee
which differs from the dimensional bound of $T_{j,L,\b}$ and $T_{j,\io}$ for
the factor $\g^{-(j-h_{L,\b})} \le \g^{h_{L,\b}}$. Note that
\be |T_{j,L,\b} - T_{j,\io}| \le \sum_{n=1}^\io\sum_{\t\in \TT_{j-1,n}}
\sum_{{\bf P}: |P_{v_0}|=m_\a} \big( \D_{j,1} + \D_{j,2} \big)\virg
\text{where}\ee
\be\D_{j,1}:= \int^* d(\xx_{v_0}/\xx_0) \left| K^{(j+1)}_{\LL,\t,\b,L,{\bf
P}}(\xx_{v_0})- K^{(j+1)}_{\LL,\t,{\bf P}}(\xx_{v_0})\right|\ee
\be\lb{Dj2} \D_{j,2}:=\int^{**}_{\L^M} d(\xx_{v_0}/\xx_0)
\lft|K^{(j+1)}_{\LL,\t,\b,L,{\bf P}}(\xx_{v_0})\rgt| + \int^{**}_{\L^M_\io}
d(\xx_{v_0}/\xx_0) \lft|K^{(j+1)}_{\LL,\t,{\bf P}}(\xx_{v_0})\rgt|\ee
where $\int^* d(\xx_{v_0}/\xx_0)$ denotes the integration over the rectangle
centered in $\xx_0$ and with sides of length $L/4$ and $\b/4$, while
$\int^{**} d(\xx_{v_0}/\xx_0)$ denotes the integration over the complementary
region.

In order to bound $\D_{j,1}$, we note that the difference between the two
kernels comes from the oscillating factors $e^{i\bar\kk_{\h\h'}\xx}$, which
appear in the $\RR$ operation written in coordinate space (obtained by
Fourier transforming \pref{2.75a} and \pref{2.75}) and from the differences
between $g^{(k)}(\xx)$ and its $\b,L\to\io $ limit $g^{(k)}_\io(\xx)$.
Regarding the first kind of contributions, we note that the difference
between the two kernels can be written as a sum over $O(n)$ terms with at
least one factor $e^{i\bar\kk_{\h\h'}\xx}-1$ associated with a tree vertex;
this factor modifies the bound by a factor $\g^{-k+h_{L,\b}}$, where $k$ is
the scale of the vertex. Since $k\ge j+1$, the dimensional bound of a single
tree is modified by a factor $\g^{-j+h_{L,\b}}$, without modifying the
dimensional properties of the sum over the tree expansion.
Regarding the second kind of contributions, if we write
$g^{(k)}(\xx)=g^{(k)}_\io(\xx)+\d g^{(k)}_\io(\xx)$, we get the dimensional
bounds:
\be|\d  g^{(k)}_\io(\xx)|\le C \g^{k}\g^{-(k-h_{L,\b})} \virg \int_{|x|\le
\b/4,\atop|x_0|\le {\b\over 4}} d\xx |\d g^{(k)}_\io(\xx)|\le C \g^{-k}
\g^{-(k-h_{L,\b})}\nn\ee
which differ from the bounds of $g^{(k)}(\xx)$ by a factor
$\g^{-k+h_{L,\b}}$, with $k\ge j$. By using the stability property (see
Remark 2 at the end of \S\ref{ss2.5}), we see that the sum over the tree
expansion is modified again by a factor $\g^{-j+h_{L,\b}}$.

In order to bound $\D_{j,2}$, we note that, given any contribution to the one
of the kernels, one can select in the spanning tree used to perform the
integration (see \pref{2.55b}) a chain of propagators  connecting $\xx_0$
with a point $\hat\xx$ at a distance greater that $\min(L,\b)$, and this
produces an extra factor much smaller than $\g^{-(j-h_{L,\b})}$ in the bound,
in an obvious way. This concludes the proof of \pref{Tj}.

We now prove that $\lim_{\min\{\b,L\}\to\io} \vv_j=\bar \vv_j$ does exist
even for $j<0$. By using the notation of \S\ref{ss2.5}, we can write, if
$h_{L,\b}<j-1$: $v_{\a,j-1}= v_{\a,j}+\b^{(j)}_{\a,L,\b}({\bf v})$ , with
$\b^{(j)}_{\a,L,\b}({\bf v})$ of the form
\be\lb{hbb1a} \b^{(j)}_{\a,L,\b}({\bf v})=\sum_{n=1}^\io\sum_{\t\in
\TT_{j,n}} \sum_{{\bf P}: |P_{v_0}|=m_\a}\int_{\L^M} d(\xx_{v_0}/\xx_0)
K^{(j+1)}_{\LL,\t,L,\b,{\bf P}}(\xx_{v_0},{\bf v })\ee
We now call $\bar v_{\a,j}$ the solution of the recurrence equation: $\bar
v_{\a,j-1}=\bar v_{\a,j}+\b^{(j)}_{\a}(\bar {\bf v})$, with
\be \b^{(j)}_{\a}(\bar {\bf v})=\sum_{n=1}^\io\sum_{\t\in \TT_{j-1},n}
\sum_{{\bf P}: |P_{v_0}|=m_\a} \int_{\L_\io^M} d(\xx_{v_0}/\xx_0)
K^{(j+1)}_{\LL,\t,{\bf P}}(\xx_{v_0},\bar {\bf v})\lb{hbb2a}\ee
where $K^{(j+1)}_{\t,{\bf P}}(\xx_{v_0},{\bf v})= \lim_{\min\{\b,L\}\to\io}
K^{(j+1)}_{\t,L,\b,{\bf P}}(\xx_{v_0},{\bf v})$. We prove by induction that,
if $j<0$ and $\l$ is small enough,
\be\lb{d_vj} |{\bf v}_j-\bar{\bf v}_j|\le  \g^{-(j-h_{L,\b})}\ee
Note that this bound is not optimal, but it is sufficient for our purposes
and very easy to prove. We can write:
\be |v_{\a,j-1}- \bar v_{\a,j-1}|\le |v_{\a,0}- \bar v_{\a,0}| +\sum_{k=j}^0
|\b^{(k)}_{\a,L,\b}({\bf v})-\b^{(k)}_{\a}(\bar {\bf v})|\ee
The bound \pref{d_vj} is an immediate consequence of this inequality and
\pref{d_v0}, if we prove that
\be\lb{v1} |\b^{(k)}_{\a,L,\b}({\bf v})-\b^{(k)}_{\a}(\bar {\bf v})| \le C
|\l| \g^{-(k-h_{L,\b})}\ee
Let us write
\be\b^{(k)}_{\a,L,\b}({\bf v})-\b^{(k)}_{\a}(\bar {\bf v})=
[\b^{(k)}_{\a,L,\b}({\bf v})-\b^{(k)}_{\a,L,\b}(\bar{\bf v})]+
[\b^{(k)}_{\a,L,\b}(\bar{\bf v})-
\b^{(k)}_{\a}(\bar {\bf v})]
\ee
The first term can be easily bounded by induction, thanks to the stability
property of the tree expansion, while the second term can be bounded as in
the proof of \pref{Tj} and taking into account the following facts. If
$\a\not=\d$, there is no term of order $1$ in $\l$ in \pref{hbb1a} and
\pref{hbb2a}, so that we can iterate the bound \pref{d_vj}, by using it only
once in the endpoints of the tree expansion of $[\b^{(k)}_{\a,L,\b}({\bf
v})-\b^{(k)}_{\a,L,\b}(\bar{\bf v})]$. This is not true if $\a=\d$; however,
in this case the only terms with $n=1$ in \pref{hbb1a} and \pref{hbb2a}
depend only on ${\bf v}_0$ and $\bar{\bf v}_0$, respectively, since they are
obtained by contracting on scale $k<0$ the irrelevant term produced on scale
$0$ by the action of the $\RR$ operator. Hence, even in this case, we get a
factor $|\l|$ in the bound, after the insertion of \pref{d_vj}.

We still have to prove that $\lim_{j\to-\io}\bar \vv_j={\vv}_{-\io}$ does
exist and satisfies \pref{2.42} and \pref{2.43}. The first claim is
essentially trivial, since it is obvious that $\bar{\bf v}$ satisfies Lemma
\ref{lm2.5}, and, in particular, this implies that, if $\bar g_{1,0}\in
D_{\e,\d}$, with $\e$ small enough (how small depending on $\d$), $\bar
g_{1,j}$ goes to $0$, as $j\to -\io$, and $\sum_{j=h}^0 |\bar g_{1,j}|^2 \le
C\d^{-1}|\l|$, uniformly in $h$. This is an easy consequence of the condition
\pref{gerr} and the condition $\hat v(2p_F)>0$; note that the power $3/2$ in
the r.h.s. of \pref{gerr} could be replaced by $2-\h$, $\h>0$, but $2$ is not
allowed. Finally, the form of the flow \pref{floweq0} implies also that $\bar
g_{2,j}$, $\bar g_{4,j}$ and $\bar\d_{j}$ converge, as $j\to-\io$, to some
limits $g_{2,-\io}$, $g_{4,-\io}$ and $\d_{-\io}$ of order $\l$, satisfying
\pref{2.42} and \pref{2.43}.\Halmos

Let us now suppose that $\l$ is a (small) positive number; the previous
bounds imply that $\bar g_{1,j}>0$, for any $j\le 0$. The following Lemma
will allow us to control the logarithmic corrections to the power law fall-off
of the correlations.

\begin{lemma}\lb{lm2.6}
There are four sequences $w_{i,h}$, $\d_{i,h}$, $i=1,2$, $h\le j_0$, such
that
\be\lb{g1sum} \sum_{j=h}^{j_0} \bar g_{1,j}= (1+ w_{1,h}) \frac1{a} \log [1+a
\bar g_{1,j_0} (j_0-h)] + \d_{1,h} \ee

\be\lb{g2diff} \sum_{j=h}^{j_0} \big[ \bar g_{2,j} - g_{2,-\io} \big] = (1+
w_{2,h}) \frac1{2a} \log [1+a \bar g_{1,j_0} (j_0-h)] + \d_{2,h} \ee
with
\be\lb{wb} |w_{i,h}| \le C\l \virg |\d_{i,h}| \le C\l^{1/2}\ee
\be\lb{wb1} |w_{i,h-1} - w_{i,h}| \le \frac{C\l}{[1+a \bar g_{1,j_0} (j_0-h)]
\log [1+a \bar g_{1,j_0} (j_0-h)]}\ee
\end{lemma}

\0{\bf Proof} - Let us put $g_0=\bar g_{1,j_0}$, and $a(s)$ the function of
$s\ge 0$, such that $a(s)=a_{j_0-n}$, if $n\le s <n+1$. Then, by using
\pref{gtilde}, \pref{gerr} and \pref{bAj}, it is easy to see that
\be\lb{wb2} \lft| \sum_{j=h}^{j_0} \bar g_{1,j} - I_{j_0-h} \rgt| \le
C\l^{1/2} \virg I_n= \int_0^n ds \frac{g_0}{1+ g_0 \int_0^s dt\, a(t)} \ee
On the other hand, \pref{bAj} also implies that $a(s)=a +\l r(s)$, with
$|r(s)|\le C$; hence
\be\nn I_n= \int_0^n ds \frac{g_0}{1+ g_0 a s} -\l \int_0^n ds \frac{g_0^2\,
\int_0^s dt\, r(t)}{[1+ g_0 \int_0^s dt\, a(t)] [1+ g_0 a s ]}\ee
implying that
\be\nn \lft| I_n - \frac1a \log(1+ a g_0 n)\rgt| \le \frac{4C\l}{a^2}
\int_0^{ag_0 n} dx \frac{x}{(1+x)^2} < \frac{4C\l}{a^2} \log(1+ a g_0 n) \ee
Hence there is a constant $\tilde w_n$ such that $I_n = (1/a + \tilde
w_n)\log(1+ a g_0 n)$, with $|\tilde w_n| \le C\l$; this bound, together with
the bound in \pref{wb2}, proves \pref{wb} for $i=1$. To prove \pref{wb1},
note that
\be\nn |I_{n+1} - I_n| \le \int_n^{n+1} ds \frac{g_0}{1+g_0 \frac{a}2 s} =
\frac2{a} \log \lft(1+ \frac{a g_0}{2+a g_0 n} \rgt)\ee
\be\nn I_{n+1} - I_n = (1/a + \tilde w_{n+1})\log \lft(1+ \frac{a g_0}{1+a
g_0 n} \rgt) + (\tilde w_{n+1} - \tilde w_n)\log(1+ a g_0 n)\ee
so that, if $\l$ is small enough,
$$|\tilde w_{n+1} - \tilde w_n| \log(1+ a g_0 n) \le \lft(\frac3{a} +
C\l\rgt) \log \lft(1+ \frac{a g_0}{1+a g_0 n} \rgt) \le \frac{4 g_0}{1+
a g_0 n}$$
To prove \pref{wb} and \pref{wb1} for $i=2$, note that, by \pref{2.46} and
Lemma \ref{lm2.5}, if $j\le j_0$,
\be\nn\bsp \bar g_{2,j} - g_{2,-\io} &= \sum_{h=-\io}^{j} \lft[\frac{a}2 +
O(\l)\rgt] g_{1,h}^2 = \lft[\frac{a}2 + O(\l)\rgt] \int_{|j|}^\io ds
\frac{g_{1,0}^2}{(1+a g_{1,0} s)^2} + O(\tilde g_{1,j}^{3/2})\\
&=\lft[\frac12 + O(\l)\rgt]  \tilde g_{1,j} + O(\tilde g_{1,j}^{3/2})\esp\ee
Hence, the proof of \pref{wb} is almost equal to the previous one, while the
proof of \pref{wb1} needs a slightly different algebra; we omit the
details.\Halmos

\subsection{The flow of renormalization constants}\lb{sec2.4}
The renormalization constant of the free measure satisfies
\bal
\lb{ffgz} {Z_{j-1}\over Z_j} &= 1+ \b_z^{(j)}(\vec
g_j,\d_j,...,\vec g_0,\d_0)+ \bar\b_z^{(j)}(\vec v_j;..,\vec v_0;\l)\;;
\eal
while the renormalization constants of the densities, for $\a=C,S_i,SC,TC_i$ and
$i=1,2$, satisfy the equations
\bal
\lb{ffg} {Z^{(i,\a)}_{j-1}\over Z^{(i,\a)}_j} &= 1+ \b_{(i,\a)}^{(j)}(\vec
g_j,\d_j,...,\vec g_0,\d_0)+ \bar\b_{(i,\a)}^{(j)}(\vec v_j;..,\vec v_0;\l)\;.
\eal
In these two formulas,
by definition, the $\b_t^{(j)}$ functions, with $t=z$ or $(i,\a)$, are
given by a sum of multiscale graphs, containing only vertices with r.c.c.  $\vec
g_h,\d_h$, $0\ge h \ge j$, modified so that the propagators $g^{(h)}_{\o}$ and the
renormalization constants $Z_h$, $Z^{(i,\a)}_h$, $0\ge h\ge j$, are replaced by
$g^{(h)}_{{\rm D},\o}$, $Z_h^{(D)}$, $Z^{(D,i,\a)}_h$ (the definition of
$Z^{(D,i, \a)}_h$ is analogue to the one of $Z_h^{(D)}$); the $\bar\b_t^{(j)}$
functions contain the correction terms together the remainder of the expansion.
Note that, by definition, the constants $Z_j^{(D)}$ are exactly those generated
by \pref{ffgz} and \pref{bal} with $\bar\b_z^{(j)}=\bar\b_\a^{(j)}=0$. Note
also that $|\bar \b_z^{(j)}|\le C \bar v_j^2 \g^{\th j}$, while
$|\bar\b_{(i,\a)}^{(j)}|\le C \bar v_j \g^{\th j}$.

By using \pref{ffgz} and \pref{ffg}, we can write
\be
{Z_{j-1}^{(1,\a)} \over Z_{j-1}}= {Z_{j}^{(1,\a)}\over Z_{j}} \lft[1+
\b_{z,(1,\a)}^{(j)}(\vec g_j,\d_j) + \hat\b_{z,(1,\a)}^{(j)}(\vec v_j;..,\vec
v_0;\l)\rgt]
\ee
with $|\hat\b_{z,(1,\a)}|\le C \bar v_j \g^{\th j}$. If we define $\tilde
\b_{z,(1,\a)}^{(j)}(\vec g,\d)$ the value of $\b_{z,(1,\a)}^{(j)}(\vec
g_j,\d_j;...;\vec g_0,\d_0)$ at $(\vec g_i,\d_i)=(\vec g,\d)$, $j\le i\le 0$
and $\tilde \b_{z,(1,\a)}^{(j,\le 1)}(\vec g,\d)$ the sum of its terms of
order $0$ and $1$ in $g_{1,h}$, it turns out that
\be \lb{beta44}|\tilde \b_{z,(1,\a)}^{(j,\le 1)}(\vec g,\d)|\le C [\max
\{|g_{1}| |g_{2}|, |g_{4}|, |\d|\}]^2 \lft[\g^{\th h} +
\g^{-(j-h_{L,\b})}\rgt]\virg \mbox{if $\a=C$} \ee
This bound, as crucial as the analogous bound \pref{beta23}, has been proved
in \cite{M007_3}; see App. \ref{appB} for some detail.
The bound \pref{beta44}, together with $\sum_{k=j}^0 |g_{1,k}|^2 \le C|\l|$ and
the fact that $Z^{(1,S_i)}_h=Z^{(1,C)}_h$ by the SU(2) spin symmetry, imply
that
\be\lb{2.48}
\left| \frac{Z_{j}^{(1,\a)}}{Z_j}-1 \right|\le C |\bar\e_j^2| \virg \a=C,S_i
\ee
Regarding the flow of the other renormalization constants, we can write
\be\lb{2.53} Z^{(t)}_j= \g^{-\h_t j} \hat Z^{(t)}_j \ee
where $Z^{(z)}_j=Z_j$ and, by definition,
\be\lb{fff}
\h_t\=\lim_{j\to -\io}\h_{t,j} :=
\log_\g \Big[ 1+ \b_t^{(0,j)}(g_{2,-\io},
g_{4,-\io},\d_{-\io};...;g_{2,-\io},g_{4,-\io} ,\d_{-\io}) \Big]
\ee
Note that the exponents $\h_t$ are functions of $\vec v_{-\io}$ only, an
observation which will play a crucial role in the following. Moreover, by an
explicit first order calculation, we see that
\be\lb{expt} \h_t=
\begin{cases}
(2\p v_F)^{-1} g_{2,-\io} + O(\l^2)& t=(2,C), (2,S_i)\\
-(2\p v_F)^{-1}  g_{2,-\io} + O(\l^2)& t=(2,SC), (2,TC_i)\\
\qquad O(\l^2) & \text{otherwise}
\end{cases}
\ee
while
\bal {\hat Z^{(t)}_{h-1}\over \hat Z^{(t)}_h} &= 1 + O(\tilde g_{1,h}\l) +
r^{(t)}_h \virg t=z,(1,\a), \a\not=TC_i\nn\\
{\hat Z^{(2,C)}_{h-1}\over \hat Z^{(2,C)}_h} &=1- a g_{1,h} + \frac{a}2
(g_{2,h}-g_{2,-\io})+O(\tilde g_{1,h}\l) + r^{(2,C)}_h\nn\\
\lb{2.56} {\hat Z^{(2,S_i)}_{h-1}\over \hat Z^{(2,S_i)}_h} &=1 +
\frac{a}2 (g_{2,h}-g_{2,-\io}) + O(\tilde g_{1,h}\l) + r^{(2,S_i)}_h\\
{\hat Z^{(2,SC)}_{h-1}\over \hat Z^{(2,SC)}_h} &=1 - \frac{a}2 g_{1,h} -
\frac{a}2 (g_{2,h}-g_{2,-\io}) + O(\tilde g_{1,h}\l) + r^{(2,SC)}_h \nn\\
{\hat Z^{(2,TC_i)}_{h-1}\over \hat Z^{(2,TC_i)}_h} &=1 +  \frac{a}2 g_{1,h} -
\frac{a}2 (g_{2,h}-g_{2,-\io}) +O(\tilde g_{1,h}\l) + r^{(2,TC_i)}_h\nn
\eal
where $a$ and $\tilde g_{1,h}$ are defined as in \pref{defa} and
\pref{gtilde}, respectively, and $\sum_{h=-\io}^0 |r^{(t)}_h| \le C|\l|^2$.
Let us define:
\be\lb{2.88a} q^{(h)}_t = {\log \hat Z^{(t)}_h\over \log (1 + a g_{1,0}|h|)}
\ee
Hence, by using \pref{g1sum}, \pref{g2diff} and \pref{wb}, we get
\be\lb{2.57} \bsp |q^{(h)}_t| \le C\l\virg
& t=z,(1,\a), \a\not=TC_i\\
|q^{(h)}_t -\frac12 \bar\z_\a| \le C\l\virg & t=(2,\a) \esp \ee
where the constants $\bar\z_\a$ are those of \pref{zalfa}.

The existence of the zero temperature and thermodynamic limit can be done
exactly as in Lemma \ref{lm2.5a}.

\subsection{Flow of $\n_h$ and calculation of $\n$.}\lb{sec2.nu}

 The sequence $\n_j$, $h_{L,\b}\le j \le 1$, must satisfy the recursive
 equation \pref{floweq0} with $\a=\n$.
 If we decompose $\hat\b^{(j)}_\n$ as in \pref{bal}, the function
 $\b_\n^{(j)}$ is exactly equal to $0$, because of the oddness of the
 propagator $g^{(h)}_{D,\o}$ and the fact that all endpoints are local (those
 with scale $2$ are excluded), hence do not contain oscillator factors.
 As concerns the function $\bar\b^{(j)}_\n$, in this case we
have to extract the contribution of the trees with at least one endpoint of
type $\n$ and we get, if $\th<\th'<1$,
\be\lb{nuh} \n_{j-1} = \g \n_j + \bar \b_\n^{(j)} = \g\n_j+  \e_j
\sum_{i=j}^1 \n_i \,\bar\b_{j,i} \g^{-\th' (i-j)} + \e_j \g^{\th' j}
\bar\b_j\virg |\bar\b_{j,i}|, |\bar\b_j| \le c_0\ee
If we iterate this equation, we get $\n_h=\g^{-h+1}[ \n_1 +\sum_{j=h+1}^1
\g^{j-2} \bar\b_\n^{(j)}]$. We want to show that it is possible to choose
$\n_1=\n$ so that the sequence $\n_h$ solves \pref{nuh} and satisfies, for
$\l$ small enough, the bound \pref{ne2} (with $\x$ large enough, see Remark
after \pref{2.34})
and $g_{1,0}\in D_{\e_0,\d}$ (see Lemma \ref{lm2.4}); these are indeed the
conditions that allowed us to control the flow of the r.c.c. and the
renormalization constants. The choice of $\n_1$ is of course not unique, at
finite $L$ and $\b$, hence we add the constraint that $\n_{h_{L,\b}} = 0$, so
that the sequence $\n_h$ must satisfy the equation
\be\lb{nuT} \n_h=-\sum_{j=h_{L,\b}+1}^h \g^{-(h-j+1)} \bar\b_\n^{(j)} :=
\bT(\un)_h\ee
if $\un$ denotes the sequence $\n_h$. Let us now consider the Banach space
$\BB_\th$ of the sequences $\un$ with norm $\|\un\|_\th = \max_j \g^{-\th j}
|\n_j|$. We want to show that the operator $\bT$ is well defined on the
closed ball $\MM_\x = \{\un: \|\un\|_\th \le \x |\l|\}$ as a bounded operator
$\bT: \MM_\x \mapsto \MM_\x$, if $\x$ is large enough and $\l$ is small
enough. This implies that the solution of our problem is a fixed point of the
operator $\bT$ in $\MM_\x$ and that this solution does exist and is unique,
if we also prove that $\bT$ is a contraction on $\MM_\x$.

Let us first prove that $\bT$ is a bounded operator of $\MM_\x$ into
$\MM_\x$. By using \pref{nuh}, we easily see that, if $\un\in \MM_\x$, $\l\in
D_{\e_0,\d}$ and $\e_j\le c_3\e_0$, then
$$\|\bT(\un)\| \le c_3 c_0 \e_0 \sum_{n=0}^\io \g^{-(1+\th'-\th)n} [1 + \x|\l|
\sum_{n=0}^\io \g^{-(\th' -\th)n}] := c_2 \e_0(1+ \bar c_2|\l|\x)$$
Hence, $\|\bT(\un)\|\le \x|\l|$, if $c_2 \bar c_2 \e_0\le 1/2$ and $\x\ge 2
c_2 \e_0/|\l|$. The proof that $\bT$ is a contraction on $\MM_\x$, if $|\l|$
is small enough, is a bit more subtle, since now we can not ignore that the
r.c.c. and the renormalization constants do depend on $\un$. Let us call
$\vv_j$ and $\vv'_j$ the r.c.c. corresponding to the sequences $\un\in\MM_\x$
and $\un'\in\MM_\x$, respectively; analogously, we shall define
$z_j=Z_{j-1}/z_j-1$ and $z'_j$. We see immediately that $\max\{|\vv_0 -
\vv'_0|,|z_0-z'_0|\} \le c_0 |\l| |\n_1-\n'_1|$; we shall prove iteratively
that there exists $c_1>c_0$ such that
\be\lb{disln} \max\{|\vv_h - \vv'_h|,|z_h-z'_h|\} \le c_1 |\l|
\|\un-\un'\|_\th \ee
The bound for $z_h-z'_h$ follows easily from that for $\vv_h - \vv'_h$; hence
we shall discuss in some detail only the bound for $\vv_h - \vv'_h$. Suppose
that \pref{disln} is satisfied for $h\ge j+1$. In order to iterate this
bound, we have to control very carefully the flow of the quantity
$\D_h:=g_{1,h}-g'_{1,h}$. The result can be easily explained, if one consider
the approximate flow of $g_{1,h}$, obtained by substituting the r.h.s. of
\pref{2.46} with $-\frac12 a g_{1,j}^2$. In this case we should get
\be \D_{h-1}= \D_h-\frac{a}2 \D_h[ g_{1,h}+g'_{1,h}]\ee
which easily implies, by using the bound \pref{gerr} (which is uniform in
$\un$), that
\be\lb{dhflow0} |\D_h| \le |g_{1,0}-g'_{1,0}| e^{-c_2\log (1+\bar a |\l||h|
|)}| \le \frac{c_0 |\l||\n_1-\n_1'|}{(1+\bar a |\l||h| |)^{c_2}}\ee
for some positive constants $c_2$ and $\bar a$. A careful analysis of the
real flow can be done by proceeding as in the proof of Lemma \ref{lm2.4} and
Lemma \ref{lm2.5}; of course it involves also the other r.c.c. and has to use
the bound \pref{disln} for $h\ge j+1$ to shows that this bound is correct for
$h=j$. One can see that the bound \pref{dhflow0} is indeed true, if one
substitutes $c_0$ with some other constant $\bar c_0>c_0$ and $|\n_1-\n_1'|$
with $\|\un-\un'\|_\th$, that is
\be\lb{dhflow} |\D_h| \le \frac{\bar c_0 |\l|\|\un-\un'\|_h}{(1+\bar a
|\l||h| |)^{c_2}}\ee
By using \pref{floweq0}, we can write
\be \vv_h - \vv'_h = \vv_0 - \vv'_0 + \sum_{j=h+1}^0 [\hat\b^{(j)} -
\hat\b^{(j)'}]\ee
with $\hat\b^{(j)}_\a = \hat\b^{(j)}_\a(\vv_j,\ldots,\vv_0;\l,\n_1)$ and
$\hat\b^{(j)'}_\a = \hat\b^{(j)}_\a(\vv'_j,\ldots,\vv'_0;\l,\n_1')$.
Moreover, by analyzing in detail the structure of the functions
$\hat\b^{(j)}_\a$ discussed in \S\ref{sec2.2} and the short memory property,
one can see that
\be |\hat\b^{(j)} - \hat\b^{(j)'}| \le c_3 \Big[ |\tilde g_{1,j} \D_j| +
(|\tilde g_{1,j}|^2 +\e_j \g^{\th j}) ( c_1|\l| +1) \|\un-\un'\|_\th\Big] \ee
Hence, if $c_1|\l|\le 1$, $|\vv_h - \vv'_h| \le |\vv_0 - \vv'_0| + c_4\e_j
\|\un-\un'\|_\th \le c_5|\l| \|\un-\un'\|_\th$. It follows that the bound
\pref{disln} is true for $c_5=c_1$ and $c_1|\l|\le 1$.

By using \pref{nuT} and \pref{nuh}, we have
\be\bsp &|\bT(\un)_h-\bT(\un')_h| \le \sum_{j=h_{L,\b}+1}^h \g^{-(h-j+1)}
|\b_\n^{(j)}-\b_\n^{(j)}| \le\\
&\sum_{j=h_{L,\b}+1}^h \g^{-(h-j+1)} [c_1 \g^{\th j} \sup_h |\vv_h-\vv'_h| +
c_2 |\l| \g^{\th j} \|\un-\un'\|_\th] \le c_3|\l| \g^{\th h} \|\un-\un'\|_\th
\esp\ee
Hence, $\bT$ is a contraction, if $c_3|\l| <1$.

\subsection{Calculation of $p_F(\bar\m,\l)$}\lb{sec2.5aa}

Let us consider the equation \pref{nuT} in the limit $L,\b\to\io$; its
solution gives the sequence $\n_h$, whose first element $\n_1$ is the unique
value of the function $\n(\m,\l)$ which allows us to fix at $p_F=\arccos \m$ the
value of the interacting Fermi momentum. We want to show that the equation
\be\lb{meq} \bar\m=\m+\n(\l,\m) \ee
can be solved with respect to $\m$ by a function $\m(\bar \m,\l)$, if $\l$ is
small enough; the interacting Fermi momentum will then be given by
$p_F(\bar\m,\l) = \arccos \m(\bar \m,\l)$. In order to prove this statement,
it is of course sufficient to prove that $|\dpr\n/\dpr\m|\le C|\l|$; since
$|\m|<1$, this is equivalent to prove that $|\dpr\n/\dpr p_F|\le C|\l|$.

We do not have an explicit expression of $\n$, but we know that it is equal
to the first element $\n_1$ of the sequence $\un$ which uniquely solves the
equation \pref{nuT}. If we make explicit the dependence of $\bT$ on $p_F$, we
have $\un = \bT(\un,p_F)$. Note that the operator $\bT(\un,p_F)$ depends on
$p_F$ explicitly through the kernels appearing in the tree expansion of the
functions $\bar\b_\n^{(j)}$ and indirectly trough the r.c.c. $\vv_j$:
\be\lb{Th} T_h(\un,p_F) = -\sum_{j=-\io}^h \g^{-(h-j+1)}\bar \b_\n^{(j)}
\big(\vv_j(\un_{\ge j},p_F),\ldots,\vv_0(\un_{\ge 0},p_F); \un_{\ge
j},p_F\big)\ee
where $\un_{\ge j}=(\n_j,\ldots,\n_1)$. Hence, we can calculate the sequence
$\uxi=\dpr\un/\dpr p_F$, by solving the equation
\be (I- \bA)\uxi = \ub \virg A_{h,i}=\frac{\dpr T_h}{\dpr \n_i} \virg b_h=
\frac{\dpr T_h}{\dpr p_F}\ee
The fact that $|\dpr\n/\dpr p_F|\le C|\l|$, for $\l$ small enough,
immediately follows from the following Lemma.

\begin{lemma}
If $\BB_{-\h}$ is defined as after \pref{nuT}, then $\ub\in \BB_{-\h}$, for
any $\h\in(0,1)$ (hence the sequence $b_h$ can diverge as $h\to -\io$) and
$\|\ub\|_{-\h} \le \bar c_\h|\l|$, with $\bar c_{\h} \to \io$ if $\th\to
0^+$. Moreover, $\bA$ is a bounded linear operator on $\BB_{-\h}$, with norm
$\|\bA\| \le c_\h |\l|$, so that, if $c_\h|\l|\le 1/2$, $\uxi\in\BB_{-\h}$
and $\|\uxi\|_{-\h} \le 2 \|\ub\|_{-\h}$.
\end{lemma}

\0{\bf Proof} - By using \pref{Th}, we get
\be b_h = -\sum_{j=-\io}^h \g^{-(h-j+1)} \left[ \frac{\dpr\bar
\b_\n^{(j)}}{\dpr p_F} + \sum_{k=j}^0 \frac{\dpr\bar \b_\n^{(j)}}{\dpr \vv_k}
\frac{\dpr \vv_k}{\dpr p_F}\right] \ee
Thanks to the short memory property, the bound \pref{2.34} is valid also for
$\dpr\bar \b_\n^{(j)}/\dpr \vv_k$ with $\e_j$ in place of $\e_j^2$. Hence, we
get, if $\th<1$ and $\un\in\BB_\th$,
\be\lb{bh} |b_h| \le \sum_{j=-\io}^h \g^{-(h-j+1)} \left[ \Big|\frac{\dpr\bar
\b_\n^{(j)}}{\dpr p_F}\Big| + C|\l|\g^{\th j} \max_{k\ge j} \Big|\frac{\dpr
\vv_k}{\dpr p_F}\Big| \right] \ee

In order to evaluate the derivatives with respect to $p_F$, note that there
is a dependence related to the dependence on $p_F$ of the single scale UV
propagators $g^{(h)}(\xx)$, $h\ge 1$ (see \pref{B.2}) and the single scale IR
propagators $g_\o^{(h)}(\xx)$, $h\le 0$ (see \pref{2.58}); it does not give
any trouble, since the bound of $\dpr g_\o^{(h)}(\xx)/\dpr p_F$ is similar to
that of $g_\o^{(h)}(\xx)$, as concerns dimensional arguments (even better in
the UV case).

In the IR scales there is also a dependence on the oscillator factors $e^{i
p_F \xx}$, which appear on the representation \pref{2.17b} of the effective
potential $\VV^{(0)}$ in terms of the $\psi^{\pm}_{\xx,\o,s}$ fields. In the
kernel of a tree this dependence will produce a bad factor $(\xx-\yy)$
multiplying the propagator of scale $j$ joining the points $\xx$ and $\yy$,
hence a factor $\g^{-j}$ in the dimensional bound. However, such oscillating
factors are not present in the local part of $\VV^{(0)}$; they only appear if
the tree has at least one endpoint of scale $+2$. It follows that
\be\lb{dbn} \Big|\frac{\dpr\bar \b_\n^{(j)}}{\dpr p_F}\Big| \le C|\l|^2
(\g^{-(1-\th)j}+\g^{\th j}) \le C|\l|^2 \g^{-(1-\th)j} \ee
In a similar way we can bound $\dpr \vv_k/\dpr p_F$. However, since $\vv_k$
depends on $p_F$ also trough $\vv_j$, $j>k$, we get a diverging contribution
also from the trees without oscillating factors. A simple analysis allows us
to show, starting from the decomposition of the Beta function \pref{bal},
that
\be \Big|\frac{\dpr \vv_{j-1}}{\dpr p_F}\Big| \le \Big|\frac{\dpr \vv_j}{\dpr
p_F}\Big| + C_1 |\l|^2 \g^{-(1-\th)j}+ C_2|\l| \max_{k\ge j} \Big|\frac{\dpr
\vv_k}{\dpr p_F}\Big| \ee
which implies the bound
\be\lb{dvjpF} \Big|\frac{\dpr \vv_j}{\dpr p_F}\Big| \le
C|\l|\g^{-(1-\th)j}\ee
If we insert this bound and \pref{dbn} in \pref{bh}, we get that $|b_h| \le
C|\l|\g^{-(1-\th)h}$.

To complete the proof, we shall now prove that, for any $\th'\in (0,1)$,
\be\lb{Ahi} |A_{h,i}| \le C|\l|\g^{-\th'|h-i|} \ee
By using \pref{Th} and the fact that $\bar\b_\n^{(j)}$ and $\vv_j$ are
independent of $\n_i$ and $\vv_i$, if $i<j$, we get
\be A_{h,i} = -\sum_{j=-\io}^{\min\{i,h\}} \g^{-(h-j+1)} \left[
\frac{\dpr\bar \b_\n^{(j)}}{\dpr \n_i} + \sum_{k=j}^i \frac{\dpr\bar
\b_\n^{(j)}}{\dpr \vv_k} \frac{\dpr \vv_k}{\dpr \n_i}\right] \ee
By proceeding as in the proof of \pref{dvjpF}, we see that, if $i\ge k$, for
any $\th\in(0,1)$,
\be\lb{dvjni} \Big|\frac{\dpr \vv_k}{\dpr \n_i}\Big| \le
C|\l|\g^{(1-\th)(i-k)}\ee
On the other hand, by using the properties of $\bar\b_\n^{(j)}$ described
before \pref{nuh} and the fact that the only term of order $1$ in $\l$ does
not depend on $\vv_j$, $j\le 0$, we get, if $i\ge j$ and $k\ge j$,
\be \left| \frac{\dpr\bar \b_\n^{(j)}}{\dpr \n_i} \right| \le
C|\l|\g^{-\th(i-j)} \virg \left| \frac{\dpr\bar \b_\n^{(j)}}{\dpr \vv_k}
\right| \le C|\l| \g^{\th j} \g^{-\th(k-j)}\ee
Hence, if $i\le h$,
\be\bsp |A_{h,i}| &\le C|\l|\g^{-(h-i)} \sum_{j=-\io}^i \g^{-(i-j)} \left[
\g^{-\th(i-j)} +\sum_{k=j}^i \g^{\th j} \g^{-\th(k-j)} \g^{(1-\th)(i-k)}
\right]\\
&\le C|\l| [1+\g^{\th i}] \g^{-(h-i)} \le C|\l|\g^{-(h-i)}\esp\ee
In the case $i>h$, we have, if $\th>1/2$,
\be |A_{h,i}| \le C|\l|\sum_{j=-\io}^h \g^{-(h-j)} \left[ \g^{-\th(i-j)}
+\g^{\th i} \g^{(1-2\th)(i-j)} \right] \le C|\l|\g^{-(2\th-1)(i-h)}\ee
Hence, the bound \pref{Ahi} is proved, with $\th'=2\th-1$. The Lemma then
follows immediately from \pref{Ahi} with $1>\th'>\h$.\Halmos

\section{Proof of Theorem \ref{th1.1}}\lb{sec2.5}


\subsection{The zero temperature and thermodynamic limit of the free energy}
\lb{sec2.5en}

Note first that the Grassmann integrals for the free energy and the Schwinger
functions are analytic in the domain $D$ \pref{2.12a}, as a consequence of
Lemmas \ref{p2.2}, \ref{p2.3}, \ref{lm2.4} and \ref{lm2.5}; therefore, by
proposition 2.1, they coincide with the free energy and Schwinger functions
of the Hubbard model.

Let us prove first the zero temperature and thermodynamic limit of the free
energy $E_{\b,L}$, which is given (see \pref{Eh} and \pref{blue1} for the
notation) by
\be E=\lim_{\b\to\io }\lim_{L\to \io} \lft[\sum_{j=h_{L,\b}}^0
(t_j^{(\b,L)}+\tilde E_j^{(\b,L)})+\sum_{j=1}^\io e_j^{(\b,L)} \rgt] \ee
We can indeed prove an even stronger result, that is the convergence under
the condition that $\min\{\b,L\} \to\io$. In fact, we shall prove that, given
$\e>0$, there exists $h_\e^*$, such that, if $\min\{\b,L\}\ge h_\e^*$, then
\be\lb{limEj} \left|\sum_{j=h_{L,\b}}^0 \tilde E_j^{(\b,L)}-\sum_{j=-\io}^0
\tilde E_j\right|\le \e \ee
where  $\tilde E_j=\lim_{\min\{\b,L\}\to\io} \tilde E_j^{(\b,L)}$ and this
limit does exist, since, by Lemma \ref{lm2.5a}, the r.c.c. involved in the
tree expansion converge in the same limit, as well as the kernels involved in
the definition of $\tilde E_j^{(\b,L)}$, by the same arguments used in the
proof of Lemma \ref{lm2.5a}.

In order to prove \pref{limEj}, we note that, given $\e>0$, there exists
$h_\e$ such that $|\sum_{j=h_{L,\b}}^{h_\e} \tilde
E_j^{(\b,L)}-\sum_{j=-\io}^{h_{\e}} \tilde E_j|\le \e/2$, as $|\tilde
E^{L,\b}_j|+|\tilde E_j|\le C \g^{2 j}$, by Lemma \ref{p2.3}, eq.
\pref{2.17aa} and Lemmas \ref{lm2.4} and \ref{lm2.5}. Moreover,
\be \tilde E_j^{(\b,L)}=\sum_{n=1}^\io\sum_{\t\in \TT_{j-1},n}
\sum_{{\bf P}: P_{v_0}=0} \int_{\L_\io^M} d(\xx_{v_0}/\xx_0)
K^{(j+1)}_{\t,{\bf P}}(\xx_{v_0},\bar {\bf v})\ee
so that, by using Lemma \ref{lm2.5a} and the procedure described in its
proof, we get;
\be \left|\sum_{j=h_\e}^0 \tilde E_j^{(\b,L)}-\sum_{j=h_\e}^0 \tilde
E_j\right|\le C|\l |\g^{-(h_\e-h_\e^*)}\le {\e\over 2} \ee
for $h_\e^*$ large enough. This argument can be repeated for $e_j$ using
that, by Lemma \ref{p2.2}, $|e_j|\le C \g^{-j}$, while the convergence of the
contribution of $t_j$ follows immediately from its very definition, see the
lines after \pref{2.77}.

\subsection{Tree expansion for the density correlations}\lb{sec2.5a}

Let us consider now the density correlations.
The tree expansion described in \S\ref{ss2.5} implies that $\O_\a(\xx-\yy)$
can be written as the sum over the values associated with all trees with $2$
special endpoints of type $J^\a$ and fixed space-time points $\xx$ and $\yy$,
a number $n\ge 0$ of normal endpoints and a root of scale $h\le 0$; moreover,
these trees must satisfy the condition that $|P_{v_0}|=0$ (no external legs
in the vertex $v_0$ of scale $h+1$ following the root), while (as always)
$|P_v|>0$ for all other vertices. We shall call $v_\xx$ and $v_\yy$ the two special
endpoints and $h_\xx+1$, $h_\yy+1$ their scale labels; moreover, we shall
denote $v_{\xx,\yy}$ the higher vertex such that $v_{\xx,\yy}< v_\xx,v_\yy$
and we shall call $h_{\xx,\yy}$ its scale.

Let un consider for definiteness $\O_C(\xx)$. The corresponding trees can be
grouped in three classes:

\0 1) the trees with both special end-points associated to the field monomial
$[Z_j^{(1,\a)}/Z_j] O_\xx^{(1,C)}$ with $j=h_\xx\le 0$ or $j=h_\yy\le 0$, see
\pref{2.20};

\0 2) the trees with both special end-points associated to the field monomial
$[Z_j^{(2,\a)}/Z_j] O_\xx^{(2,C)}$;

\0 3) the other trees, that is those which have at least one special endpoint
of scale $+2$ and those which have both special endpoint of scale $\le 1$,
associated one to $O_\xx^{(1,C)}$ and the other to $O_\xx^{(2,C)}$.

If one extracts from the first two classes the trees with no normal endpoints
and substitutes in their values the propagators $g^{(h)}_\o(\xx)$ with their
asymptotic expressions $g^{(h)}_{{\rm D},\o}(\xx)$, see \pref{gjth}, one gets
the following expression
\bal \O_C(\xx) &= \O^{(1,C)}(\xx) + \cos(2p_F x)
\O^{(2,C)}(\xx)+\O^{(3,C)}(\xx)\lb{2.175}\\
\O^{(1,C)}(\xx) &= 2\sum_\o \sum_{h,h'=h_{L,\b}}^0 \frac{\lft[
Z^{(1,C)}_{h\vee h'} \rgt]^2}{Z_h Z_{h'}} g^{(h)}_{{\rm D},\o}(\xx)
g^{(h')}_{{\rm D},\o}(\xx)+\sum_{h=h_{L,\b}}^0 \left[\frac{Z^{(1,C)}_h}{Z_h}
\right]^2 R_1^{(h)}(\xx)\lb{2.176}\\
\O^{(2,C)}(\xx) &= 4 \sum_{h,h'=h_{L,\b}}^0 \frac{\lft[ Z^{(2,C)}_{h\vee h'}
\rgt]^2}{Z_h Z_{h'}} g^{(h)}_{{\rm D},+}(\xx) g^{(h')}_{{\rm D},-}(\xx)+
\sum_{h=h_{L,\b}}^0 \left[\frac{Z^{(2,C)}_{h}}{Z_h}\right]^2
R_2^{(h)}(\xx)\lb{2.177}\\
\O^{(3,C)}(\xx) & =\sum_{h=h_{L,\b}}^1 \frac{Z^{(1,C)}_{h}
Z^{(2,C)}_{h}}{Z_h^2} R_3^{(h)}(\xx)\nn \eal
where $h\vee h'=\max\{h,h'\}$ and the definition of $Z^{(i,C)}_{h}$ has been
extended to $h=1$ as $Z^{(i,C)}_1 =1$. $R_i^{(h)}(\xx-\yy)$ is defined, for
$i=1,2$ as the sum over all trees of the class $i$ with $n\ge 1$ normal
endpoints, such that $h_{\xx,\yy}=h$ and $h_r=h_{L,\b}-1$, if $h_r$ is the
scale of the root, plus the corrections to the terms with no normal
endpoints. $R_3^{(h)}(\xx-\yy)$ is the sum over all trees of the class $3$,
such that $h_{\xx,\yy}=h$ and $h_r=h_{L,\b}-1$.

The functions $R_i^{(h)}(\xx)$, $i=1,2,3$ have the role of corrections, since
we can show that
\be\label{pr}  |R_i^{(h)}(\xx-\yy)|\le C_N (|\l| + \g^{\th h}) {\g^{2h}\over
1+[\g^h|\xx-\yy|]^N}\ee
\be\label{pr1} |R_3^{(h)}(\xx-\yy)|\le C_N {\g^{2h}\g^{\th h}\over
1+[\g^h|\xx-\yy|]^N}\ee

In order to prove \pref{pr}, let us consider a tree in the tree expansion of
$R_i^{(h)}(\xx-\yy)$ and note that, given a fixed spanning tree graph $T$
defined as in \pref{B.17}, there is a unique path $\CC_{\xx,\yy}\in T$
joining $v_\xx$ with $v_\yy$; for each line $l$ of this path, there is a
propagator of scale $h_l \ge h=h_{\xx,\yy}$. If one takes into account the
effects of the regularization procedure, some of these propagators are
derived and join some interpolated points in place of the space-time points
associated to the endpoints following $v_{\xx,\yy}$; however, by using the
fact that $|g^{(j)}_\o(\xx)| \le C_N \g^j [1+ (\g^j |\xx|)^N]^{-1}$, one can
show (see \S5.9 of \cite{BM001}) that one can extract from these chain of
propagators at least a decaying factor $C_N (2n+1)^N [1+ (\g^j
|\xx|)^N]^{-1}$, where $n$ is the number of normal endpoints; note that this
bound is trivial in absence of regularization. After this operation, one can
bound the sum over all trees as in the proof of \pref{2.17aa}, with two main
differences. First of all, one has to perform the sum over the scale indices
by fixing $h$ in place of $h_r$, but this does not change nothing, since the
scaling dimensions of the non trivial vertices are all positive, except that
of $v_0$, which is $0$. The second difference is that one has to take into
account that now two of the space-time points are fixed; hence, in order to
perform the integrals over the other points, one can still use the
propagators in the spanning tree graphs, but one has to neglect one of them;
since the path $\CC_{\xx,\yy}$ always contains at least one propagator of
scale $h$, this implies that one gain a factor $\g^{2h}$ with respect to the
bound leading to \pref{2.17aa}, which has to be also deprived of the volume
factor $L\b$ and multiplied by the decaying factor. As concerns the factors
$Z_{h_\xx}^{(i,C)}/Z_{h_\xx}$ and $Z_{h_\yy}^{(i,C)}/Z_{h_\yy}$, one can use
\pref{ass} to ``change'' their scale to $h$, at the price of a innocuous
factor $e^{c_1\e_0 [(h_\xx-h) + (h_\yy-h)]}$, which can be distributed along
the paths joining $v_\xx$ and $v_\yy$ with $v_{\xx,\yy}$, by slightly
modifying the factors $\g^{-(h_v-h_{v'})[D_v+z(|P_v|,1)]}$ associated to the
corresponding non trivial vertices, see \pref{B.18caa}. In the case of the
corrections to the leading term, coming from the trees with no endpoints, one
has also a factor $\g^{\th h}$ coming from the bound \pref{2.30} and Lemma
\ref{lm2.5a}.

The proof of \pref{pr1} is very similar. One has only to remark that all the
trees involved in the tree expansion of $R_3^{(h)}(\xx-\yy)$ must have at
least an endpoint of scale $+2$. This follows from the observation that all
field monomials associate to normal endpoints of scale less than $2$ contain
an even number of field $\psi_\o$, $\o=\pm 1$; hence, it is not possible to
build a Feynmann graph with no external lines and two source terms, one
proportional to $O_\xx^{(1,C)}$, which has two fields with the same $\o$, the
other proportional to $O_\xx^{(2,C)}$, which has two fields of opposite $\o$.

\subsection{Zero
temperature and thermodynamic limit for the density correlations}\lb{sec2.5den}

Using the above tree expansion, we can prove the existence of the zero
temperature and the thermodynamic limit for the density correlations. Let us
consider for definiteness the second term in \pref{2.176} and let us indicate
explicitly its $\b,L$ dependence
\be\sum_{h=h_{L,\b}}^0 \left[\frac{Z^{(\b,L)(1,C)}_h}{Z^{(\b,L)}_h}
\right]^2 R_1^{(\b,L)(h)}(\xx)
\ee
We want to show that, given $\e>0$, there exists $h_\e^*$ such that, if
$\min\{\b,L\}\ge h_\e^*$, then
\be\left| \sum_{h=h_{L,\b}}^0 \left[\frac{Z^{(\b,L)(1,C)}_h}{Z^{(\b,L)}_h}
\right]^2 R_1^{(\b,L)(h)}(\xx)- \sum_{h=-\io}^0 \left[\frac{\bar Z^{(1,C)}_h}{\bar Z_h}
\right]^2 R_1^{(h)}(\xx)\right|\le \e
\ee
where, with a notation similar to that used in Lemma \ref{lm2.5a}, we write
$\lim_{\min\{\b,L\}\to\io} Z_h^{(\b,L)}=\bar Z_h$ and
$\lim_{\min\{\b,L\}\to\io} Z_h^{(\b,L)(1,C)}=Z_h^{(1,C)}$, while $
R_1^{(h)}(\xx)=\lim_{\min\{\b,L\}\to\io} R_1^{(\b,L)(h)}(\xx)$; all these
limits do exist for the same arguments used in \S\ref{sec2.5en}. By using
that, by \pref{pr} and the analysis of the ren.c.'s given in \S 2.7,
$\frac{\bar Z^{(1,C)}_h}{\bar Z_h} R_1^{(h)}(\xx)$ and
$\frac{Z^{(\b,L)(1,C)}_h}{Z^{(\b,L)}_h} R_1^{(\b,L)(h)}(\xx)$ are bounded by
$C\g^{2h}$, there exists $h_\e$ such that
\be\left| \sum_{h=h_{L,\b}}^{h_\e} \left[\frac{Z^{(\b,L)(1,C)}_h}{Z^{(\b,L)}_h}
\right]^2 R_1^{(\b,L)(h)}(\xx)- \sum_{h=-\io}^{h_\e} \left[\frac{\bar Z^{(1,C)}_h}{\bar Z_h}
\right]^2 R_1^{(h)}(\xx)\right|\le {\e\over 2}
\ee
Moreover, by using the tree expansion described before \pref{2.175} and Lemma
\ref{lm2.5a}, together with the procedure described in its proof, we get

\be\left| \sum_{h=h_\e}^0 \left[\frac{Z^{(\b,L)(1,C)}_h}{Z^{(\b,L)}_h}
\right]^2 R_1^{(\b,L)(h)}(\xx)- \sum_{h=h_\e}^0 \left[\frac{\bar
Z^{(1,C)}_h}{\bar Z_h} \right]^2 R_1^{(h)}(\xx)\right|\le C
\sum_{h=h_\e}^0\g^{2h}\g^{-(h-h_\e^*)} \le {\e\over 2} \ee
for $h_\e^*$ large enough. We can proceed similarly for all the terms
appearing in \pref{2.175},\pref{2.176},\pref{2.177} and this concludes the
proof of the existence of the zero temperature and thermodynamic limits for
the density correlations

\subsection{Asymptotic behavior of the density correlations}\lb{sec2.5b}

We want now to discuss how we can derive from the form of the leading terms
in \pref{2.176} and \pref{2.177} the leading asymptotic behavior, as
described in \pref{asymp}. The idea is that, since $|g^{(h)}_{{\rm
D},\o}(\xx)| \le C_N \g^h [1+ (\g^h |\xx|)^N]^{-1}$, if $|\xx|\ge 1$, in the
sums over $h,h'$ of \pref{2.176} and \pref{2.177} the main contribution is
given by the terms with $|h|$ and $|h'|$ of the same size as $\log_\g |\xx|$.
Hence, one expects that the asymptotic behavior of $\O^{(i,C)}(\xx)$,
$i=1,2$, is the same of the function $\bar\O^{(i,C)}(\xx)$, obtained by the
substitutions of $\g^{-h}$ and $\g^{-h'}$ with $|\xx|$ in the asymptotic
expressions of the renormalization constants, given by \pref{2.53} and
\pref{2.56}, that is
\be\lb{substZ} \frac{[Z^{(i,C)}_{h\vee h'}]^2}{Z_h Z_{h'}} \rightarrow
|\xx|^{2(\h_{i,C}-\h_z)} \Big[ 1+f(\l) \log |\xx| \Big]^{2\lft(
q^{(h_\xx)}_{i,C}-q^{(h_\xx)}_z\rgt)} \ee
where the coefficients $q^{(h)}_t$ are defined as in \pref{2.88a}, $h_\xx=
\inf \{h:\g^h |\xx|\ge 1\}$, and, by \pref{init}, \pref{defa}, \pref{gtilde}
and Lemma \ref{lm2.6},
\be\lb{2.57a} f(\l)=\frac{a g_{1,j_0}}{\log\g} = \frac{2\l \hat v(2p_F)}{\p
v_F} + O(\l^{3/2}) \ee
In order to justify the substitution \pref{substZ}, let us put
$\h_i=2(\h_{i,C}-\h_z)$ and $q_i(\xx)$ any continuous interpolation between
$2[ q^{(h_\xx)}_{i,C}-q^{(h_\xx)}_z]$ and $2[ q^{(h_\xx-1)}_{i,C}
-q^{(h_\xx-1)}_z]$. Note that, thanks to the bounds \pref{wb} and \pref{wb1},
$q_i(\xx)$ is a bounded function of order $\l$, defined up to fluctuations
bounded, for $|\xx|\ge 1$, by $C\l [L(\xx) \log L(\xx)]^{-1}$, with $L(\xx)=
1+f(\l) \log |\xx|$; hence, its precise definition modifies the following
expressions only for a factor $1+O(\l)$. Let us now note that
\be\lb{2.61} \bsp &|\O^{(i,C)}(\xx) - \bar\O^{(i,C)}(\xx)| \le C_N
|\xx|^{\h_i-2} [1+f(\l) \log |\xx|]^{q_i(\xx)} \sum_{h,h'} \frac{\g^h |\xx|}{
1+ (\g^h |\xx|)^N} \frac{\g^{h'}
|\xx|}{ 1+ (\g^{h'} |\xx|)^N}\;\cdot\\
&\cdot\; \left| \frac{(\g^h |\xx|)^{\h_z} (\g^{h'} |\xx|)^{\h_z}} {(\g^{h\vee
h'}|\xx|)^{2\h_{i,C}}} \left[ \frac{L(|\xx|)}{L(\g^{|h|})}
\right]^{q^{(h)}_z} \left[ \frac{L(|\xx|)}{L(\g^{|h'|})} \right]^{q^{(h')}_z}
\left[ \frac{L(|\xx|)}{L(\g^{|h\vee h'|})} \right]^{-2q^{(h\vee h')}_{i,C}}
\frac{c_h c_{h'}}{\tilde c_{h\vee h'}^2} -1\right| \esp \ee
where
$$L(t)=1+f(\l)\log t \virg c_h = L(\g^{|h|})^{q^{(h)}_z}/\hat Z_h^{(z)} \virg \tilde
c_h = L(\g^{|h|})^{-q^{(h)}_{i,C}}/\hat Z_h^{(i,C)}$$
By \pref{2.56}, \pref{g1sum} and \pref{g2diff}, $c_h = 1 + O(\l^{1/2})$ and
$\tilde c_h =1 + O(\l^{1/2})$. On the other hand, if $r>0$ and $t\not=0$,
$$|r^t -1| \le |t \log r| (r^t + r^{-t})$$
and, if $q\not=0$,
$$\left| \left[ \frac{L(|\xx|)}{L(\g^{|h|})} \right]^{q} -1\right|
\le C_q \left[ |f(\l)\log(\g^h|\xx|)| + |f(\l)\log(\g^h|\xx|)|^{|q|+1}\right]$$
These two bounds, together with the bound
$$\sum_{h=-\io}^0 \frac{(\g^h r)^\a |\log(\g^h r)|^\b}{1+(\g^h r)^N} \le
C_{N,\a,q}$$
valid for any $\b$, $r>0$, $a>0$ and $N>\a$, imply that
\be\lb{2.183} |\O^{(i,C)}(\xx) - \bar\O^{(i,C)}(\xx)| \le C_N \l^{1/2}
|\xx|^{\h_i-2} [1+f(\l) \log |\xx|]^{q_i(\xx)} \ee
By the remark after \pref{gtilde}, the factor $\l^{1/2}$ can be improved up
to $\l^{1-\th}$, $\th<1$.

In order to complete the proof of \pref{asymp} in the case $\a=C$, we have
only to calculate $\bar\O^{(1,C)}(\xx)$ and $\bar\O^{(2,C)}(\xx)$. By using
\pref{2.48}, we see that $\h_{1,C}=\h_z$ and $q^{(h)}_{1,C}=q^{(h)}_z$, so
that, if we define $X_{2,C} =1-\h_{2,C}-\h_z$ and $\z_C(\xx)= 2[q_{2,C}(\xx)
- q_z(\xx)]$, we get
\be\bsp
\bar\O^{(1,C)}(\xx) &= 2\sum_\o g_{{\rm D},\o}(\xx) g_{{\rm D},\o}(\xx)\\
\bar\O^{(2,C)}(\xx) &= 4 |\xx|^{2(1-X_{2,C})} [1+f(\l) \log
|\xx|]^{\z_C(\xx)} g_{{\rm D},+}(\xx) g_{{\rm D},-}(\xx) \esp \ee
where $g_{{\rm D},\o}(\xx) = \sum_{h=-\io}^0 g^{(h)}_{{\rm D},\o}(\xx)$. On
the other hand, it is easy to see that, for any $N\ge 2$,
$$g_{{\rm D},\o}(\xx) = \frac{1}{2\p} \frac1{v_F x_0 + i\o x} + O(|\xx|^{-N})$$
It follows that, up to terms of order $|\xx|^{-2-\th}$ (as those coming from
$\O^{3,C}$),
\be \bar\O^{(1,C)}(\xx) = \frac1{\p^2 \tilde\xx^2} \bar\O_0(\xx) \virg
\bar\O^{(2,C)}(\xx) = \frac{L(\xx)^{\z_C(\xx)}}{\p^2 |\tilde\xx|^{2X_C}} \ee
where the functions $\bar\O_0(\xx)$ and $L(\xx)$ are defined as in Theorem
\ref{th1.1}. The functions $R_C(\xx)$ and $\tilde R_C(\xx)$, appearing in
\pref{asymp}, are defined in an obvious way in terms of the contributions of
order greater than $0$ in $\l$, which have the same asymptotic behavior of
the zero order terms, starting from \pref{pr} and \pref{2.183}. Hence, by
using \pref{2.42}, \pref{expt} and \pref{2.57a}, we get \pref{asymp} for
$\a=C$, together with the fact that $\z_C(\xx)=-3/2 + O(\l)$, in agreement
with \pref{zalfa}, and
$$X_C=1-c\l +O(\l^2) \virg
c=\lim_{\l\to 0} \frac{g_{2,-\io}}{2\p v_F\l} = \frac{2 \hv(0)- \hv(2p_F)}{2\p v_F}$$
Note also that, in Theorem \ref{th1.1}, we
have modified the function $f(\l)$ by erasing the terms of order greater than
$1$ in $\l$; the only effect of this modification is a change of the function
$\tilde R_C(\xx)$, which does not change its bound.

The proof of \pref{asymp} in the other cases is done in the same way. In
particular, in the case $\a=S_i$ we have to use again the bound \pref{2.48},
while the fact that there is no oscillating contribution to the leading term of
$\O_{TC_i}$ is due to the fact there is no local marginal term which can
produce it, by the remark after \S\ref{op2}.

\subsection{The two-point function}\lb{sec2.5g}

Let us now consider the two-point function $S_2(\xx-\yy)$. The proof of
\pref{s2x} can be done by using the same strategy. In this case, we have to
select the trees with two special endpoints of type $\h$ and fixed space-time
points $\xx$ and $\yy$, the first one associated to the $\h_\xx^-$ field, the
second to the $\h_\yy^+$ field; all the other properties, in particular the
definition of $v_\xx$, $v_\yy$ and $v_{\xx,\yy}$ are the same as before. Such
trees can be grouped in two classes: the first class contains the trees with
both special endpoints of scale $\le 1$, the second class contains the
remaining trees. As before, one can see that the second class is associated
with terms which decay faster than the leading ones; hence we analyze in
detail the trees of the first class and we shall call $\bar S_2(\xx-\yy)$
their contribution.

If one extracts from the first class the trees with no normal endpoints and
substitutes in their values the propagators $g^{(h)}_\o(\xx)$ with their
asymptotic expressions $g^{(h)}_{{\rm D},\o}(\xx)$, see \pref{gjth}, one gets
the following expression:
\be\lb{bS2x} \bar S_2(\xx-\yy) = \sum_\o e^{-i\o p_F(x-y)}\sum_{h=h_{L,\b}}^0
\frac1{Z_h} g^{(h)}_{D,\o}(\xx-\yy) + \sum_{h=h_{L,\b}}^0 \frac1{Z_h}
R^{(h)}(\xx-\yy)\ee
where ${Z_h}^{-1} R^{(h)}(\xx-\yy)$ is defined as the sum over all trees with
$n\ge 1$ normal endpoints, such that $h_{\xx,\yy}=h$ and $h_r=h_{L,\b}-1$, if
$h_r$ is the scale of the root, plus the corrections to the terms with no
normal endpoints. By proceeding as in the proof of \pref{pr}, we can show
that
\be\lb{prS} |R^{(h)}(\xx-\yy)|\le C_N (|\l| + \g^{\th h}) {\g^h\over
1+[\g^h|\xx-\yy|]^N}\ee
The ``extraction'' of the decaying factor $[1+[\g^h|\xx-\yy|]^N]^{-1}$ is
performed exactly as in the proof of \pref{pr}. After this operation, one can
bound the sum over all trees as in the proof of \pref{2.17aa}, by taking into
account that, in the crucial bound \pref{B.18caa}, the dimensional factor
$\g^{-D_{v_0}h_r}=\g^{-h_r}$ has to be multiplied by a factor $\g^{2h}$ to
compensate the ``missing integration'' (as in the case of $\O_C(\xx)$, see
above). Since $\g^{-h_r+2h}= \g^h \prod_{v_0<v\le v_{\xx,\yy}} \g^1$, this
implies that the bound \pref{B.18caa} has to be modified by substituting the
factor $\g^{-D_{v_0}h_r}$ with $\g^h$ and by adding $-1$ to the scaling
dimension of all vertices belonging to the path which connects $v_0$ with
$v_{\xx,\yy}$. Since the dimension of these vertices is $\ge 2$, we can
perform without any problem the sum over the scale indices by fixing $h$ in
place of $h_r$. As concerns the factors $1/\sqrt{Z_{h_\xx}}$ and
$1/\sqrt{Z_{h_\yy}}$ associated to the two special endpoints, one can use
\pref{ass} to ``change'' their scale to $h$, at the price of a innocuous
factor $e^{\frac12 c_1\e_0^2 [(h_\xx-h) + (h_\yy-h)]}$, which can be
distributed along the paths joining $v_\xx$ and $v_\yy$ with $v_{\xx,\yy}$,
by slightly modifying the factors $\g^{-(h_v-h_{v'})D_v}$ associated to the
corresponding non trivial vertices, see \pref{B.18caa}. In the case of the
corrections to the leading term, coming from the trees with no endpoints, one
has also a factor $\g^{\th h}$ coming from the bound \pref{2.30}. By using
the same arguments as in the case of $\O_C(\xx)$, is an easy exercise to show
that \pref{bS2x} can be rewritten in the form \pref{s2x}. Finally the proof of the existence
of the zero temperature and thermodynamic limit of $\bar S_2(\xx-\yy)$ can be done exactly as
for the density correlations.

\subsection{Borel summability}\lb{sec2.6}
First consider  the free energy $E(\l)$ \pref{fe}; we can
decompose it as $E(\l)= \sum_{h=-\io}^0 E_h(\l)$, where $E_h(\l)$
is the contribution of the trees whose root has scale $h$, hence
depends only on the running couplings $\vec v_j$ with scale $j>
h$. We will show that $E_h(\l)$, $h\le 0$ is analytic in the set
\be\lb{domh} D^{(h)}_{\e,\d}:= D_{\e,\d} \bigcup \lft\{\l\in\CCC:|\l| <
\frac{c_0} {1+|h|}\rgt\}\ee
and such that \be\lb{Lesn} |E_h(\l)| \le c_1 e^{-\k|h|} \ee
By using the Lemma at pag. 466 of \cite{Le087}, this property implies that
the perturbative expansion of $E(\l)$ satisfies the Watson Theorem, see pag.
192 of \cite{Hardy049}. Hence it is Borel summable in the usual meaning.

The tree expansion implies that there exists $\e_0$, such that, if
\be\lb{cond00} \bar\l_h = \max_{j\ge h}|\vec v_j|\le \e_0 \ee
then $|E_h| \le c_2 \g^{2h} \e_0$, with $c_2$ independent of $h$. The
analysis of \S\ref{sec2.2} implies that, given $\d\in(0,\p/2)$, there exists
$\e$ such that, if $\l\in D_{\e,\d}$, the condition \pref{cond00} is verified
uniformly in $h$; then it is easy to see that $E(\l)$ is analytic in $
D_{\e,\d}$ and continuous in its closure. The domain of analyticity of
$E_h(\l)$ is in fact larger; the form of the beta function immediately
implies that there exist two constants $c_3$ and $\bar c$ such that $\bar\l_0
\le c_3|\l|$ and, if $\bar\l_h\le \e_0$, then $\bar\l_{h-1} \le \bar\l_h +
\bar c \bar\l_h^2$; hence, if $c_3|\l| \le \min \{\e_0/2, 1/[4 \bar c
(|h|+1)]\}$, then, if $j>h$ and $\bar\l_j \le 2\bar\l_0$,
$$\bar\l_{j-1} \le \bar\l_0 + |j| \bar c \bar\l_j^2 \le \bar\l_0(1+ 4|j|\bar c \bar\l_0)
\le 2\bar\l_0$$
It follows that $E_h(\l)$ is analytic in the set \pref{domh}, with
$c_0=c_3^{-1} \min\{\e_0/2, 1/(4\bar c)\}$, and that $|E_h(\l)| \le c_1 \g^{2
h}$, with $c_1=c_2 \e_0$; hence $E(\l)$ satisfies \pref{Lesn} with
$\k=2 \log \g$.

Let us now consider the 2-point Schwinger function $S_2(\xx)$. The discussion
in \S\ref{sec2.5g} about its asymptotic behavior implies that we can write
$S_2(\xx) = \sum_{h=-\io}^0 S_{2,h}(\xx)$, where $S_{2,h}(\xx)$ is the
contribution of the trees whose root has scale $h$, and we can prove that, if
\pref{cond00} is verified (possibly with a smaller $\e_0$), then, for any
$N>0$,
\be
|S_{2,h}(\xx)| \le c_N\sum_{\bar h=h+1}^0 \g^{-\frac{\bar h-h}{2}}
\frac{\g^{\bar h}}{Z_{\bar h}} \frac1{1+(\g^{\bar h} |\xx|)^N}\virg
\lft|\frac1{Z_h} \rgt| \le \g^{|h|\over 4}
\ee
with $c_N$ independent of $h$. Hence, if we define $h_{\xx}$ so
that $\g^{h_\xx}|\xx|\in [1,\g)$, then, if $h_\xx>h$
\be |S_{2,h}(\xx)| \le c_2 \lft[ \sum_{\bar h=h+1}^{h_\xx} \g^{-{\bar
h\over 2}}\g^{{3\over 4}\bar h}\g^{h\over 2}+\sum_{\bar
h=h_{\xx}}^{0} \g^{-{\bar h\over 2}}\g^{{3\over 4}\bar
h}\g^{h\over 2}\g^{2(h_\xx-\bar h)} \rgt]\le \tilde c_2
\, \g^{\frac{h}2}\g^{\frac{h_\xx}4} \ee
and a similar bound holds for $h_\xx<h$ so that
\be |S_{2,h}(\xx)| \le c_s \g^{\frac{h}2} (1+|\xx|)^{-1/4} \ee
and we can proceed as in free energy case, so proving the analogous of
\pref{Lesn} for $S_2(\xx)$, with $c_1=c_s (1+|\xx|)^{-1/4}$ and
$\frac{\k}{\log \g}=\frac12$ (this value could be improved up to any value
smaller than $1$, at the price of lowering $\e_0$ down to $0$).

A similar argument can be used for the response functions.


\appendix

\section{Proof of \pref{ch}}
\lb{app0}

We have to prove that the r.h.s. of \pref{ch} and \pref{limN} are equal; thanks to
the antiperiodic condition, we can suppose that $|\t|\le \b/2$. Since the
sum over $k$ is a finite sum and $\chi(\g^{-M} k_0) = 1$ for $|k_0|\le \g^M$, it is sufficient
to prove that the function
\be \D_\b(\t) \= \frac1{\b} \sum_{k_0\in \DD_\b:|k_0|\ge \g^M} \chi(\g^{-M} k_0)
\frac{e^{-i k_0\t}}{-i k_0 +e(k)}
\ee
goes to $0$ as $M\to\io$, if $|\t|\le \b/2$. Since $\chi(t) =\chi(-t)$, we can write
\be\bsp
\D_\b(\t) &= \frac2{\b} \sum_{k_0\ge \g^M} \chi(\g^{-M} k_0)
\frac{k_0 \sin(k_0\t)}{k_0^2 +e(k)^2} +
\frac2{\b} \sum_{k_0\ge \g^M} \chi(\g^{-M} k_0)
\frac{e(k)\cos(k_0\t)}{k_0^2 +e(k)^2}\\
&\= \D_{\b,1}(\t) + \D_{\b,2}(\t)
\esp\ee
Note that $|\D_{\b,2}(\t)| \le C \int_{\g^M}^{\g^{M+1}} dk_0 \, k_0^{-2} \le C\g^{-M}$;
hence, $lim_{M\to\io} \D_{\b,2}(\t)=0$. Moreover, $\D_{\b,1}(0)=0$ and, if $\t=\pm\b/2$,
$\sin(k_0\t)=\pm(-1)^n$, if $n=(\b k_0)/(2\p)-1/2$. Hence, if we put
\be
F_M(k_0)\= \chi(\g^{-M} k_0) \frac{k_0}{k_0^2 +e(k)^2} \virg h\= \frac{2\p}{\b}
\ee
we get
\be |\D_{\b,1}(\pm\b/2)| \le
C \int_{\g^M}^{\io} dk_0 \left|\frac{F(k_0)-F(k_0+h)}{h}\right| \le C\g^{-M}
\ee

To get a similar bound for $\D_{\b,1}(\t)$, $\t\not=0,\pm \b/2$,
we have to use the oscillation properties of $\sin(k_0\t)$. Note that, if
$\sin(h\t)\not=0$,
\be \sin(k_0\t) = \frac{\cos(k_0\t -h\t) - \cos(k_0\t +h\t)}{2\sin(h\t)} \ee
On the other hand, if $k_0\in\DD_\b$, the same is true for $k_0\pm h$. Hence,
if we put $\bar k_0 \= \min \{k_0\ge \g^M\}$, we can write
\be\bsp
\D_{\b,1}(\t) &= \frac{h}{2\p \sin(h\t)} \Bigg[ \sum_{k_0\ge \g^M} \cos(k_0\t)
[F(k_0+h) -F(k_0-h)] + \\
&F(\bar k_0)\cos((\bar k_0-h)\t) - F(\bar k_0-h)\cos(\bar k_0\t)\Bigg]
\esp\ee
so that $|\D_{\b,1}(\t)| \le C h [2\p \sin(h\t)]^{-1} \g^{-M}$.\Halmos

\section{The $g_1$ map}
\lb{appA}

Let us consider the following map on the complex plane:
\be\lb{A1} g_{n+1}=g_n - a_n g_n^2 \ee
where $a_n$ is a sequence depending on $g_0$, such that, if $|g_0|$ is small
enough,
\be
\lb{A1a} a_n = a+\s_n\virg |\s_n|\le c_0 |g_0|\;,
\ee
for some positive constants $a$ and $c_0$. We want to study the trajectory of
the map \pref{A1}, under the condition that
\be g_0\in D_{\e,\d}=\{z\in\CCC: |z|<\e, |\text{Arg } (z)|\le \p-\d
\} \virg \d\in (0,\p/2) \ee
We shall first study the properties of a sequence $\tilde g_n$, which turns out
to be a good approximation of $g_n$.
Let us define:
\be\lb{A1d} A_n= \frac1n \sum_{k=0}^{n-1} a_k \ee

\begin{lemma}\lb{lmA1}

Given $\d \in (0,\p/2)$, there exists $\e_0(\d)$ such that, if $\e\le\e_0(\d)$
and $g_0\in D_{\e,\d}$, the sequence
\be\lb{Agn} \tilde g_n = \frac{g_0}{1+g_0 n A_n} \ee
at any step $n\ge 0$ is well defined  and does not exit the  larger domain
$D_{\e_1,\d_1}$, for $\e_1=2\e/(\sin\, \d)$ and $\d_1=\d/2$.
\end{lemma}

\0{\bf Proof} - First of all, we choose $\e$ so that
\be\lb{A4} c_0\e\le a/2\quad \Rightarrow\quad a/2\le \Re\, a_n \le 3a/2 \virg
|\Im\, a_n| \le c_0|g_0|
\ee
where $c_0$ is the constant defined in \pref{A1a}; we can write
\be A_n=\a_n + i \b_n \virg \a_n\ge a/2 \virg |\b_n| \le c_0|g_0|\;.
\ee
Define
$\tilde z_n := 1+g_0 n A_n := 1+g_0 n \a_n + \tilde w_n$; then, if $g_0\in
D_{\e,\d}$,
\be\lb{AA4} |1+g_0 n \a_n| \ge  \max \left\{\sin\d,\ \frac{\sin\ \d}{3}(1+|g_0| n\a_n)
\right\}\ee
In fact, it is trivial to show that $|1+g_0 n \a_n|\ge \sin\d$; on the other
hand, if $|g_0| n\a_n \ge 2$,

$$|1+g_0 n \a_n|\ge |g_0| n \a_n - 1 = (|g_0| n \a_n  + 2|g_0| n \a_n - 3)/3
\ge (|g_0| n \a_n  + 1)/3$$
By using \pref{AA4}, we get
\be\lb{A5} \frac{|\tilde w_n|}{|1+g_0 n \a_n|} \le \frac{6c_0}{a\sin\d}
|g_0|\;. \ee
It follows that, if $\e$ is small enough,
\be\lb{A3} |\tilde z_n| \ge  \frac12 \sin\d \ee
so that, in particular, the definition \pref{Agn} is meaningful.

Now we want to prove that $\tilde g_n\in D_{\e_1,\d_1}$, with $\e_1=2\e/(
\sin\,\d)$ and $\d_1 = \d/2$, if $\e$ is small enough. Let $g_0=\r_0 e^{i
\theta_0}$; by using \pref{AA4} and \pref{A5}, we see that, if $\e$ is small
enough,
\be\lb{A3b} |\tilde g_n| \le \frac{2\,|g_0|}{|1+\a_n g_0 n|} \le
\frac{2\e}{\sin \d}\;; \ee
besides it is easy to see that
$$\lft| \text{Arg } \left( \frac{g_0}{1+ \a_n g_0 n}
\right)\right| = \lft| \text{Arg } \left( \frac{\r_0}{e^{-i\theta_0}+ \a_n \r_0 n}
\right)\right|\le |\theta_0| \le \p-\d\;.
$$
Then, since $\tilde g_n = \frac{\r_0}{e^{-i\theta_0}+ \a_n \r_0 n} (1 +
w_n)$, with $w_n$ of order $g_0$, for $\e$ small enough,
\be\lb{A6} |\text{Arg } (\tilde g_n)|\le \p-\d/2
\ee
\Halmos

\begin{proposition}\lb{propA2}

Given $\d \in (0,\p/2)$, there exists $\e_0(\d)$, such that, if $\e\le\e_0(\d)$
and $g_0\in D_{\e,\d}$, then
\be\lb{A2c} g_n\in D_{\e_2,\d_2} \virg \e_2=\frac{3 \e}{\sin\, \d}
\virg \d_2=\frac{\d}4 \ee
Moreover, if $\tilde g_n$ is defined as in \pref{Agn},
\be\lb{A2b} |g_n - \tilde g_n| \le |\tilde g_n|^{3/2} \ee
\end{proposition}

\0{\bf Proof} - We shall proceed by induction on the condition \pref{A2b},
which is true for $n=0$. Suppose that it is true for $n\le N$; then, by using
\pref{A3b} and \pref{A6}, we see that, if $\e$ is small enough and $n\le N$,
\be\lb{A8}|g_n|\le 3|\tilde g_n|/2 \le 3\e/\sin \d\virg |\text{Arg }
(g_n)|\le \p-\d/4
\ee
which proves \pref{A2c}. Moreover, by \pref{A1}, if $\e$ is small enough,
\be\lb{A7} |g_{N+1}| \le 2|g_N| \le 3|\tilde g_N|\ee

Note now that
\be\lb{A11}
\frac{1}{g_{n+1}} - \frac{1}{g_n} = \frac{a_n}{1-a_n g_n} = a_n + a_n^2 g_n +
\D_n =\frac{1}{\tilde g_{n+1}} - \frac{1}{\tilde g_n} + a_n^2 g_n + \D_n
\ee
where $\D_n$ is a quantity which can be bounded by $c_1 |g_n|^2$, for some
constant $c_1$. We can rewrite \pref{A11} in the form
\be\lb{A10}
\frac{1}{g_{n+1}} - \frac{1}{\tilde g_{n+1}} = \frac{1}{g_n} - \frac{1}{\tilde
g_n} + a_n^2 g_n + \D_n
\ee
By using \pref{A4}, \pref{AA4}, \pref{A5}, \pref{A8}, \pref{A7} and \pref{A10},
we get, if $\e$ is small enough,
\be
\bsp &|g_{N+1} - \tilde g_{N+1} | = |g_{N+1}|\,|\tilde g_{N+1}|
\left| \frac1{g_{N+1}} - \frac1{\tilde g_{N+1}} \right| \\
\le 3|\tilde g_N|\,|\tilde g_{N+1}| &\sum_{n=0}^N [6 a^2 |\tilde g_n| + \frac94
c_1|\tilde g_n|^2] \le c_2|\tilde g_N|^{3/2} \frac{|g_0|^{1/2}}{(1+\frac{a}2
|g_0| N)^{1/2}} \sum_{n=0}^N \frac{|g_0|}{1+\frac{a}2 |g_0| n}\\
&\le |\tilde g_N|^{3/2} \frac{c_3|g_0|^{1/2}}{(1+\frac{a}2 |g_0| N)^{1/2}} \log
\lft(1+\frac{a}2 |g_0| N\rgt) \le |\tilde g_N|^{3/2}\esp
\ee
where $c_2$ and $c_3$ are two suitable constants.\Halmos

\section{Proof of the partial vanishing of the Beta function}\lb{appB}

In this appendix we want to prove the crucial bounds \pref{beta23} and
\pref{beta44}. This result will be achieved by comparing the beta function of
the Hubbard model at the IR scales with that of a reference model, which will
be studied in detail in the companion paper \cite{BFM012_2}. This model is
built as a perturbation of a Grassmannian-valued Gaussian measure with a
two-dimensional continuous field, whose propagator is of the same form as the
propagator \pref{gjth} on the IR scales, with $\xx$ varying on a continuous
square torus; the perturbation is given by an interaction which produces an
effective potential with a local part of the same form as that of the Hubbard
model, see \pref{blue}, with $\n_j=0$. The point is that, for certain values
of the parameters, we can control the beta function of this model by
exploiting carefully the {\em local gauge invariance} of its interaction.
This will be proved in \cite{BFM012_2}; here we discuss how we can use the
results of this paper together with the global symmetries of the model to
prove \pref{beta23} and \pref{beta44}. This strategy is a way to implement
the concept of {\it emerging symmetries} in a rigorous mathematical setting.

The effective model is expressed in terms of the following
Grassmann integral:
\be\lb{vv1} \bsp e^{\WW_{[l,N]}(\h,J)} &= \int\! P_Z(d\psi^{[l,N]})\; \exp
\left\{ -\tilde V(\sqrt{Z}\psi^{[l,N]}) + \sum_{\o,s} \int_{\L} d\xx\;
J_{\xx,\o,s}
\psi^{[l,N]+}_{\xx,\o,s} \psi^{[l,N]-}_{\xx,\o,s}\right.\\
& + \left. \sum_{\o,s} \int_{\L} d\xx\; \lft[\psi^{[l,N]+}_{\xx,\o,s}
\h^-_{\xx,\o,s} + \h^+_{\xx,\o,s} \psi^{[l,N]-}_{\xx,\o,s}\rgt]\right\}\;,
\esp \ee
where $\L$ is a square subset of of size $\tilde L$, $P_Z(d\psi^{[l, N]})$ is
the fermion measure with propagator
\be\lb{gth} g^{[l,N]}_{{\rm D},\o}(\xx)={1\over Z}{1\over \tilde
L^2}\sum_{\kk}e^{i\kk\xx}{\chi_{l,N}(|\tilde \kk|)\over -ik_0+\o c k} \virg
\tilde\kk = (ck, k_0) \virg c=v_F(1+\d)\ee
where $Z>0$ and $\d$ are two parameters and $\chi_{l,N}(t)$ is a smooth
compact support function defined for $t\ge 0$, equal to $1$ for $\g^{l}\le t
\le \g^{N}$ and vanishing for $t \le \g^{l-1}$ or $t \ge \g^{N+1}$; $\g^l$ is
the {\em infrared cut-off} and $\g^N$ is the {\em ultraviolet cut-off}. The
limit $N\to\io$, followed by the limit $l\to-\io$, will be called the {\it
limit of removed cut-offs}. We choose the cut-off function such that
\be\lb{sumscale} \chi_{l,N}(|\tilde \kk|)=\sum_{j=l}^N f_j(|\tilde
\kk|)\ee
with $f_j(|\tilde \kk|)=\chi(\g^j|\tilde
\kk|)-\chi(\g^{j+1}(\tilde\kk)$ and $\chi(t)$ is $C^\io(\RR^+)$
and such that $\chi(t)= 1$ if $t<1/\g$ and $=0$ if $|t|>1$;
therefore $f_j(|\tilde \kk|)$ has non vanishing support in
$\g^{j-1}\le |\tilde\kk|\le \g^{j+1}$.
The interaction is
\be\lb{fs1} \tilde V(\ps)=g_{1,\perp} V_{1,\perp}(\ps)+ g_\arp
V_\arp(\ps) + g_\erp V_\erp(\ps) + g_4 V_4(\ps) \ee
with
\be\lb{five1} \bsp V_{1,\perp}(\ps) &= \frac12 \sum_{\o,s}\int_{\L} d\xx d\yy
h_{\tilde L}(\xx-\yy)
\ps^+_{\xx,\o,s}\ps^-_{\xx,\o,-s} \ps^-_{\yy,-\o,s} \ps^+_{\yy,-\o,-s}\\
V_\arp(\ps) &= \frac12 \sum_{\o,s}\int_{\L} d\xx d\yy h_{\tilde L}(\xx-\yy)
\ps^+_{\xx,\o,s}\ps^-_{\xx,\o,s}\ps^+_{\yy,-\o,s}\ps^-_{\yy,-\o,s}\\
V_\erp(\ps) &= \frac12 \sum_{\o,s}\int_{\L} d\xx d\yy h_{\tilde L}(\xx-\yy)
\ps^+_{\xx,\o,s}\ps^-_{\xx,\o,s}\ps^+_{\yy,-\o,-s}\ps^-_{\yy,-\o,-s}\\
%
%
V_4(\ps) &= \frac12 \sum_{\o,s}\int_{\L} d\xx d\yy h_{\tilde L}(\xx-\yy)
\ps^+_{\xx,\o,s}\ps^-_{\xx,\o,s}\ps^+_{\yy,\o,-s}\ps^-_{\yy,\o,-s} \esp \ee
where $h_{\tilde L}(\xx)$ is defined in the following way. Let us take a
smooth function $\hat h(\pp)$, defined on $\RRR^2$ and rotational invariant,
such that $|\hat h(\pp)|\le C e^{-\m |\pp|}$ for some positive $C$ and $\m$,
and $\hat h(0)=1$; moreover, let us call $\DD_{\tilde L}$ the set of
space-time momenta $\kk=(k,k_0)$, with $k={2\pi\over \tilde L}n$ and
$k_0={2\pi\over \tilde L} n_0$. Then
\be\lb{defh} h_{\tilde L}(\xx) :={1\over \tilde L^2}\sum_{\pp\in \DD_{\tilde
L}} \hat h(\pp) e^{i\pp\xx}\ee
We write
\be g^{[l,N]}_{{\rm D},\o}(\xx)=\sum_{j=l}^N g^{(j)}_{{\rm
D},\o}(\xx) \ee
where $g^{(j)}_{\o}(\xx)$ is defined as $g^{[l,N]}_{\o}(\xx)$ with
$\chi_{l,N}(|\tilde \kk|)$ replaced by $f_j(|\tilde \kk|)$, see
\pref{sumscale}.

\*

The multiscale analysis of \cite{BFM012_2} shows that, even if the free
propagator has the same UV singularity of the Thirring model, the integration
of the UV scales is not problematic, since the interaction is not local.
As concerns the integration of the infrared scales, it can be
done in a way similar to the one in the Hubbard model described in
\S\ref{sec2}, which we shall refer to for the notation.

However, before starting the multiscale IR integration, we have to perform
some technical operations, which will make possible to compare the flow of
the running couplings with that of the Hubbard model. After the integration
of the UV scales up to $j=1$, the free measure propagator is given by
$g^{[l,1]}_{{\rm D},\o}(\xx)$, defined as in \pref{gth} with $N=1$. In this
expression, the velocity $c$ has the role of the Fermi velocity $v_F$ of the
Hubbard model. In order to match the asymptotic behavior of the two models,
we can not choose $c=v_F$; for this reason we introduced the parameter $\d$.
However, it is not possible to compare the RG flows of the two models if the
two velocities are different; hence, we have to move from the free measure to
the interaction the term proportional to $\d$. Moreover, since also the
cutoff function $\chi^{[l,1]}(|(k_0, c k)|)$ depends on $\d$, we have to
``modify'' it in $\chi_{[l,1]}(|(k_0, v_F k)|)$. A simple way to perform
these operations without introducing spurious singularities is described in
\cite{BFM012_2}; we shall omit the technical details, which are not important
in the following discussion. The final result is that, up to negligible
differences for $j=0$ and $j=l$, the effective potential is the same we
should get if the propagator of $\psi^{[l,0]}$ were equal to
$${1\over Z}{1\over L^2} \sum_{\kk} e^{i\kk\xx} {\chi_{l,0}(|(k_0, v_F k)|)\over
-ik_0+\o v_F k}$$
so that the renormalized single scale propagator will have the form
corresponding to the leading behavior of the single scale propagator in the
Hubbard model, see \pref{ne1}.

Let us now analyze in more detail the RG flow of the effective model for
$j\le 0$. The main difference with respect to the Hubbard model is that
\pref{rel} has to be replaced by
\be\lb{fs2a} \bsp \tilde V^{(j)}(\sqrt{Z_j}\ps) &= g_{1,\perp,j}
F_{1,\perp}(\sqrt{Z_j}\ps)+
g_{\arp,j} F_\arp(\sqrt{Z_j}\ps) +\\
+ g_{\erp,j} F_\erp(\sqrt{Z_j}\ps) &+ F_4(\sqrt{Z_j}\ps) + \d_j
V_\d(\sqrt{Z_j}\ps)\esp \ee
where the functions $F_\a(\psi)$ are defined as the functions $V_\a(\psi)$ of
\pref{five1}  with $\d(\xx-\yy)$ in place of $h_{L}(\xx-\yy)$; the absence of
local terms proportional to $\psi^+\psi^-$ is a consequence of the oddness in
$\kk$ of the free propagator. The running couplings verify equations of the
form
\be\lb{bf} \bsp g_{\a,h-1}-g_{\a,h} &= B_{\a}^{(h)}\lft(\vec g_h, \d_h,..\vec
g_0,
\d_0, \vec g, \d\rgt)\\
\d_{h-1}- \d_h &= B_{\d}^{(h)}\lft(\vec g_h, \d_h,..\vec g_0, \d_0, \vec g,
\d\rgt)\esp \ee
where $\a=(1,\perp),\arp,\erp,4$ and $\vec g_j=(g_{1,\perp,j}, g_{\arp,j},
g_{\erp,j}, g_{4,j})$. Note that the functions $B_{\a}^{(h)}$ and
$B_{\d}^{(h)}$ are of the second order in their arguments; in the case of
$B_{\d}^{(h)}$, this follows from the structure of $\tilde V(\psi)$ (see
\pref{five1}), which does not allow us to build Feynmann graphs of the first
order in $\vec g$. For the same reason
\be \d_0=\d + O(\e_0^2)\virg \e_0=\max\{|\vec g|,|\d|\}\ee
and this relations can be inverted, if $\e_0$ is small enough.


There are some symmetries which is important to exploit. For notational
simplicity, we will write $(G_{1,\perp}, G_\arp,G_\erp, G_4, \D)$ or $(\vec
G,\D)$ in place of $\lft(\vec g_h, \d_h,..\vec g_0, \d_0, \vec g, \d\rgt)$ .

\bd \item{a.}  {\it Spin U(1).} Both the free measure and the interaction are
invariant under the transformation
$$
\ps^\e_{\xx,\o,s}\to e^{i\e\a_s}\ps^\e_{\xx,\o,s}
$$
where $\a_s$ is a spin-dependent angle. This means that the local part of the
effective interaction only contains terms which have as many $\ps^+_s$ as
$\ps^-_s$, for each given $s$. Moreover, it is clear from the symmetries
$\o\to -\o$ and $s\to -s$ that all the terms must occur in the same linear
combinations of \pref{five1}.

\item{b.} {\it Vector-Axial Symmetry.} Both the free measure and the
    interaction are
    invariant under the transformation
\be\lb{u1} \ps^\e_{\xx,\o,s}\to e^{i\e\th_{\o,s}}\ps^\e_{\xx,\o,s} \ee
with $\a_{\o,s}$ dependent on $\o$ and $s$. All the interaction terms in
\pref{five1} are invariant but $V_{1,\perp}$. However, if $g_{1,\perp}=0$, it
is easy to see, by a graph by graph analysis, that a term of this type can
not be generated by the other ones; hence, the function $B_{1,\perp}^{(h)}$
must be odd in $g_{1,\perp}$:
\be\lb{r13} B_{1,\perp}^{(h)}(G_{1,\perp},G_\arp,G_\erp, G_4, \D) =
G_{1,\erp} \bar B^{(h)}(G^2_{1,\perp},G_\arp,G_\erp, G_4, \D)\ee
where $G_\a^2$ denotes the tensor $\{g_{\a,j} g_{\a,j'}\}_{j.j'\ge h}$ and
$G_\a \bar B_\a^{(h)}$ is a shorthand for $\sum_{j\ge h} g_{\a,j}
B_\a^{(h,j)}$. In particular, this implies that, if $g_{1,\perp}=0$, then
$g_{1,\perp,j}=0$, that is the surface $\CC_1=\{\vec g,\d :g_{1,\perp}=0\}$,
in the space of the interaction parameters $(\vec g,\d)$, is invariant.

In the same manner, it is easy that the other $B_\a^{(h)}$ functions are even
in $g_{1,\perp}$:
\be\lb{ralfa} B_\a^{(h)}(G_{1,\perp},G_\arp,G_\erp, G_4, \D) = \bar
B_\a^{(h)}(G^2_{1,\perp},G_\arp,G_\erp, G_4,\D) \virg \a=\arp,\erp,4,\d\ee

\item{c.} {\it Spin SU(2).} It is convenient to rewrite the interaction as
\be\lb{fs2} \bsp \tilde V(\ps) &= g_{1,\perp}
\lft(V_{1,\perp}(\ps)-V_\arp(\ps)\rgt) +
(g_\arp+g_{1,\perp}-g_\erp) V_\arp(\ps) +\\
&+g_\erp \lft(V_\erp(\ps)+ V_\arp(\ps)\rgt) + g_3 V_3(\ps) + g_4
V_4(\ps) + \d_h V_\d(\ps)\esp \ee
It is evident that $V_{1,\perp}-V_\arp$, $ V_\erp+ V_\arp$, $V_3$
$V_4$ and $V_\d$, as well as the free measure, are invariant under
the transformation of the fields
$$
\hp^-_{\kk,\o,s}\to \sum_{s'}U_{s,s'}\hp^-_{\kk,\o,s'} ,\quad
\hp^+_{\kk,\o,s}\to \sum_{s'}\hp^+_{\kk,\o,s'} U^\dag_{s',s}
$$
for $U\in SU(2)$. While $V_\arp$ isn't: if $g_\arp+g_1-g_\erp=0$ it will
remain zero. Thus we find two others invariant surfaces:
$$
C_{1,+}=\{\vec g,\d:g_{1,\perp}=g_{\erp}-g_{\arp}\} \virg
C_{1,-}=\{\vec g ,\d:-g_{1,\perp}=g_{\erp}-g_{\arp}\}
$$

\ed

\vspace{.3cm}

Finally we consider the flow of $Z_h$ and the renormalization
constant $Z^{(1)}_h$ associated with the density operator
$\r_{\xx,\o,s}=\psi^+_{\xx,\o,s} \psi^-_{\xx,\o,s}$ in the
generating functional \pref{vv1}; $Z^{(1)}_h$ is defined as
$Z^{(1,C)}_h$ in \pref{2.20}. It is easy to see, by using the
symmetry properties of the model as before, that $Z_{h-1}/Z_h = 1+
B_z^{(h)}(\vec G, \D)$ and $Z^{(1)}_{h-1}/Z^{(1)}_h =
1+B_\r^{(h)}(\vec G, \D)$, with $B_\a^{(h)}(\vec G, \D) = \bar
B_\a^{(h)}(G^2_1,G^2_3,G_\arp,G_\erp, G_4,\D)$ for $\a=z,\r$.
Hence
\be \frac{Z^{(1)}_{h-1}}{Z_{h-1}} = \frac{Z^{(1)}_h}{Z_h} [1 +
\tilde B^{(h)}(\vec G,\D)] \ee
with \be\lb{ralpha1} \tilde B^{(h)}(\vec G,\D)=\bar
B^{(h)}(G^2_{1,\perp},G^2_3,G_\arp,G_\erp, G_4,\D) \ee

Let us now consider the Hubbard model. In \pref{bal} we have written its Beta
function as sum of two terms, the second of which is asymptotically
negligible, by \pref{2.34}; the first term, denoted in \pref{bal} by
$\b^{(j)}_\a(\vec g_j,\d_j;...;\vec g_0,\d_0)\equiv
\b^{(j)}_\a(G_{1},G_{2},G_{4},\D)$ coincides with the Beta function of the
effective model on the invariant surface $\CC_{1,+}$, if we subtract from it
the contribution of the trees containing endpoints of scale grater than $0$
and we interpret everywhere the integrals over the space-time variables,
which in the Hubbard model case are a shorthand for $\sum_{x\in\CC}
\int_{-\b/2}^{\b/2}$ as the integrals over $\L$, with $\tilde L^{-1} = \max
\{L^{-1}, \b^{-1}\}$. As discussed in \S4.6 of \cite{BM001}, this
modification produces an error of order $\e_j^2 \g^{\th j}$.

Hence, by using \pref{ralfa} and \pref{r13}, we get
\be\lb{C.17} \bsp \b_1^{(j)}(G_1,G_2,G_4,\D) &= G_1 \bar
\b_{1\perp}^{(j)}(G_1^2,G_2-G_1,G_2,G_4,\D) + O(\e_j^2 \g^{\th j})\\
\b_2^{(j)}(G_1,G_2,G_4,\D) &= \bar
\b_{\perp}^{(j)}(G_1^2,G_2-G_1,G_2,G_4,\D) + O(\e_j^2 \g^{\th j})\\
\b_4^{(j)}(G_1,G_2,G_4,\D) &= \bar
\b_4^{(j)}(G_1^2,G_2-G_1,G_2,G_4,\D) + O(\e_j^2 \g^{\th j})\\
\b_\d^{(j)}(G_1,G_2,G_4,\D) &= \bar \b_\d^{(j)}(G_1^2,G_2-G_1,G_2,G_4,\D) +
O(\e_j^2 \g^{\th j}) \esp \ee
where $\bar \b_\a^{(j)}(G_1^2,G_2-G_1,G_2,G_4,\D)$ denotes the value of $\bar
B_\a^{(j)}(G_1^2,G_2-G_1,G_2,G_4,\D)$, after the subtraction of the trees
containing endpoints of scale grater than $0$.

Therefore, if $\a\not=1$, the contributions of order $0$ and $1$ in $G_1$ of
$\b_\a^{(j)}(\vec G,\D)$ are the same as the contributions of the same order
of $\bar \b_\a^{(j)}(0, G_2-G_1, G_2, G_4, \D)$. On the other hand, in the
following paper, see (4.44) of \cite{BFM012_2}, we will prove that, if we
call $b_{\a}^{(j)}(\bar g_\arp, \bar g_\erp, \bar g_4, \bar\d)$ the value
taken by $\bar \b_\a^{(j)}(0,G_\arp,G_\erp, G_4,\D)$ when $(g_{\arp,j},
g_{\erp,j}, g_{4,j}, \d_j)=(\bar g_\arp, \bar g_\erp, \bar g_4, \bar\d)$ for
any $j$, then, in the limit $\tilde L,N=\io$,
\be\lb{van} \lft|b_{\a}^{(j)}(\bar g_\arp, \bar g_\erp, \bar g_4, \bar\d)
\rgt| \le C [\max \{ |\bar g_\arp|, |\bar g_\erp|, |\bar g_4|, |\bar \d|
\}]^2 \g^{\th j}\virg \a=\arp, \erp,4,\d \ee
The contributions of order $i=0,1$ of the functions $b_{\a}^{(j)}(\vv)$
coincide with the functions $b_{\a,i}^{(j)}(\vv)$ of the Hubbard model, up to
the corrections described in \pref{C.17}, if we take the limit $L,\b\to\io$.
It is easy to see that this implies a difference of order $|\vv|^2
\g^{-(j-h_{L,\b})}$, so that we get the bound \pref{beta23}.

In we define in a similar way the function $\tilde b^{(j)}(g_\arp, g_\erp,
g_4, \d)$ in terms of $\bar B^{(j)}(0,G_\arp,G_\erp, G_4,\D)$, in
the following paper, see (4.44) 0f \cite{BFM012_2}, it is also proved that
\be\lb{vanZ} \lft|\tilde b^{(j)}(g_\arp, g_\erp, g_4, \d) \rgt|
\le C [\max \{ |g_\arp|, |g_\erp|, |g_4|, |\d| \}]^2 \g^{\th j}\ee
which implies \pref{beta44}.

Note that, in order to prove \pref{beta23}, which is a crucial ingredient in
the proof of the boundedness of the flow of the spin-symmetric Hubbard model,
we need information from a non spin symmetric model; in fact, we have derived
\pref{beta23} from the model \pref{vv1} with $g_{\erp}\not=g_{\arp}$ and
$g_{1,\perp}=0$.

\vskip.3cm {\it Acknowledgements} G.B. and V.M. acknowledge the
financial support of MIUR, PRIN 2008. V.M. gratefully acknowledges
also the financial support from the ERC Starting Grant
CoMBoS-239694.

\end{document}